\documentclass[journal]{IEEEtran}

\usepackage{amsmath,mdwlist,amssymb,amsfonts,amsthm,mathrsfs}
\usepackage{bm,cite}

\usepackage{epsfig} 
\usepackage{graphicx,latexsym} 
\usepackage{xcolor,color,soul}
\usepackage{threeparttable}
\usepackage{tabularx,multirow}

\begin{document}
\title{High-Sensitivity Electric Potential Sensors for Non-Contact Monitoring of Physiological Signals}

\author{Xinyao Tang,~\IEEEmembership{Student Member,~IEEE,}
        Wangbo Chen,~
        Soumyajit Mandal,~\IEEEmembership{Senior Member,~IEEE,}\\
        Kevin Bi,~
        and~Tayfun Ozdemir,~\IEEEmembership{Senior Member,~IEEE}
\thanks{X. Tang and W. Chen are with the Department
of Electrical, Computer, and Systems Engineering, Case Western Reserve University, Cleveland, OH 44106, USA. e-mail: \{xxt81, wxc342@case.edu\}@case.edu.}
\thanks{S. Mandal is with the Department
of Electrical and Computer Engineering, University of Florida, Gainesville, FL 32611, USA. e-mail: s.mandal@ufl.edu.}
\thanks{K. Bi and T. Ozdemir are with Virtual EM, Inc., Ann Arbor, MI 33174, USA. e-mail: \{kbi, tayfun\}@virtualem.com.}
\thanks{This work was funded by an U.S. Air Force STTR Phase I Contract awarded to Virtual EM Inc. (Contract No. FA8650-19-P-6132).}
\thanks{Manuscript received November 25, 2020.}}

\markboth{IEEE Transactions on Measurements and Instrumentation,~Vol.~X, No.X, XX~2020}%
{Tang \MakeLowercase{\textit{et al.}}: Highly-Sensitive Electric Potential Sensors}
\maketitle

\begin{abstract}
The paper describes highly-sensitive passive electric potential sensors (EPS) for non-contact detection of multiple biophysical signals, including electrocardiogram (ECG), respiration cycle (RC), and electroencephalogram (EEG). The proposed EPS uses an optimized transimpedance amplifier (TIA), a single guarded sensing electrode, and an adaptive cancellation loop (ACL) to maximize sensitivity (DC transimpedance $=150$~G$\Omega$) in the presence of power line interference (PLI) and motion artifacts. Tests were performed on healthy adult volunteers in noisy and unshielded indoor environments. Useful sensing ranges for ECG, RC, and EEG measurements, as validated against reference contact sensors, were observed to be approximately 50~cm, 100~cm, and 5~cm, respectively. ECG and RC signals were also successfully measured through wooden tables for subjects in sleep-like postures. The EPS were integrated with a wireless microcontroller to realize wireless sensor nodes capable of streaming acquired data to a remote base station in real-time.
\end{abstract}


%
\IEEEpeerreviewmaketitle

\section{Introduction}
\IEEEPARstart{C}{ontact}-based wearable physiological sensors, such as skin-coupled electrocardiogram (ECG) monitors, respiration belts, electroencephalogram (EEG) headbands, and smart watches~\cite{apple}, offer great promise for personalized healthcare~\cite{majumder2017wearable,heikenfeld2018wearable}. Meanwhile, the contacted-based sensing systems usually provide a good signal-to-noise ratio (SNR), simply due to its directly contacting the monitored subject, thus reducing the interface circuit complexity. However, they suffer from challenges during long-term use due to issues with patient comfort, security and privacy, engagement and interaction, and psychological burden~\cite{dunn2018wearables,baig2017systematic}. For example, wearables can induce subjects to change their behavior for the worse, become more anxious about their health, self-diagnose problems, become addicted to the device, or place too much trust in its data~\cite{schukat2016unintended}. Contact sensors can also interfere with natural sleep and thus bias the results of sleep studies~\cite{fallmann2019computational}. One example is the well-known ``first-night'' effect on polysomnography (PSG)~\cite{newell2012one}, in which normal sleep structure is significantly altered during the first few nights of a study due to the discomfort of the electrodes, movement limitations due to gauges and cables, and the psychological consequences of being under scrutiny~\cite{le2001first,le2000mild}. In some cases, contact sensors can even damage the skin, for example while monitoring infants in the NICU~\cite{atallah2014unobtrusive}. Finally, contact sensors are not useful when the person to be monitored is not wearing the device, e.g., during battery charging or in a search-and-rescue situation where the goal is to detect survivors. Thus, a variety of methods have been proposed for unobtrusive non-contact physiological monitoring~\cite{steffen2007mobile,shao2015noncontact,uenoyama2006non}. We here define the non-contact sensing as measuring physiological signals with an off-body air gap and classify the contact sensing as signal detection using gel, spiky, dry (directly skin contact but gel-less), and insulated electrodes\cite{chi2010dry}. One limitation for the non-contact method is that sensors need to be placed in different locations, such as bed, wheelchair, and automobiles in order to monitor health state in people's daily life. Therefore, low power, low cost, easy deployment and relative large sensing range is a necessity for ubiquitous monitoring.


Key metrics for non-contact monitoring include safety, power consumption, useful sensing distance, sensitivity to subject position and environmental conditions, ability to simultaneously sense multiple subjects, and installation/maintenance costs. Table~\ref{table:summaryEPS} summarizes qualitative values of these metrics for the main non-contact sensing modalities used in cardiopulmonary monitoring, which is a major application due to the widespread and growing prevalence of cardiovascular diseases (CVDs) worldwide~\cite{world2016cardiovascular,hong2019wearable}. The relative performance of the proposed modalities can be compared as follows: (i) \underline{Active Energy Injection}: Sensors that actively inject energy into the subject, such as radar/laser or impedance-based methods, must be assessed for safety and operating risks. Passive methods are intrinsically safe; they also have advantages of portability and low power consumption. (ii) \underline{Distance}: PPG, BCG/SCG, and cECG require direct mechanical coupling with the subject, so the sensing range is limited to a few mm~\cite{Gao2020,Ozkan2020}. However, electric potential sensors (EPS) and impedance methods operate up to the cm range, while radar/laser, video, and thermography-based analysis can measure cardiopulmonary signals up to the meter range. (iii) \underline{Position Sensitivity}: All non-contact modalities suffer from fluctuations in signal quality due to subject motion (variable position and/or orientation) and environmental changes, which limits their diagnostic utility. Non-directional methods such as thermography and EPS are less sensitive to these effects. (iv) \underline{Cost}: PPG, cECG, and EPS sensors have the lowest cost, since they use relatively simple low-power circuits compared to radar/laser, video motion, and thermography-based systems.

\begin{table*}[ht]
    \centering
	\caption{Summary of prior work on non-contact sensing of cardiopulmonary signals}
	\begin{tabularx}{0.90\linewidth}{ c l c l l l c}
	\hline
	\hline
    \textbf{Method} & \textbf{Sensing principle} & {\textbf{Active energy}} & {\textbf{Distance}} & \textbf{Position Sensitivity} & \textbf{\# of monitored} & \textbf{Costs} \\
    \hline
    PPG~\cite{allen2007photoplethysmography} & Photon absorption & Yes & $<$mm & o & 1 & $\downarrow$ \\
    Radar/Laser~\cite{xiao2007portable,aubert1984laser} & Velocity, Displacement &  Yes & m & $\downarrow$ & $>$1 & $\uparrow$ $\uparrow$\\
    Impedance~\cite{griffiths2001magnetic} & Conductivity, Permittivity & Yes & cm & o & 1 & o \\
    Video Motion~\cite{nakajima1997detection} & Displacement & No & m & $\downarrow$ & $>$1 & $\uparrow$ $\uparrow$ \\
    BCG/SCG~\cite{bruser2011adaptive} & Force, Displacement & No & m & $\downarrow$ & 1 & o \\
    Thermography~\cite{fujimasa2000converting} & Temperature & No & cm & $\downarrow$ $\downarrow$ & $>$1 & $\uparrow$ $\uparrow$ \\
    cECG~\cite{chi2010dry,chen2019400} & Bioelectricity & No & $<$mm & $\uparrow$ $\uparrow$ & 1 & $\downarrow$ \\
    EPS~\cite{harland2001electric,tang2019indoor} & Bioelectricity, Displacement & \bf{\textcolor{blue}{No}} & \bf{\textcolor{blue}{cm}} & \bf{\textcolor{blue}{$\downarrow$ $\downarrow$ }}& 1 & \bf{\textcolor{blue}{$\downarrow$}} \\
    \hline
    \hline
    \end{tabularx}
    \vskip 0.5ex
    {\textbf{PPG}:~Photoplethysmography;~\textbf{BCG}:~Ballistocardiography;~\textbf{SCG}:~Seismocardiography; \textbf{cECG}:~Capacitive Electrocardiography;\\
    \textbf{EPS}: Electric Potential Sensing;~\textbf{o}: median; $\uparrow$: high; $\downarrow$: low. \par} 
\label{table:summaryEPS}
\end{table*}

The discussion above suggests that passive EPS have the advantages of low power, relatively long range, low position sensitivity, and low cost. We have earlier demonstrated low-power EPS for reliable capacitive measurement of respiration rate (RR) at distances up to $\sim$1.5 m in unshielded indoor environments~\cite{tang2019indoor}. The sensors can also be used for accurate motion estimation, human localization, and tracking at distances up to several meters. In this paper, we build on our prior work by developing passive EPS with improved sensitivity for non-contact monitoring of multiple physiological signals. Besides non-contact respiration cycle (RC) and RR sensing, the proposed EPS can also detect electrocardiograms (ECG) at distances up to $\sim$0.5~m in noisy unshielded rooms, thus enabling non-contact measurements of heart rate (HR) and HR variability (HRV). By contrast, the current state-of-art non-contact EPS can only detect ECG up to 0.1 $\sim$ 0.3~m in electromagnetically-shielded environments~\cite{harland2001electric,harland2002remote}. This makes it difficulty for the ubiquitous monitoring in daily life, where power line interference (PLI) becomes a critical issue. Our sensors can also be used for non-contact detection of other electrical biosignals such as electromyograms (EMG) and electroencephalograms (EEG). Such capabilities are of interest since, by contrast with the wide range of non-contact cardiopulmonary monitoring methods shown in the table, non-contact detection of EMG and EEG is much less common. In fact, earlier efforts have been restricted to distances less than 0.3~cm from the body surface~\cite{harland2002remote,chi2010wireless,pino2018wearable}. Similarly to contacted-based biopotential sensing, our proposed EPS suffer motion artifacts issues. The small coupling capacitance and high relative movement between skin and electrode lead to more difficulties in sensing electrode and analog front end (AFE) design than wet/dry contact. We summarize the paper's main contributions as follows:


The rest of the paper is organized as follows. Section~\ref{sec:sensor_design} describes the design of the proposed multi-functional EPS for non-contact measurements of both respiration and electrical biosignals. Experimental results on human subjects are presented in Section~\ref{sec:expt_results}, while Section~\ref{sec:conclusion} concludes the paper.

\section{Sensor Design}
\label{sec:sensor_design}

\subsection{System Overview}
\begin{figure}[htbp]
	\centering
	\includegraphics[width=0.27\columnwidth]{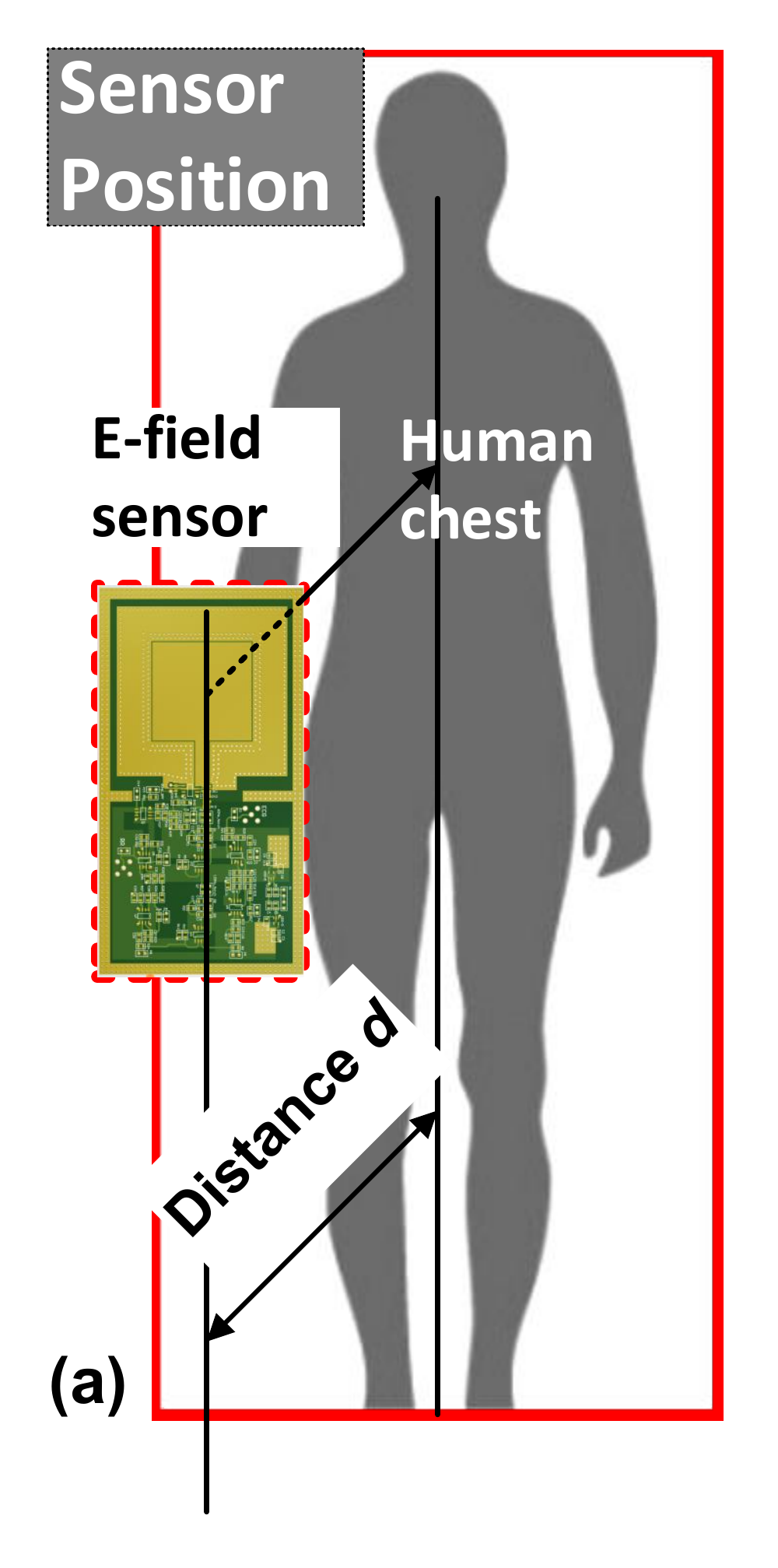}
	\includegraphics[width=0.71\columnwidth]{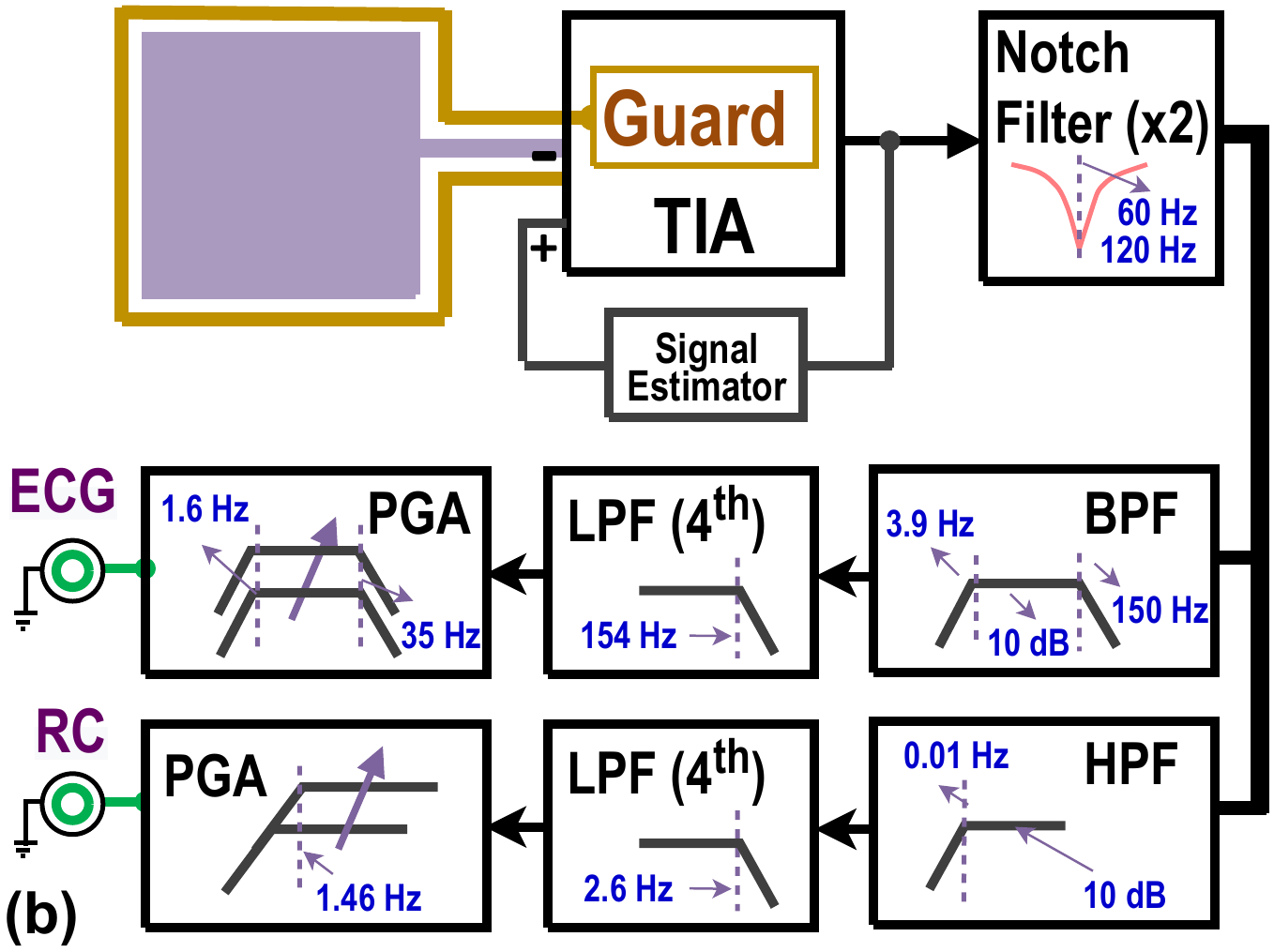}
	\caption{{(a) Typical geometry used for non-contact electric potential sensing (EPS) of cardiopulmonary signals.} (b) System overview of the proposed EPS, including active PLI cancellation at the TIA stage.}
	\label{fig:remote_block_a}
\end{figure}

The proposed not-contact sensing system uses a network of low-cost wireless nodes based on passive EPS. Each node also contains i) an off-the-shelf microcontroller (MCU) for on-board signal processing and power management, and ii) a low-power radio for wireless networking with a remote base station. By default, the plane of the EPS electrode is aligned with the chest surface and placed a distance $d$ away from it for cardiopulmonary sensing, as shown in Fig.~\ref{fig:remote_block_a}(a). In addition to physiological signals, the EPS can also sense indoor occupancy patterns and human crowd dynamics~\cite{tang2019indoor}. 

For convenience, let us denote the local power line interference (PLI) frequency (50 or 60 Hz) as $f_0$. Fig.~\ref{fig:remote_block_a}(b) shows a high-level block diagram of a single non-contact EPS, which consists of the following main parts: i) sensor with active guard electrode; ii) transimpedance amplifier (TIA) with feedback band-pass filter (BPF) tuned to $f_0$; iii) notch filters tuned to $f_0$ and $2f_0$; and iv) additional BPFs, adjustable gain stages (VGAs), and low-pass filters (LPFs) optimized to measure the two main output signals of interest, namely the ECG and RC. Note that ECG and RC signals are read out through parallel signal channels to enable their gains and bandwidths to be individually optimized. In particular, the LPFs define the final bandwidth of each channel; these are typically set to 154~Hz for ECG and 2.6~Hz for RC, respectively. {Clearly, PLI can be readily filtered out from the RC signal} (as long as it is not large enough to saturate the TIA or other stages), but will be a problem for the ECG. The wireless communication system has been designed to support two such channels.

\subsection{Choice of TIA Architecture}
\begin{figure}[htbp]
    \centering
    \includegraphics[width=0.99\columnwidth]{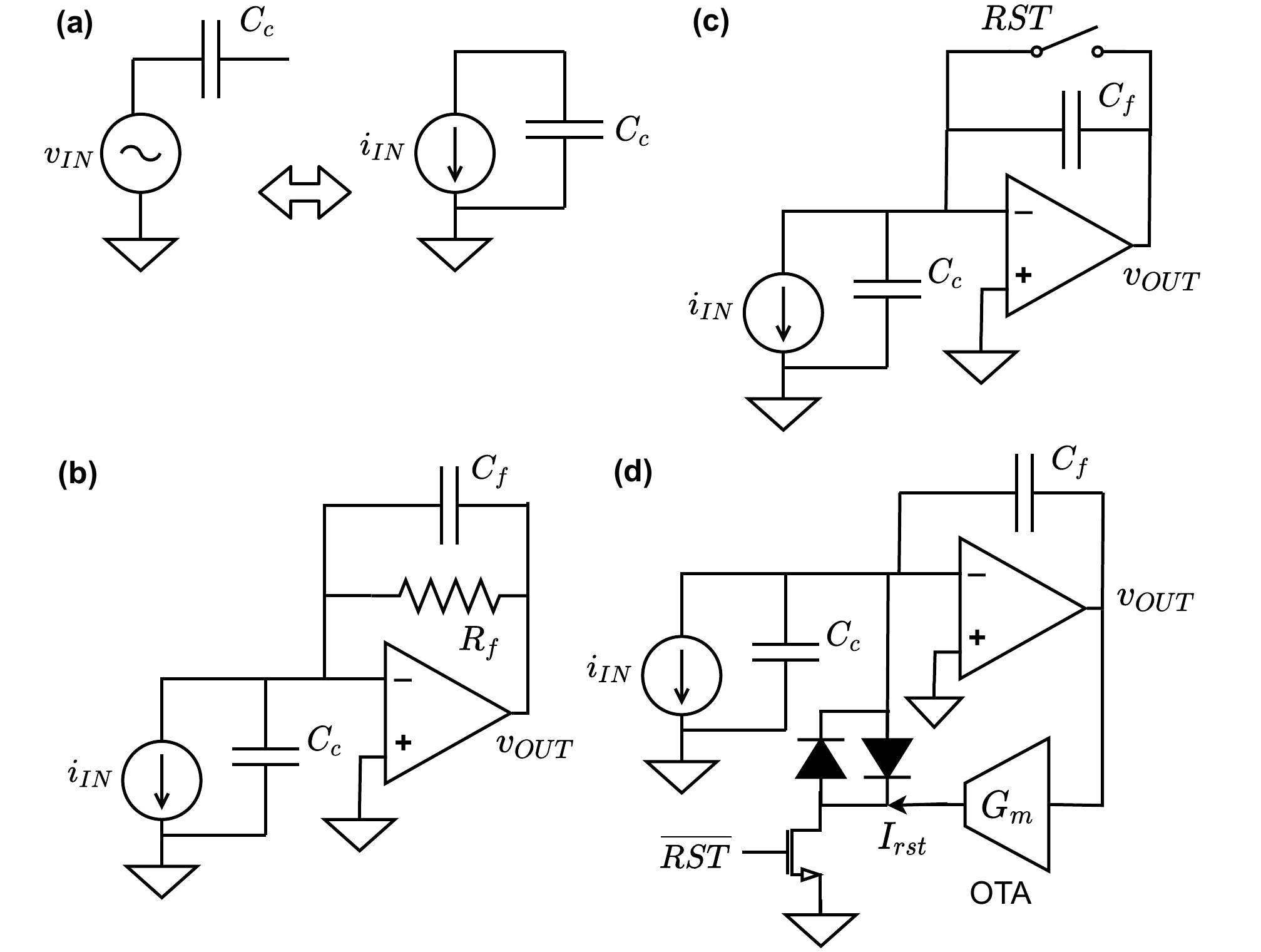}
    \caption{(a) Simplified Thevenin and Norton equivalent circuit models of non-contact EPS-based biophysical measurements. (b)-(d) Transimpedance amplifier (TIA) circuits for EPS: (b) continuous-time, (c) integrator (charge amplifier) with reset, resulting in a switched integrator, and (d) alternate switched integrator with a current switch. In each case $i_{IN}$ denotes the input current source, and $Z_s=1/\left(sC_c\right)$ is the source impedance.}
    \label{fig:TIA_circuits}
\end{figure}

Fig.~\ref{fig:TIA_circuits}(a) shows simplified Thevenin and Norton equivalent circuit models for EPS-based biophysical measurements. The Thevenin equivalent consists of a voltage source $v_{IN}$ in series with a capacitor $C_{c}$ which models electrostatic coupling between the body and sensing electrode, while the Norton equivalent consists of a source current $i_{IN}=sC_{c}v_{IN}$ in parallel with $C_c$. In either case, the frequency-dependent source impedance is $Z_{s}=1/(sC_{c})$. Our EPS design is based on current sensing using a TIA, so the Norton equivalent is more appropriate and will be used throughout this paper.

The performance of any low-current sensor is largely determined by the input TIA stage that converts the input current $i_{IN}$ into a voltage $V_{OUT}$. A conventional continuous-time TIA uses a resistor $R_f$ in negative feedback across an op-amp for this purpose (see Fig.~\ref{fig:TIA_circuits}(b)); a small capacitor $C_f$ is often required in parallel with $R_f$ to ensure stability. The resulting transimpedance is simply $Z_T=R_f||1/\left(sC_{f}\right)$. This simple circuit suffers from a fundamental trade-off: while high values of $R_f$ are required to reduce the input-referred current noise PSD $4kT/R_f$, they also make the circuit more likely to saturate from unwanted sources such as PLI.

An alternative is to remove $R_f$, resulting in a purely capacitive transimpedance of $Z_{T}(s)=1/\left(sC_f\right)$. This circuit is known as a current integrator or charge amplifier. The low-pass nature of its transimpedance provides some suppression of unwanted high-frequency sources, but additional circuitry is required to stabilize the DC operating point.

In one popular approach, pure charge integration is replaced by a correlated double sampling (CDS) scheme in which we i) sample the output voltage twice with an interval of $T_s$, and ii) take the difference between the readings. CDS is also convenient to implement since the capacitor can be reset before the next pair of samples is measured, thus eliminating the need for adding DC feedback across $C_f$ to prevent saturation. The transimpedance of such a \emph{switched integrator}~\cite{mountford2008development} is
\begin{equation}
    Z_{T,CDS}(s)=\frac{v_{OUT}(s)}{i_{IN}(s)}=\frac{1-e^{-sT_s}}{sC_f}.
\end{equation}
Firstly, at low frequencies where $\omega T_s\ll 1$, we can assume that $e^{-sT_s}\approx 1-sT_s$. This is known as the switched-capacitor (SC) regime, and results in a resistive transimpedance of value
\begin{equation}
    Z_{T,CDS}(s)=\frac{v_{OUT}(s)}{i_{IN}(s)}\approx\frac{T_s}{C_f}.
\end{equation}
Secondly, the subtraction operation removes voltage offsets (e.g., due to charge injection), thus improving accuracy. Finally, $Z_{T,CDS}=0$ at frequencies where $\omega T_s = 2\pi N$ where $N$ is an integer. These nulls, which occur at frequencies $f=N/T_s$ for which an integer number of cycles fits within the integration period $T_s$, are useful for eliminating unwanted signals (e.g., all the PLI harmonics if $T_s=1/f_0$)\footnote{This principle is widely used in low-noise test equipment, where it is often known as synchronous integration.}.

Switched integrators can be implemented in two main ways. Conventionally, a switch is placed in parallel with $C_f$ (see Fig.~\ref{fig:TIA_circuits}(c)) to periodically reset the output voltage. The resulting noise performance is ultimately limited by the leakage current $I_{rst}$ of the reset switch. Commercial CMOS or BiCMOS analog switches have $I_{rst}>10$~pA, making them unsuitable for measuring very low currents\footnote{Much lower values of $I_{rst}$ are possible using custom low-leakage CMOS switches, but this requires fabrication of a custom IC.}. Electromechanical switches (reed relays or MEMS switches) have much lower leakage currents. However, they also have several disadvantages: i) slow switching (turn-on times $>10$~$\mu$s), limiting the achievable TIA bandwidth; ii) limited reliability ($\sim$10$^9$ switching cycles); iii) high power consumption in the on-state, and iv) unpredictable switching transients due to mechanical bounce.

Fig.~\ref{fig:TIA_circuits}(d) shows an alternate switched integrator that replaces the switch across $C_f$ by a shunt current switch at the negative input terminal of the op-amp~\cite{gardner1987improved}. During the on-state, this switch carries a current $G_{m}v_{OUT}$ where $G_m$ is the transconductance of an operational transconductance amplifier (OTA) placed in feedback, thus discharging $C_f$. During the off-state, the input current integrates on $C_f$ and the input-referred noise is again limited by the switch leakage current $I_{rst}$. However, the switch can now be realized using back-to-back ultra-low-leakage diodes (e.g., PAD-1, $I_{rst}\approx 100$~fA at room temperature), thus minimizing input-referred noise.

\subsection{Choice of Op-Amp}
The choice of op-amp is critical for TIA performance. Key requirements include i) low input bias current $I_{b}$ and current noise PSD $\overline{i_{n}^{2}}$, ii) low voltage noise PSD $\overline{e_{n}^{2}}$, and iii) low offset voltage $V_{OS}$. The traditional solution adds high-performance discrete JFETs to the input of a general-purpose op-amp~\cite{auer2007low}, but this limits $I_{b}$ to the $\sim$1~pA range at room temperature. MOSFET-input op-amps are needed to obtain even lower input bias currents. Two good examples are the LMP7721 (Texas Instruments) and the ADA-4530-1 (Analog Devices).

Apart from the op-amp itself, the board- and package-level leakage resistance $R_{leak}$ at its input terminals can be another important source of offset (mean leakage current $I_{leak}$) and noise. Fortunately, proper material selection and layout techniques can reduce $I_{leak}=\Delta V/R_{leak}$ to the fA level, where $\Delta V$ is the voltage across $R_{leak}$. The keys are i) to increase $R_{leak}$ by using a high-quality dielectric material for the PCB (e.g., Rogers 4350B)\footnote{Here, high-quality implies both low dissipation factor (loss) to reduce noise, and also low soakage (dielectric relaxation) to ensure fast settling.}; and ii) to eliminate $\Delta V$ by surrounding the sensitive node (negative input terminal of the op-amp) with a guard terminal that is actively-driven to the same potential. The guard ensures that $\Delta V\approx 0$ for the possible leakage paths (e.g., through the PCB dielectric), thus minimizing $I_{leak}$. The ADA-4530-1 has a built-in guard driver circuit for this purpose. It also has lower input capacitance and lower $I_{b}$ than the LMP7721, and so is used for this design.

\subsection{TIA Noise Analysis}
\subsubsection{Continuous-time TIA}
If the op-amp's input-referred voltage and current noise terms are uncorrelated, the input-referred current noise PSD for a continuous-time TIA is
\begin{align}
\nonumber \overline{i_{n,tot}^{2}}&=\frac{4kT}{R_{leak}}+\overline{i_{n}^{2}}+\frac{\overline{e_{n}^{2}}}{\left|Z_{in}||Z_{s}||Z_{f}\right|^2}+\frac{4kT}{R_{f}}\\
&=2qI_{b,eff}\left(1+\frac{R_{n}^{2}}{\left|Z_{in}||Z_{s}||Z_{f}\right|^2}\right)+\frac{4kT}{R_{f}||R_{leak}},
\end{align}
where $Z_{in}$ is the op-amp's input impedance, $Z_{s}$ is the source impedance (including any parasitic capacitance at the inverting input terminal), $Z_{f}$ is the feedback impedance, $I_{b,eff}\equiv \overline{i_{n}^{2}}/2q$ is the effective op-amp input bias current for estimating shot noise, and $R_{n}\equiv\sqrt{\overline{e_{n}^{2}}/\overline{i_{n}^{2}}}$ is the noise matching resistance. The input-referred noise PSD is a function of frequency because i) $\overline{e_{n}^{2}}$ exhibits $1/f$ noise, and ii) $I_{in}$, $Z_{in}$, $Z_{s}$, and $Z_{f}$ are frequency-dependent. In many cases $I_{b,eff}\gg I_{b}$ since op-amps generally achieve low input bias current by canceling two uncorrelated current sources (e.g., from two reverse-biased diodes), such that $I_{b}=I_{b,1}-I_{b,2}$. While this process subtracts the mean (i.e., DC) currents, their variances add to generate a shot-noise PSD corresponding to $I_{b,eff}=I_{b,1}+I_{b,2}$. Similarly, guarding suppresses the DC current through $R_{leak}$ (thus removing its effect on $Z_{in}$) but does not affect its noise.

For the ADA-4530-1, $\sqrt{\overline{i_{n}^{2}}}=0.07$~fA/Hz$^{1/2}$ at 0.1~Hz, corresponding to $I_{b,eff}=15.3$~fA. While very low, this is larger than the specified input bias current of $I_{b}< 1$~fA, as expected\footnote{This increase may also be due to a $1/f$ component in the current noise, but this is unlikely since the measured PSD is white from 4~mHz to 1~Hz.}. Also, $\sqrt{\overline{e_{n}^{2}}}=14$~nV/Hz$^{1/2}$ at high frequencies with a $1/f$ corner frequency of $f_{c}\approx 300$~Hz. The input impedance is $Z_{in}=R_{in}||\frac{1}{j\omega C_{in}}$ with $R_{in}>100$~T$\Omega$ and $C_{in}=8$~pF. The input time constant $\tau_{in}=R_{in}C_{in}>800$~sec, so in practice the capacitive component dominates and $Z_{in}\approx\frac{1}{j\omega C_{in}}$.


\subsubsection{Switched-integrator TIA}
The same analysis is valid for a CDS-based switched-integrator TIA, except for the fact that noise from the feedback resistor $R_{f}$ is replaced by shot noise from the leakage current of the reset switch. Thus, the input-referred noise PSD is modified by setting $R_f\rightarrow\infty$ (assuming a lossless feedback capacitor) and adding a new term:
\begin{equation}
\overline{i_{n,tot}^{2}}=2qI_{b,eff}\left(1+\frac{R_{n}^{2}}{\left|Z_{in}||Z_{s}||Z_{f}\right|^2}\right)+4qI_{rst}+\frac{4kT}{R_{leak}},
\end{equation}
where $I_{rst}$ is the saturation current of a reverse-biased PN junction within the reset switch (assumed to be a semiconductor device). If an electromechanical switch is used instead, $4qI_{rst}$ is replaced by a thermal noise term $4kT/R_{rst}$ where $R_{rst}$ is the resistance between the switch and control terminals in the off-state (generally dominated by dielectric loss).
\subsubsection{Total Input-Referred Noise}
Based on the analysis above, we can estimate the total input-referred voltage noise $v_{n,in}$ of the EPS as a function of source and feedback impedances ($Z_s$ and $Z_f$, respectively) and amplifier temperature:
\begin{equation}
v_{n,in}^{2}=\int_{f_1}^{f_2}{\overline{v_{n,tot}^{2}(f)}df} = \int_{f_1}^{f_2}{\overline{i_{n,tot}^{2}(f)}\left|Z_s(f)\right|^2df}, 
\end{equation}
where $v_{n,tot}$ is the input-referred voltage noise PSD and $\left[f_{1},f_{2}\right]$ is the measurement bandwidth. We consider two possible designs: i) a continuous-time TIA with $R_f=150$~G$\Omega$ and $C_{f}=0.1$~pF, and ii) a switched integrator with $C_{f}=10$~pF. Both are assumed to measure electrical biosignals of interest (ECG, EEG, and EMG) for various values of $C_{c}$ using op-amp and switch parameters for the ADA-4530-1 and PAD-1, respectively. For simplicity, the source impedance was assumed to be capacitive, i.e., $Z_s\approx 1/\left(sC_c\right)$; any resistive component $R_s$ would degrade the signal-to-noise ratio (SNR) by contributing an input-referred current noise term $4kT/R_s$.

\begin{figure}[htbp]
    \centering
    \includegraphics[width=0.49\columnwidth]{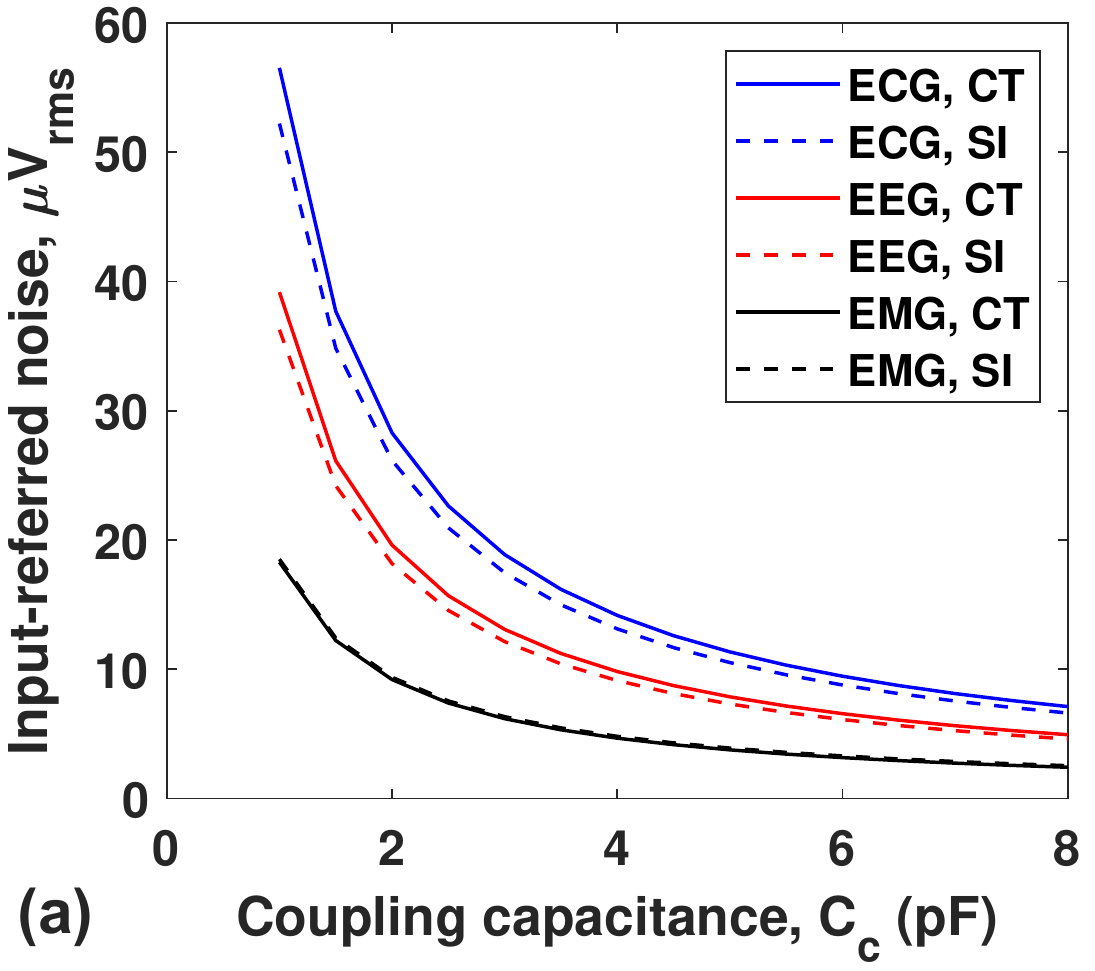}
    \includegraphics[width=0.48\columnwidth]{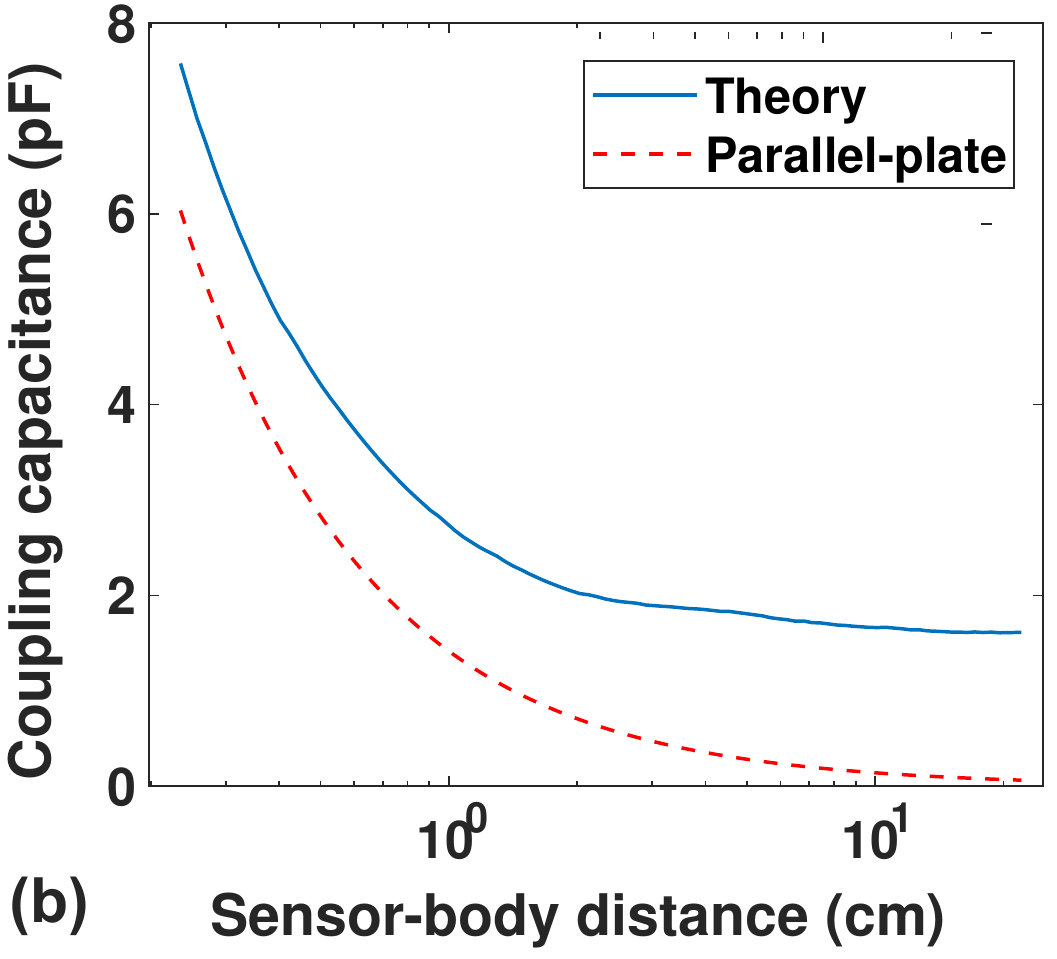}
    \caption{(a) Simulated input-referred voltage noise for continuous-time (CT) and switched integrator (SI) TIA designs for recording three important electrical biosignals (ECG, EEG, and EMG) as a function of the coupling capacitance $C_{c}$. (b) Theoretical coupling capacitance $C_{c}$ between the body and a circular sensing electrode of radius $a=22.6$~mm (chosen to have the same area as a square electrode of side $w=40$~mm) as a function of coupling distance. The parallel-plate capacitance result is also shown for comparison.}
    \label{fig:SNR_calc}
\end{figure}

Fig.~\ref{fig:SNR_calc}(a) summarizes the estimated input-referred noise $v_{n,in}$ for typical ECG (1-150~Hz), EEG (2-40~Hz), and EMG (10-500~Hz) measurements with $C_{c}$ between 1-8~pF. As expected, $v_{n,in}$ decreases with $C_{c}$ (mainly because $\left|Z_s(\omega)\right|\approx 1/\left(\omega C_c\right)$ decreases). The continuous-time and switched integrator designs have similar sensitivity, and EMG has the lowest $v_{in,in}$ in both cases. This is because the EMG band occurs at higher frequencies than ECG or EEG, where $\left|Z_s(\omega))\right|$ (and thus the voltage noise PSD) is lower. The results in the figure can be used to estimate the minimum coupling capacitance $C_{c}$ required to obtain acceptable SNR. For example, the typical ECG signal amplitude is $\sim$0.5~mV, so $SNR>10$ can be achieved whenever $C_{c}>1$~pF. By contrast, the typical EEG signal amplitude is $\sim$20~$\mu$V, for which $SNR>2$ requires $C_{c}>4$~pF. Note that these SNR values can be improved by post-processing. For example, EEG recordings can be decomposed into smaller sub-bands prior to further analysis. 

\subsection{Optimized Design of the Sensing Electrode and TIA}
\label{sec:opt_design}
\subsubsection{Sensing Electrode}
A square sensing electrode (size $w=l=40$~mm, gold-coated copper) on the surface of a printed circuit board (PCB) was surrounded by an active guard structure (driven by the TIA) to minimize leakage current at the TIA input terminals. Coupling capacitance $C_c$ can be analytically calculated as a function of electrode-body distance $d$ for circular electrodes if we assume that the body surface is an equipotential~\cite{borkar1975capacitance,iossel1971calculation}. We utilize this result by replacing our square electrode with a circular electrode of the same area (radius $a=w/\sqrt{\pi}$) during the analysis; the result of this approximation is to slightly underestimate $C_{c}$.

The estimated dependence of $C_{c}$ on coupling distance $d$ is shown in Fig.~\ref{fig:SNR_calc}(b). The result transitions between the parallel-plate formula $C_{c} = \epsilon_0\pi a^2/d$ (valid for $d\ll a$) and the self-capacitance $C_{c,min}=8\epsilon_0 a$ (valid for $d\gg a$). In the latter limit, $C_c$ becomes independent of $d$, suggesting that i) long-range EPS-based sensing is possible; and ii) the electrode size can be reduced if needed (e.g., to miniaturize the sensor) since $C_{c}$ is a relatively weak function of electrode size in this regime (i.e., $C_{c}\propto a$). Unfortunately, this simplified analysis has ignored the inevitable presence of other conducting objects in the environment; in practice, as $d$ increases, more and more of the field lines terminate on these objects rather than the sensing electrode. As a result, the effective value of $C_{c}$ does not saturate to $C_{c,min}$, but keeps decreasing. This field sharing effect ultimately limits the useful sensing range.

\subsubsection{TIA Architecture}
The analysis above shows that while switched-integrator TIAs can efficiently remove PLI (via synchronous integration), they also require an ultra-low-leakage reset switch. Given the difficulties in selecting an appropriate switch, we designed a continuous-time TIA using the ADA-4530-1 for this application. A surface-mounted feedback resistor of value $R_f=150$~G$\Omega$ (realized as three 50~G$\Omega$ resistors in series) was used to maximize sensitivity; the resulting shunt capacitance $C_f\approx 0.1$~pF. {The feedback impedance is dominated by $C_f$ at frequencies greater than $\omega_{1}=1/(R_f C_f)\approx 2\pi\times 10$~Hz. Thus, the voltage transfer function $v_{OUT}/v_{IN}$ resembles a first-order high-pass filter with a $-3$~dB frequency of $\omega_1$ and a high-frequency gain of $v_{OUT}/v_{IN}\approx -C_c/C_f$. Since $\omega_1$ is relatively low, the voltage gain is constant over most of the ECG bandwidth, which minimizes distortion in the non-contact ECG waveforms.}

\begin{figure}[htbp]
	\centering
	\includegraphics[width = 0.86\columnwidth]{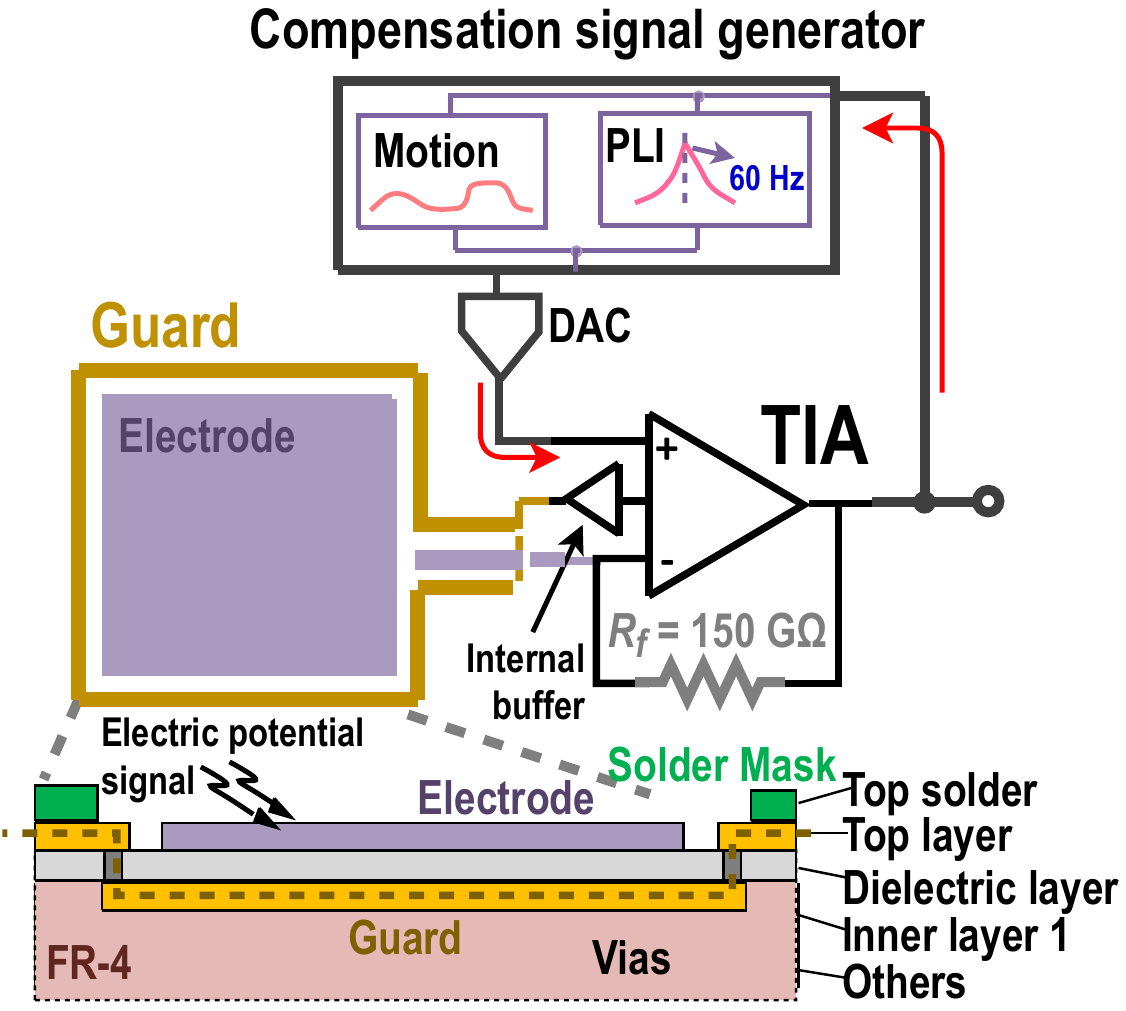}
	\caption{Schematic and physical cross-section of the optimized sensing electrode and TIA, including a DSP-based signal estimator for the adaptive cancellation loop (ACL). DAC = digital-to-analog converter.}
	\label{fig:remote_block_b}
\end{figure}

Fig.~\ref{fig:remote_block_b} shows a detailed view of the optimized sensing electrode and TIA. The sensing electrode and top portion of the guard electrode (which surrounds it) reside on the top layer of the PCB. The guard also includes another portion beneath the sensing electrode, connected through vias. The sensing electrode is separated from the surface portion of the guard by an isolation trench (gap $=0.406$~mm) on the top layer to minimize the leakage current. Ground shields are laid out around the whole sensor, as shown in Fig.~\ref{fig:remote_block_b}.

The EPS uses a TIA-based adaptive cancellation loop (ACL) to cancel unwanted signal components such as PLI and motion artifacts. The loop uses either an analog circuit or a digital signal processing (DSP) algorithm running on the on-board MCU to estimate the unwanted components in real-time; a digital version is shown in the figure. The estimated signal is fed back to the (normally unused) positive input terminal of the TIA. Since the TIA input stage is differential, the feedback signal cancels the unwanted signal components, as analyzed in detail in the next subsection. Thus, $R_f$ can be significantly increased before the TIA output saturates. Since the sensitivity of the EPS is directly proportional to $R_f$, the net result is improved sensitivity. While a DSP-based signal estimator is more flexible, our prototype EPS uses a two-stage analog estimator for simplicity, consisting of i) an auxiliary surface loop, and ii) a BPF (tuned to $f_0$) and adjustable attenuator.

\subsubsection{Minimizing Leakage Currents}
EPS sensitivity can be improved by minimizing leakage currents at the TIA input terminal, i.e., maximizing $R_{leak}$. The value of $R_{leak}$ is sensitive to moisture, i.e., the relative humidity (RH) of the air. This dependence is due to two mechanisms: \emph{adsorption} and \emph{absorption}. Adsorption is a rapid process in which thin films (e.g., water molecules) adhere to the surface of the PCB and reduce its insulation resistance. The resulting leakage current can be eliminated by guard rings since it flows near the surface. By contrast, absorption is a slow process in which molecules diffuse through the PCB, thus affecting its bulk conductivity. Guard rings do not reduce absorption-related currents because they flow through the bulk. In fact, even 3-D guard structures do not eliminate bulk leakage. Fortunately, experiments show that both effects are small for RH$<50$\%, so the recommended EPS operating range is 0-50\% RH. 

In addition to humidity effects, $R_{leak}$ can be substantially decreased if the board is contaminated by solder flux, body oils, dust, or dirt. Non-ionic contaminants form leakage paths near the PCB surface, so the resulting currents are suppressed by guarding. However, ionic contaminants act as electrolytes in the presence of humidity, thus forming weak batteries. Solder flux residue and body oils are particularly effective at creating such parasitic batteries, and the resulting leakage currents $V_{batt}/R_{batt}$ cannot be suppressed by guarding. Thus, the test boards were periodically cleaned to remove contaminants. 

\subsection{Performance of the Other Circuit Blocks}
\label{sec:pli}
\subsubsection{PLI Cancellation}
Our ACL design uses a small auxiliary ground loop placed over the surface of the main sensing electrode for initial PLI sensing and cancellation. In the next step, a BPF at $f_0$ is connected in feedback around the TIA, thus creating the inverse function (a notch filter) in the closed-loop response. Denoting the BPF transfer function (which should be inverting to ensure stability) as $-H_{BPF}(s)$, the closed-loop transfer function of the TIA at the low frequencies of interest (where the op-amp is nearly ideal) is given by
\begin{equation}
\label{eq:remote_eq1}
H(s) = \frac{V_{out}}{V_{in}} =
\frac{-H_{0}(s)}{H_{BPF}(s)[1+H_{0}(s)]+1},
\end{equation}  
where $H_{0} = \dfrac{sC_{c}R_{f}}{sC_{f}R_{f}+1}$ is the closed-loop transfer function in the absence of the BPF. This has the form of a high-pass filter with cut-in frequency of $\omega_{1} = 1/(R_{f}C_{f})$ and high-frequency gain of $C_{c}/C_{f}$. Note that since $|H_{0}(s)|$ increases with frequency, so does sensitivity. Intuitively, this is because the impedance of the coupling capacitor decreases with frequency as $1/(j\omega C_{c})$, resulting in increased current flow.

\begin{figure}[htbp]
	\centering
	\includegraphics[width = 0.49\columnwidth]{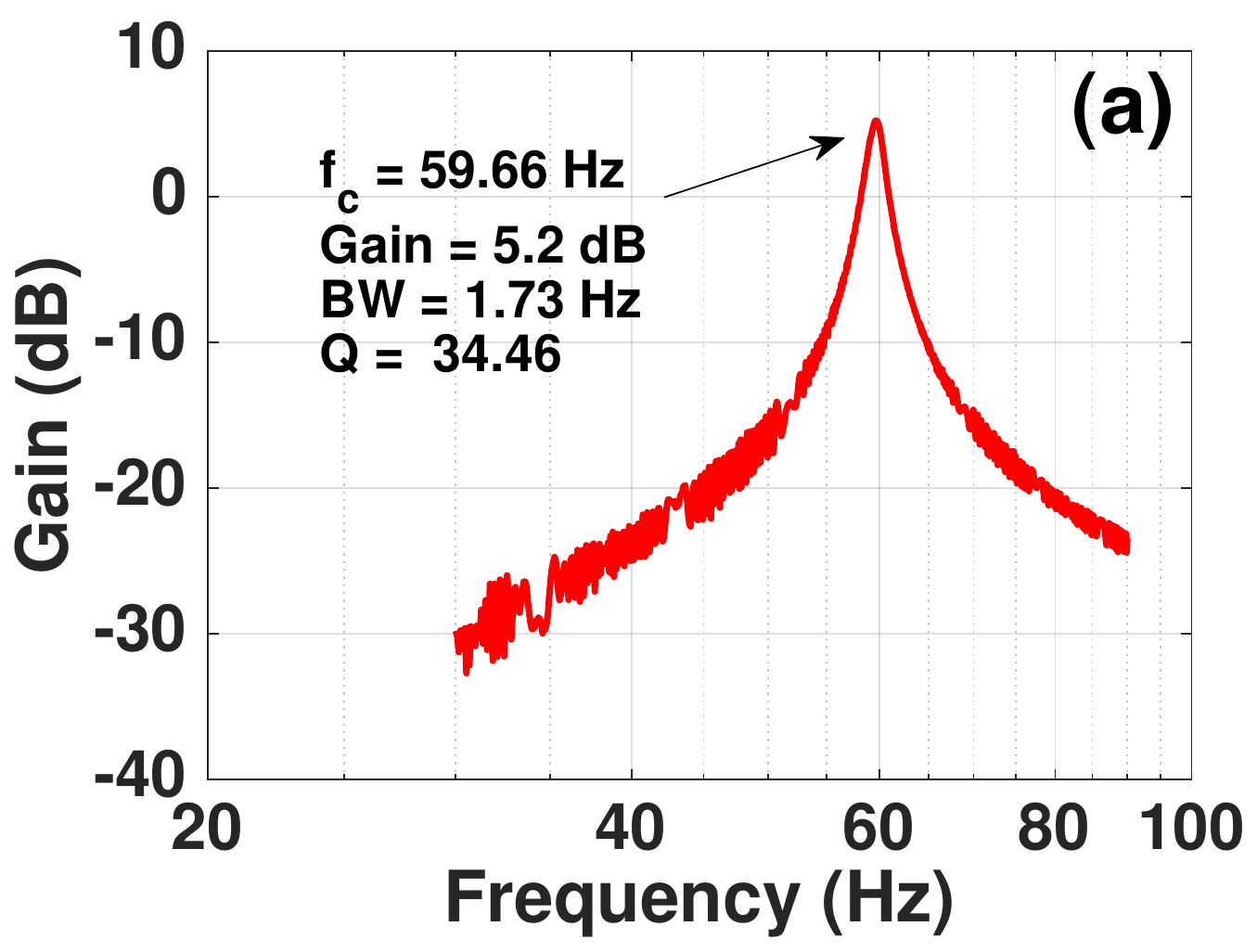}
	\includegraphics[width = 0.49\columnwidth]{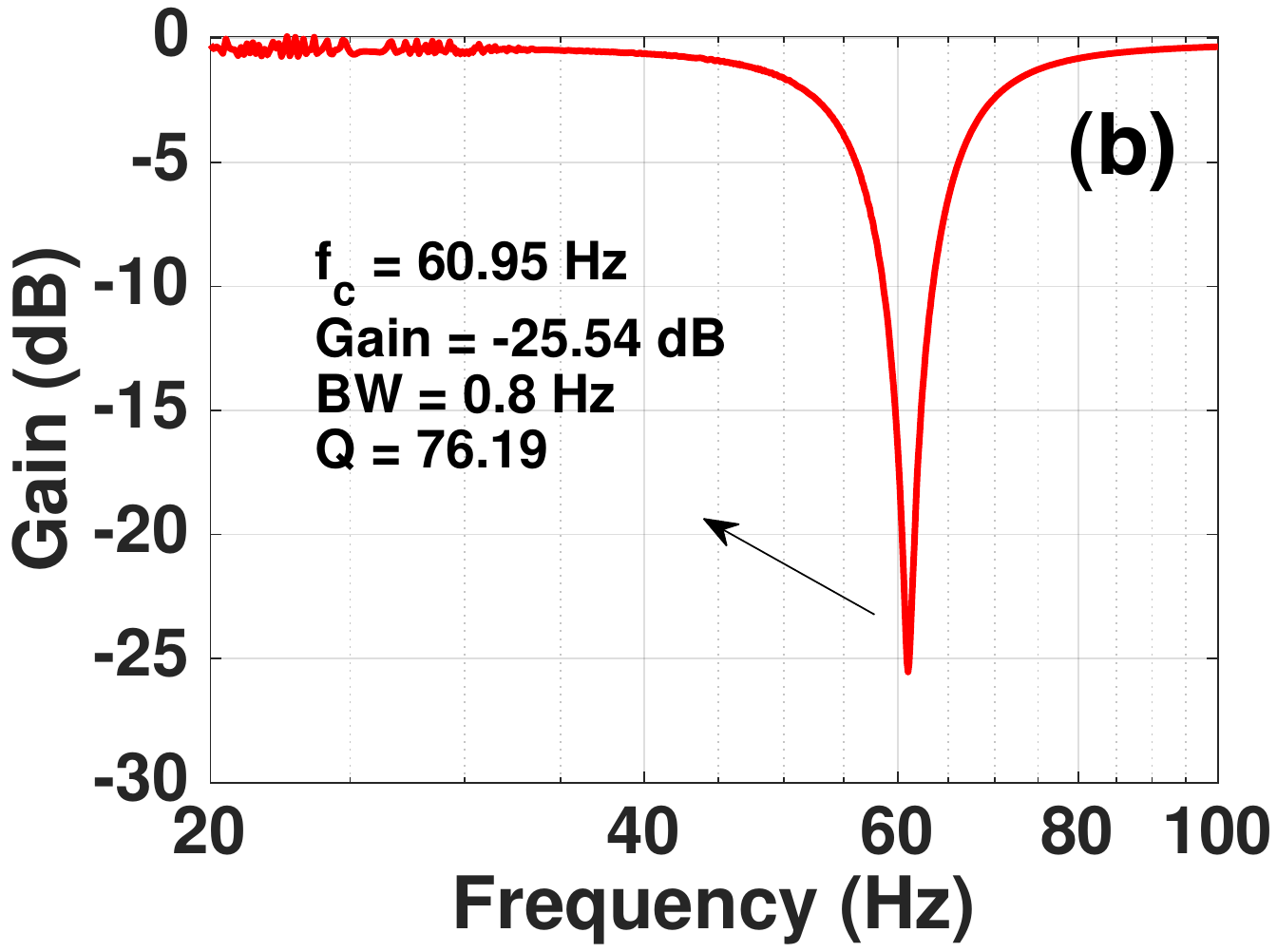}
	\caption{(a) Band pass filter (BPF) tuned to $f_0$ for active PLI cancellation at the TIA stage; (b) Measured spectrum of the notch filter tuned to $f_0$.}
	\label{fig:remote_60Hz_bpf}
\end{figure}

In practice $C_{c}\ll C_{f}$ (i.e., the sensor is weakly coupled to the body), so $|H_{0}(s)|\ll 1$. In this case, eqn.~(\ref{eq:remote_eq1}) simplifies to
\begin{equation}
\label{eq:remote_eq2}
H(s) = \frac{ V_{out}}{V_{in}} \approx
\frac{-H_{0}(s)}{H_{BPF}(s)+1}.
\end{equation}
Thus, the closed-loop voltage gain at the resonant frequency of the feedback BPF is reduced by the factor ($1+A_{BPF}$), where $A_{BPF}$ is the BPF’s peak voltage gain. Thus, PLI can be cancelled by tuning the BPF’s resonant frequency to $f_0$. Clearly, the BPF must have relatively high quality factor Q (i.e., narrow bandwidth) to ensure that the closed-loop gain is only reduced around $f_0$, thus minimizing distortion of the received ECG signal. We used a single op-amp design with $A_{BPF}\approx 2$ and $Q\approx 35$ to implement the feedback BPF, resulting in a cancellation amplitude and bandwidth of 3 ($\sim$10 dB) and 2~Hz, respectively. Fig.~\ref{fig:remote_60Hz_bpf}(a) shows the measured frequency response of the feedback BPF.

\subsubsection{Additional Gain and Filter Stages}
Fig.~\ref{fig:remote_60Hz_bpf}(b) shows the measured frequency response of the $f_0$ notch filter that follows the TIA. This stage uses a standard twin-T topology to provide additional PLI suppression. The figure shows that $\sim$25~dB of suppression is obtained over a bandwidth of $\sim$0.8~Hz. A $2f_0$ notch filter was added for further PLI suppression. The rest of the ECG channel consists of a band-selector (BPF), VGA, and LPF. The response is close to simulations, with i) a mid-band gain of $\sim$24~dB, and ii) an upper cut-off frequency of $\sim$154~Hz. Measurements of the RC channel also agree with simulations; the upper cut-off frequency is $\sim$5~Hz.

\section{Experimental Results}
\label{sec:expt_results}

\subsection{Experimental Setup}
\begin{figure}[htbp]
	\centering
	\includegraphics[width = 0.86\columnwidth]{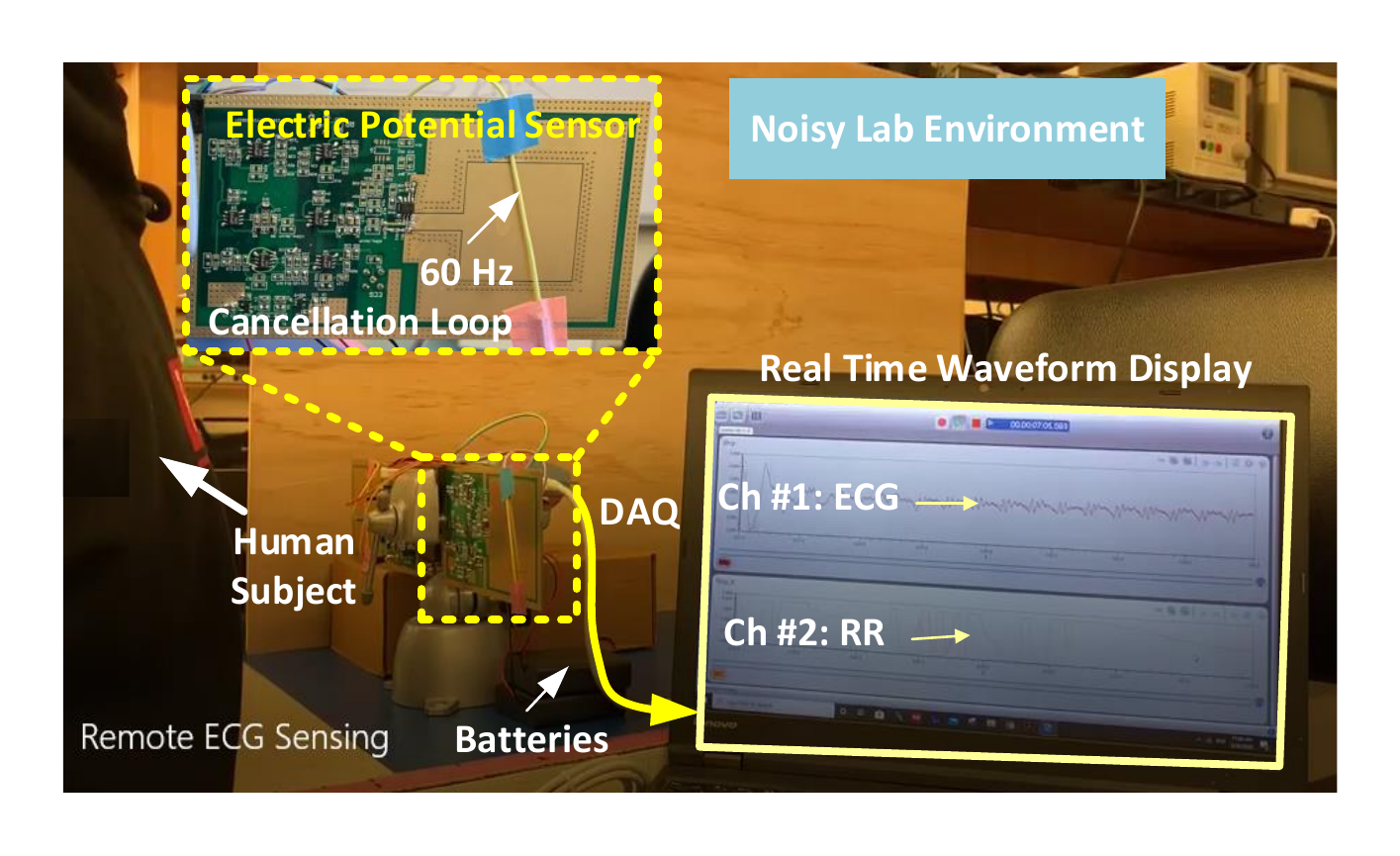}
	\caption{{Experimental setup for non-contact cardiopulmonary sensing using a DAQ (USB-1608-FS, Measurement Computing) interfaced with a battery-powered PC (left) and a zoomed-in view of two sensing boards (right).}}
	\label{fig:remote_Setup1}
\end{figure}

This section is focused on experimental verification of the custom EPS on healthy human volunteers. The inset in Fig.~\ref{fig:remote_Setup1} shows a photograph of a fully-assembled sensor board, while the rest of the figure shows the typical experimental setup used for non-contact sensing. Data was acquired by a USB-based data acquisition system (DAQ) (USB-1608-FS, Measurement Computing) connected to a battery-powered personal computer (PC) to minimize PLI. Two identical sensor boards were simultaneously interfaced with the DAQ to allow evaluation of differential sensing. Non-contact sensing results from single boards are presented in the following subsections.

{Signals were measured from a total of 4 healthy adult volunteers, who wore similar indoor clothing during the measurements. These test results were validated using three off-the-shelf contact sensors (BITalino (r)evolution, NeuLog NUL-236, and OpenBCI Ganglion) that have themselves been validated in earlier studies~\cite{da2014bitalino,batista2017experimental,peterson2020feasibility}. Major properties of the contact sensors and our non-contact EPS are listed in Table~\ref{table:comparision}.}

\begin{table}[htbp]
	\caption{{Parameters for reference contact sensors and the EPS}}
	\label{table:comparision}
	\centering
	\begin{tabular}{@{}l|cccc@{}}
		\hline
		\textbf{Parameters} & \textbf{(r)evolution} &\textbf{NUL-236} & \textbf{Open BCI} & \textbf{E-field} \\
		& \cite{BITalino} & \cite{Rbelt} & \cite{OpenBCI} & \\
		\hline
		Manufacturer & BITalino & NeuLog & Ganglion & Custom\\
		Sensor type & Foam & Pressure & Comb & ENIG\\
		 & electrodes & belt & electrodes & electrode\\
		Meas. type & Contact & Contact & Contact & Non-contact\\
		Sampling rate  & 1~kHz & 100~Hz & 200~Hz & 1~kHz\\
		Signals & ECG & RC & EEG & ECG/RC/EEG\\
		Meas. duration & 30~s & 60~s & 2~mins & 30/60~s \\
		Data transfer & Bluetooth & USB & Bluetooth & USB \\
		\hline
	\end{tabular}
	\vskip 0.5ex
    {Note: 4 healthy participants: age (21~$\sim$~32), weight (140~$\sim$~170~lbs), height (175~$\sim$~185~cm).\par}
\end{table}

\subsection{Adaptive Cancellation Loop (ACL)}
\begin{figure}[htbp]
	\centering
	\includegraphics[width = 0.482\columnwidth]{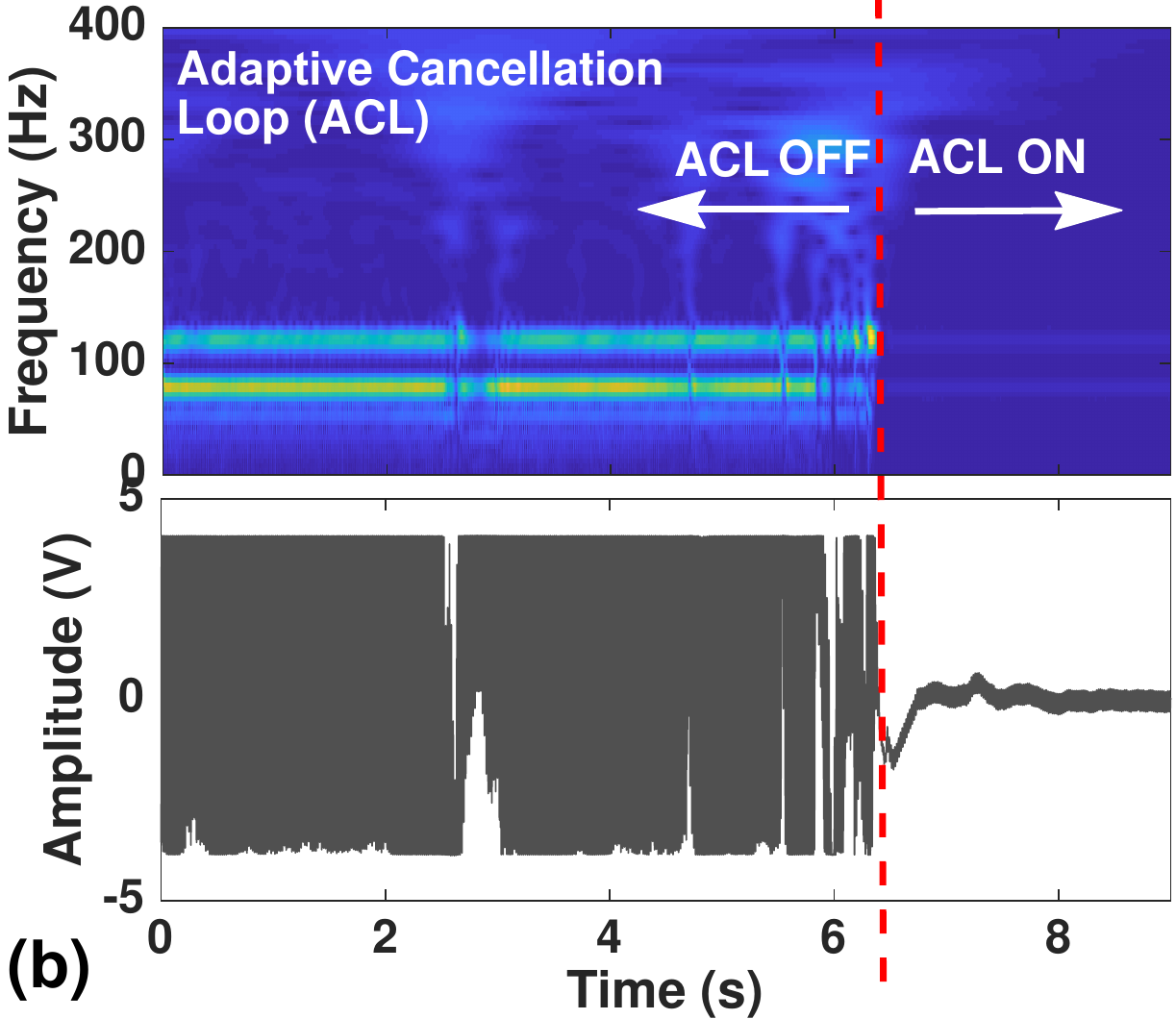}~~
	\includegraphics[width = 0.495\columnwidth]{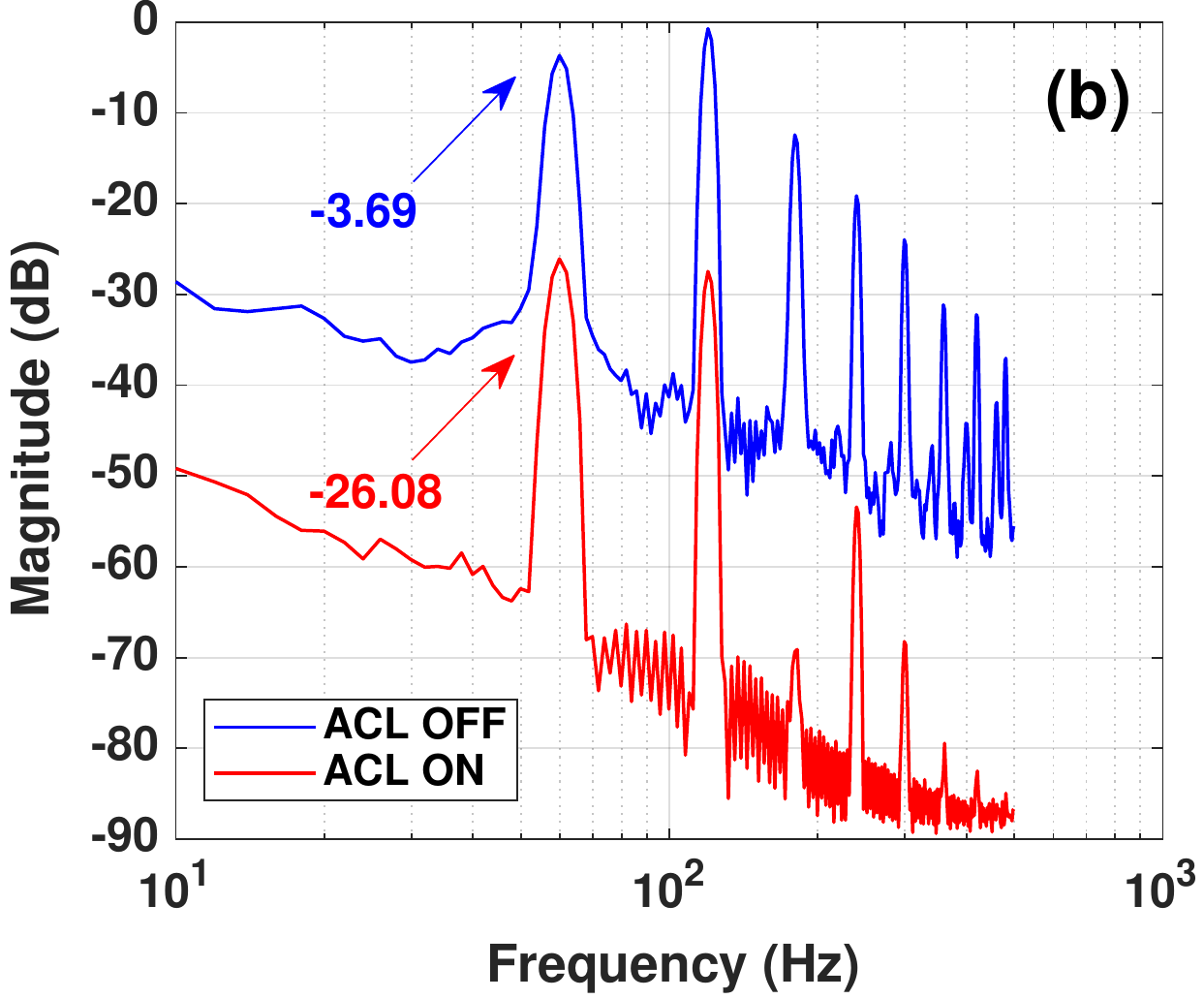}
	\caption{Measured PLI cancellation performance of the auxiliary loop within the ACL: (a) time and time-frequency domains; (b) signal spectra when the loop is ON and OFF, respectively; 22~dB PLI suppression is obtained at $f_0$.}
	\label{fig:remote_60Hz_sp}
\end{figure}

In addition to the PLI cancellation methods described in Section~\ref{sec:pli} (auxiliary loop, feedback BPF, and notch filters), the sensor board also included electrostatic shields on the sides and bottom surfaces. The overall amount of PLI suppression (ACL $+$ notch filters) is estimated to be $\sim$60~dB. Fig.~\ref{fig:remote_60Hz_sp} shows that the auxiliary loop alone provides $>20$~dB of suppression.

\subsection{Motion Cancellation Loop (MCL)}
\begin{figure}[htbp]
	\centering
	\includegraphics[width = 0.49\columnwidth]{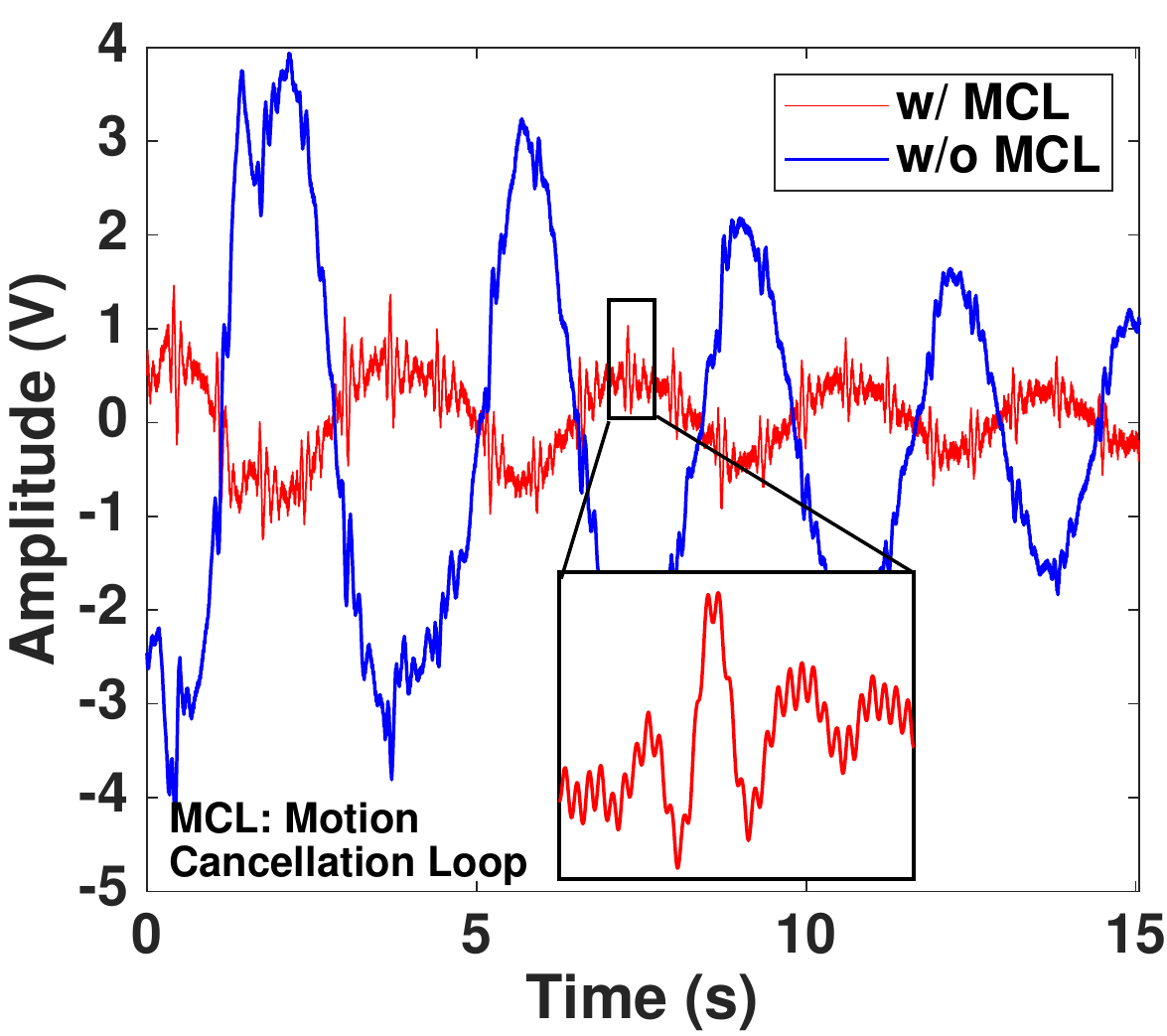}~~
	\includegraphics[width = 0.48\columnwidth]{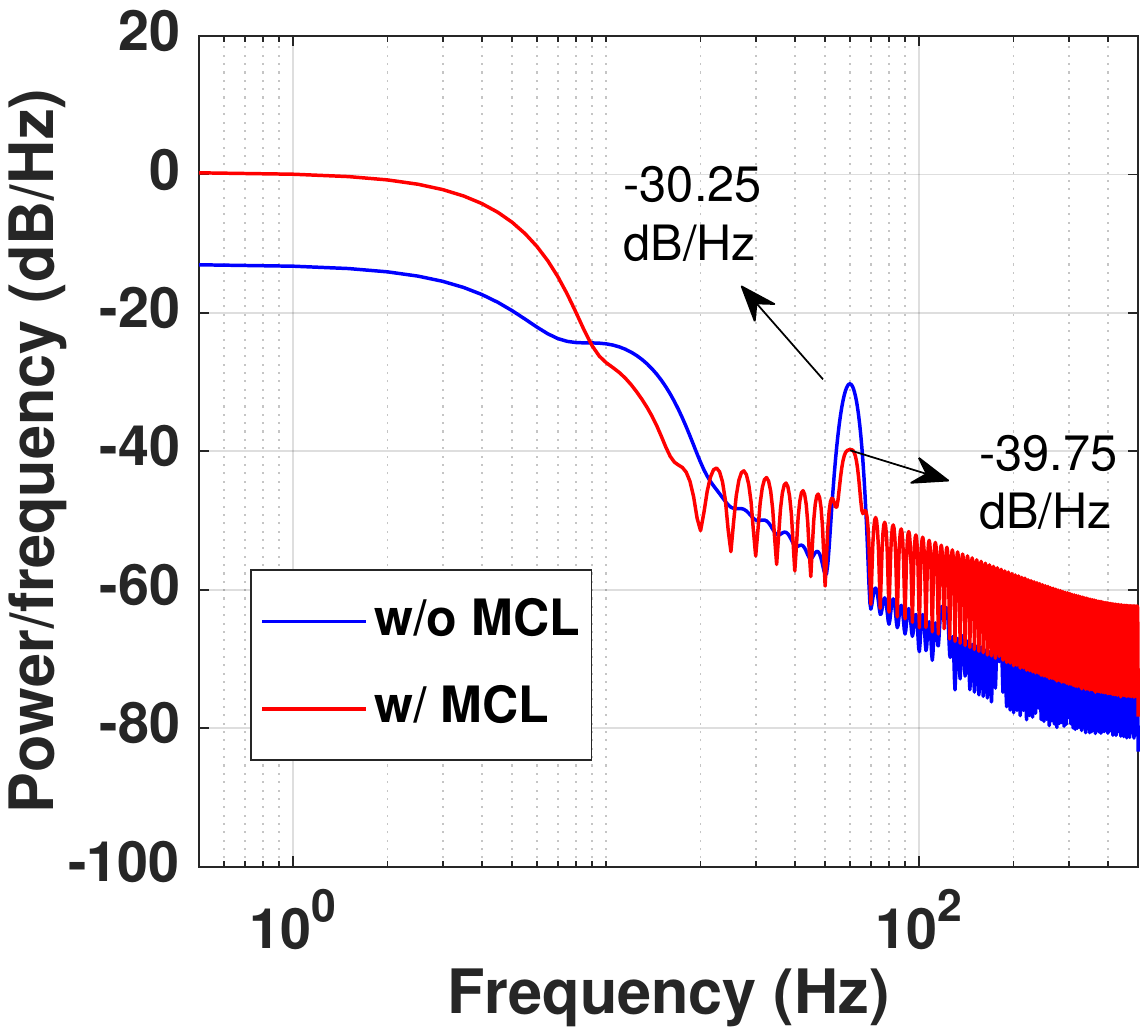}
	\caption{Simulated performance of active motion cancellation at the TIA (using pre-recorded data): (a) time domain; (b) signal spectra when MCL is ON and OFF, respectively; 9.5~dB suppression of motion artifacts is obtained.}
	\label{fig:remote_motion_sp}
\end{figure}

In addition to PLI, large-amplitude signals from gross body motion (e.g., walking) can also saturate the TIA. Fig.~\ref{fig:remote_motion_sp} illustrates a typical example of improved dynamic range (DR) obtained by using an adaptive motion cancellation loop (MCL) to cancel such motion artifacts. The MCL is based on a real-time signal estimator, as shown in Fig.~\ref{fig:remote_block_b}. In our initial implementation, we simply used a LPF for motion estimation and fed its output (after attenuation to ensure stability) back to the TIA's positive terminal. A typical motion suppression level of 9.5~dB was observed. More sophisticated signal estimators can be used to obtain greater amounts of suppression.

\subsection{Non-contact Electrocardiogram (ECG) Sensing}

\subsubsection{Measured Signal Properties}
\begin{figure}[htbp]
	\centering
	\includegraphics[width = 0.46\columnwidth]{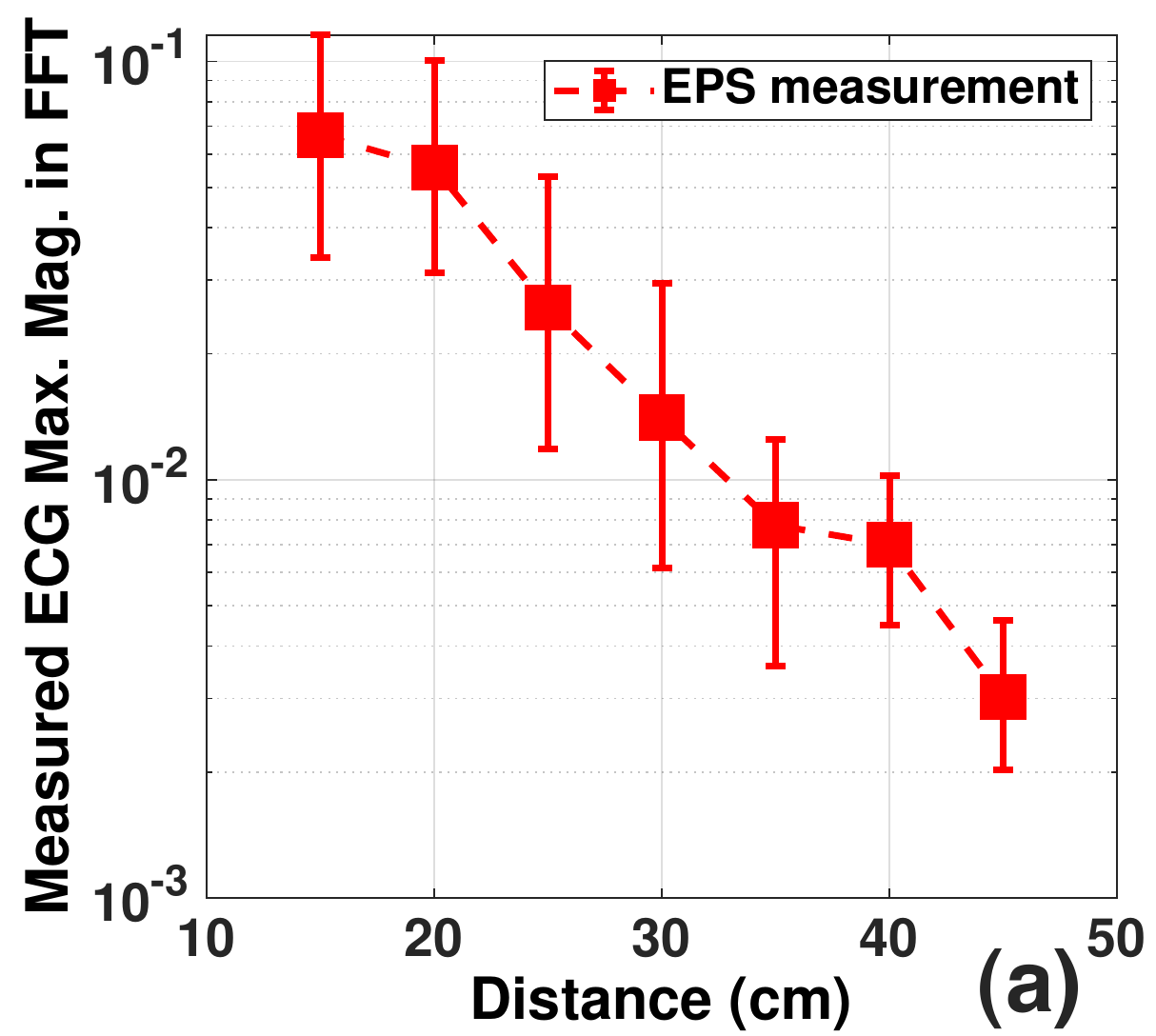}
	\includegraphics[width = 0.48\columnwidth]{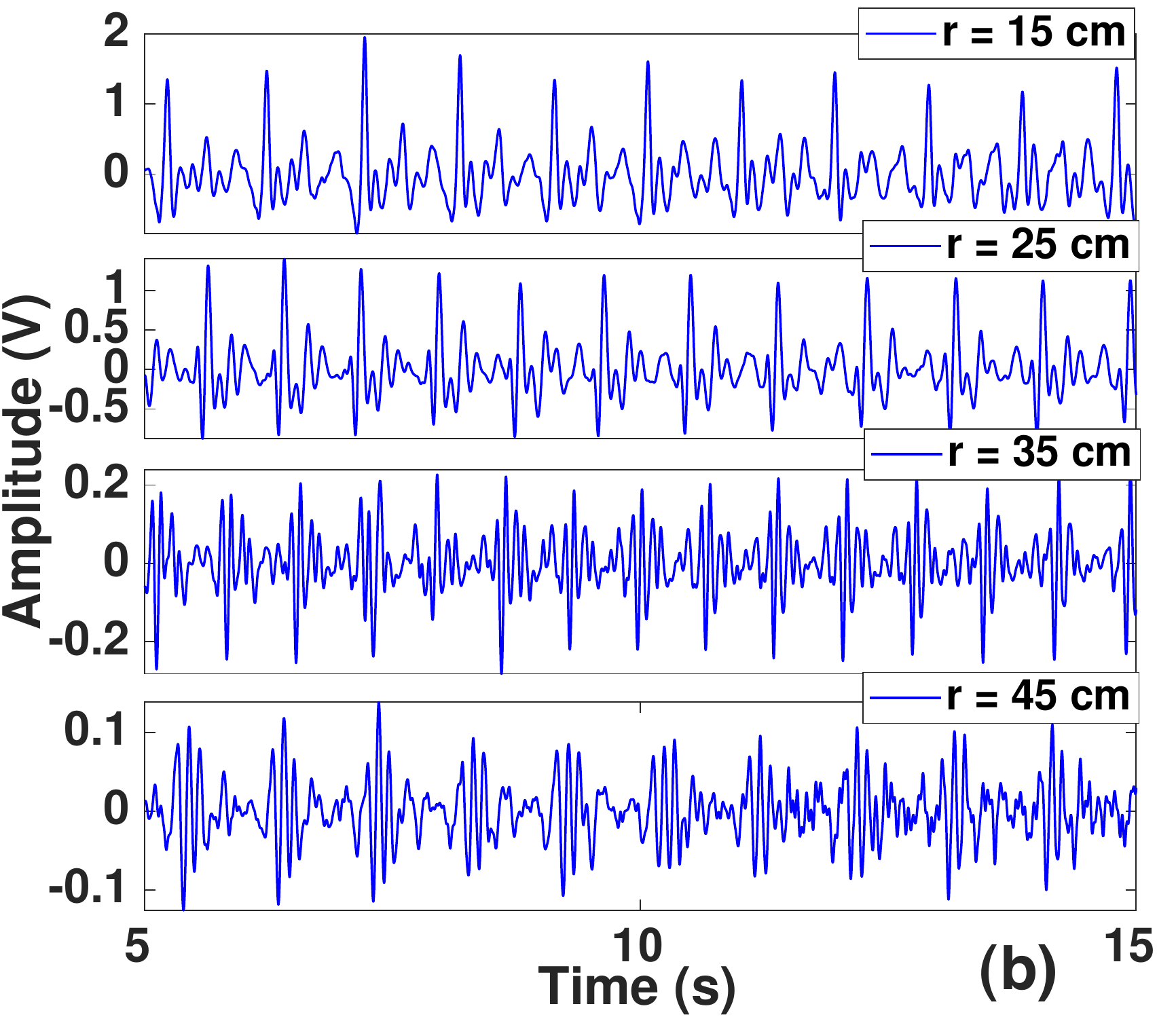}
	\caption{(a) Measured ECG amplitudes at different sensor-subject distances; (b) continuous wavelet transform (CWT) filtered ECGs in the time domain, showing that the signal amplitude decreases with sensing distance.}
	\label{fig:remote_ecg_ex}
\end{figure}

As shown in Fig.~\ref{fig:remote_Setup1} the subject generally sat (in this case, $\sim$30~cm from the sensor) in a very noisy and unshielded laboratory environment during the experiments. In most cases ECG waveforms could be clearly observed without any pre-filtering, as visible on the PC screen in the figure. To quantify the maximum non-contact measurement distance, we measured ECG signals at different sensor-subject distances. Specifically, the subject was located at distances of 15~cm to 50~cm (with a step of 5~cm) from the surface of the electrode. Each experimental setting was repeated 6 times to reduce the sensor-subject distance variations during experiments. Fig.~\ref{fig:remote_ecg_ex}(a) shows that the measured ECG signal amplitude decays quickly as the distance increases; note that the power spectra used to create this plot were obtained using a fast Fourier transform (FFT).

The raw time-domain data was filtered using a continuous wavelet transform (CWT) with the Morlet wavelet before further analysis. Fig.~\ref{fig:remote_ecg_ex}(b) shows the filtered time-domain waveforms, where the ECG can be clearly observed at each distance. The shapes of the ECG waveforms shown in Fig.~\ref{fig:remote_ecg_ex}(b) are clearly distance-dependent; they resemble typical capacitively-coupled ECG for distances up to $\sim$30~cm, beyond which the QRS complex becomes harder to detect. H. Prance~\cite{harland2001electric,harland2002remote} mentioned that ECG cannot be measured by electrodes spaced by an air gap because the signal off-body is not purely electric cardiac act ivies that includes the movement signal cased by the arterial pulse moving the chest wall as well. Therefore, the completely non-contact detection~\cite{Yu2014} does have different morphology as contact measurement. It is likely that the measured signals are actually a combination of ECG and ballistocardiogram (BCG) components, as analyzed in recent work~\cite{uguz2020physiological}. The BCG component arises from periodic mechanical motion of triboelectrically-induced charges on the body surface during the cardiac cycle, which results in current flow through the time-varying coupling capacitance $C_{c}$ between the body and the sensing electrode. It appears to decay more slowly with distance than the ECG component, and thus dominates at large sensing distances.

\begin{figure}[htbp]
	\centering
	\includegraphics[width = 0.665\columnwidth]{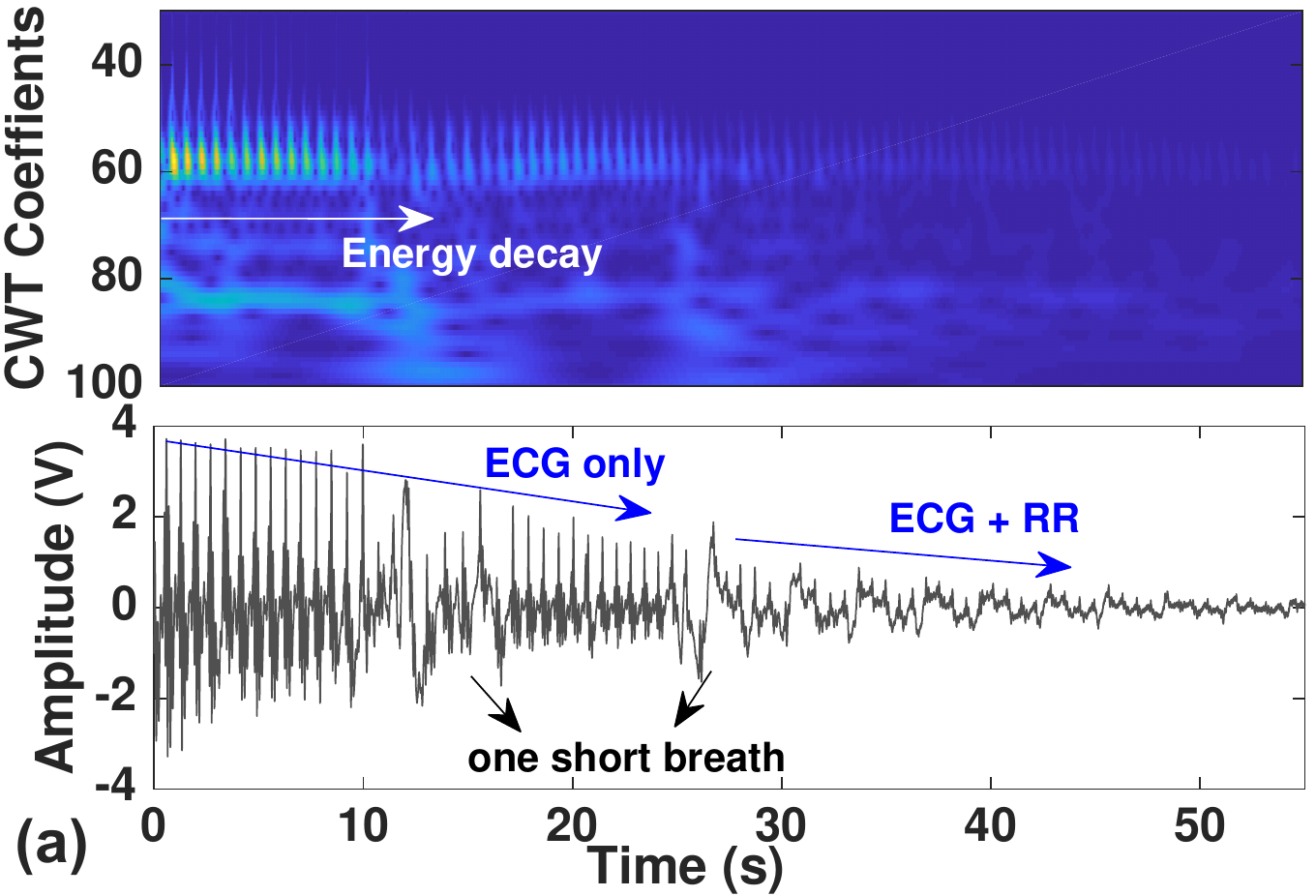}
	\includegraphics[width = 0.32\columnwidth]{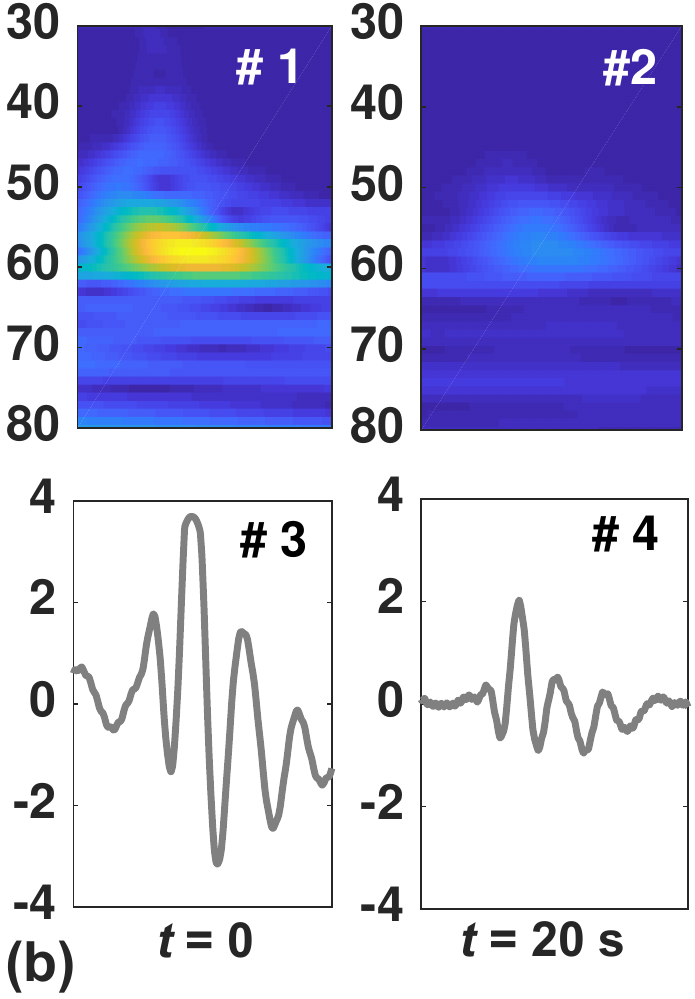}
	\caption{(a) Time-domain waveform (bottom) and its spectrum (top) measured after the subject generated a large amount of triboelectric charge at $t=0$; (b) zoomed-in views of the time- and frequency-domain signals for a single cardiac cycle centered around $t=0$ (left) and $t=20$~sec (right).}
	\label{fig:remote_ecg_tribo}
\end{figure}

The potential importance of the BCG component is illustrated in Fig.~\ref{fig:remote_ecg_tribo}. In this experiment, the subject generated triboelectric charge by mechanically rubbing his outer clothing layer (a woollen sweater). The resulting charge increased the measured cardiac signal amplitude by approximately $40\times$ (from $\sim$200~mV$_{pp}$ to 8~V$_{pp}$) before dissipating (in an exponential manner) with a time constant $\tau\approx 25$~sec, as shown in Fig.~\ref{fig:remote_ecg_tribo}(a). Zoomed-in time- and frequency-domain waveforms for a single cardiac cycle centered around $t=0$ and 20~sec are shown in Fig.~\ref{fig:remote_ecg_tribo}(b). The plot confirms that the waveform gradually transitions from BCG-like (at $t=0$) to ECG-like as the triboelectric charge dissipates. From now on, we use ``ECG'' as a short-hand for the measured cardiac signal, while remembering that it also includes a BCG component.

The measured decay rate of the combined cardiac signal (defined as the peak of the FFT spectrum) follows $s(d) \approx k/d^{2.5}$, where $d$ is the body-electrode distance and $k$ is a constant. This relatively rapid decay is due to two effects. The first is decay of coupling capacitance $C_{c}$ with distance due to the field sharing effect (discussed in Section~\ref{sec:opt_design}). The second is the relatively localized distribution of ECG potential (and charge) on the body surface, which has been estimated to have a half-maximum width of $\sim$10~cm~\cite{he1992body}. As a result, the assumption that the body surface is an equipotential, which was used to derive $C_{c}$, breaks down as $d$ increases This effect also causes $C_{c}$ (and thus the detected ECG signal amplitude) to decay faster with distance than in our idealized model. 

\begin{figure}[htbp]
	\centering
	\includegraphics[width = 0.475\columnwidth]{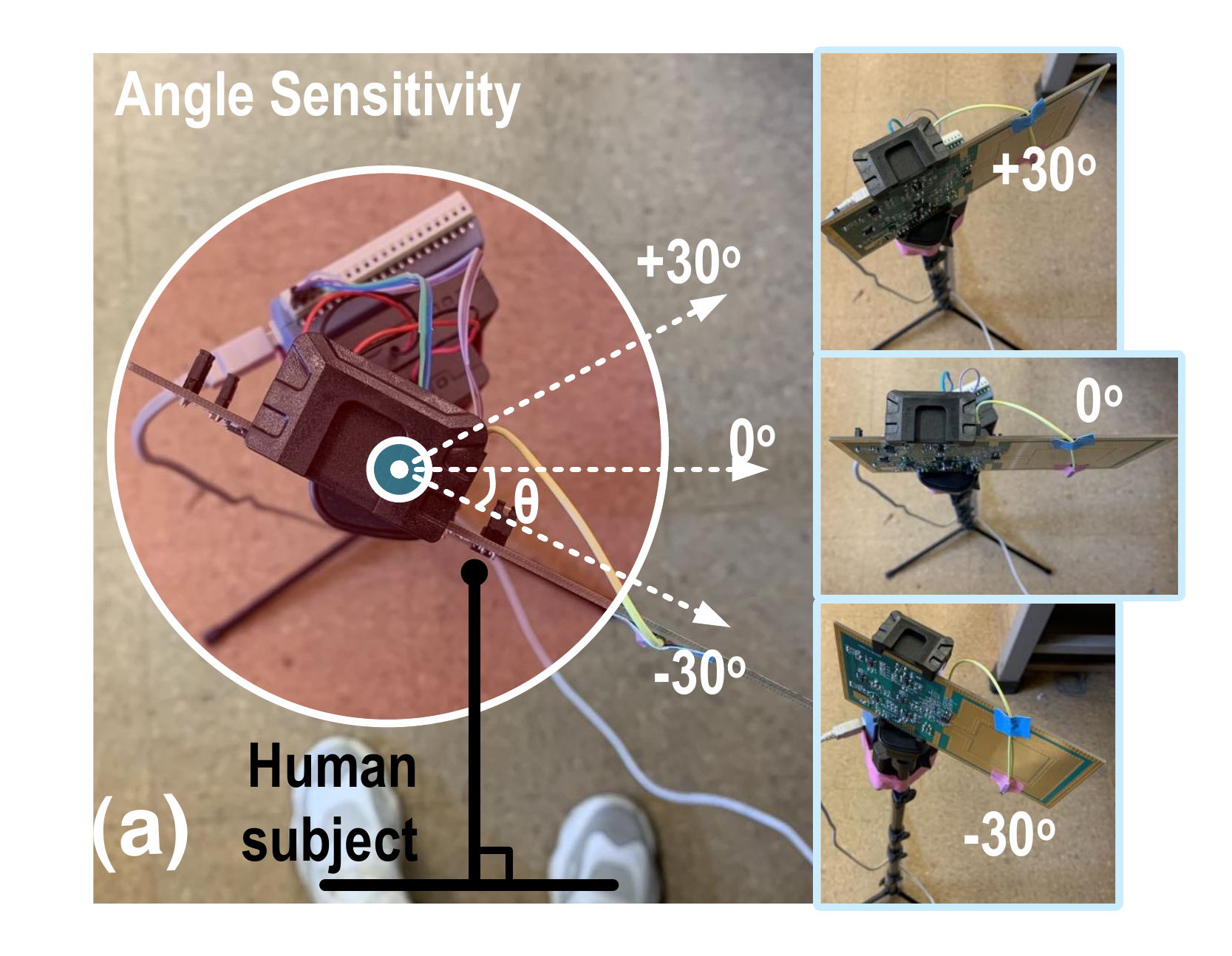}
	\includegraphics[width = 0.51\columnwidth]{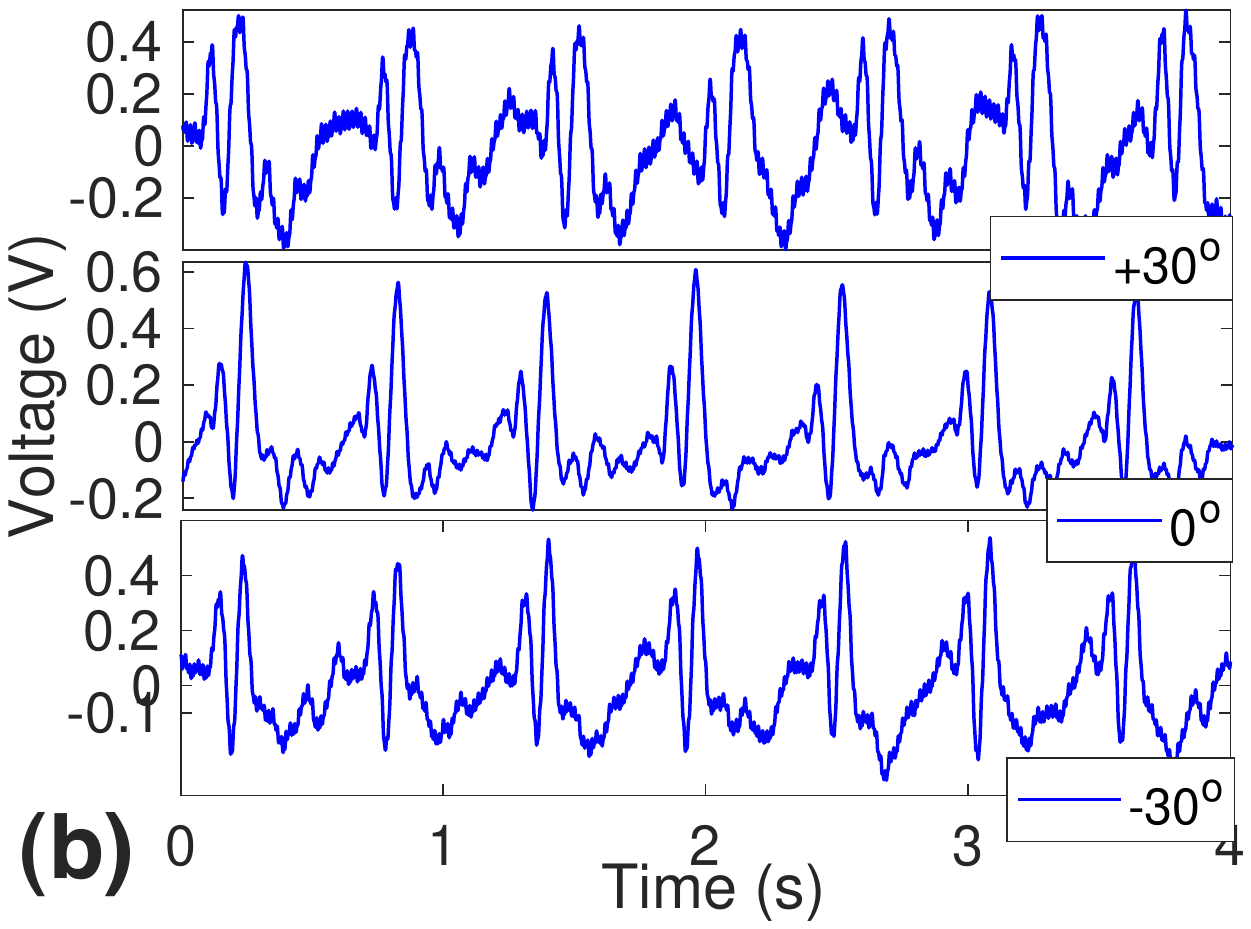}
	\caption{{(a) Experimental setup used for studying the angle sensitivity of the proposed EPS; (b) Non-contact sensing of ECG activity when the sensor is positioned along $+30^{\circ}$, $0^{\circ}$, and $-30^{\circ}$.}}
	\label{fig:angle_sen}
\end{figure}

{Fig.~\ref{fig:angle_sen}(a) shows the experimental setup used to study the angular dependence of the proposed EPS. During this test, the plane of the EPS electrode was rotated to different angles ($\theta$) while the human subject sat at a relatively constant sensor-subject distance ($\sim$30~cm). Fig.~\ref{fig:angle_sen}(b) shows the corresponding measured ECG signals, which can be clearly observed at each angle ($+30^{\circ}$, $0^{\circ}$ and $-30^{\circ}$). However, the waveforms are angle-dependent. Notably, the amplitude of the ECG-like component decreases with $|\theta|$, while that of the BCG-like component remains relatively constant. Finally, we confirmed that the EPS can simultaneously detect ECG waveforms from multiple subjects. However, developing the necessary source separation algorithms is beyond the scope of this paper.}

\subsubsection{Synchronization with Contact ECG Recordings}
The ECG measurements discussed above were further confirmed using synchronization experiments. For this purpose, the non-contact EPS was validated using the reference contact sensor (BITalino wireless physiological recording platform using conventional patch electrodes). To synchronize the two recordings, we introduced signal artifacts from large body motions during the experiments and aligned them during post-processing (using MATLAB) before further data analysis.

\begin{figure}[htbp]
	\centering
	\includegraphics[width = 0.49\columnwidth]{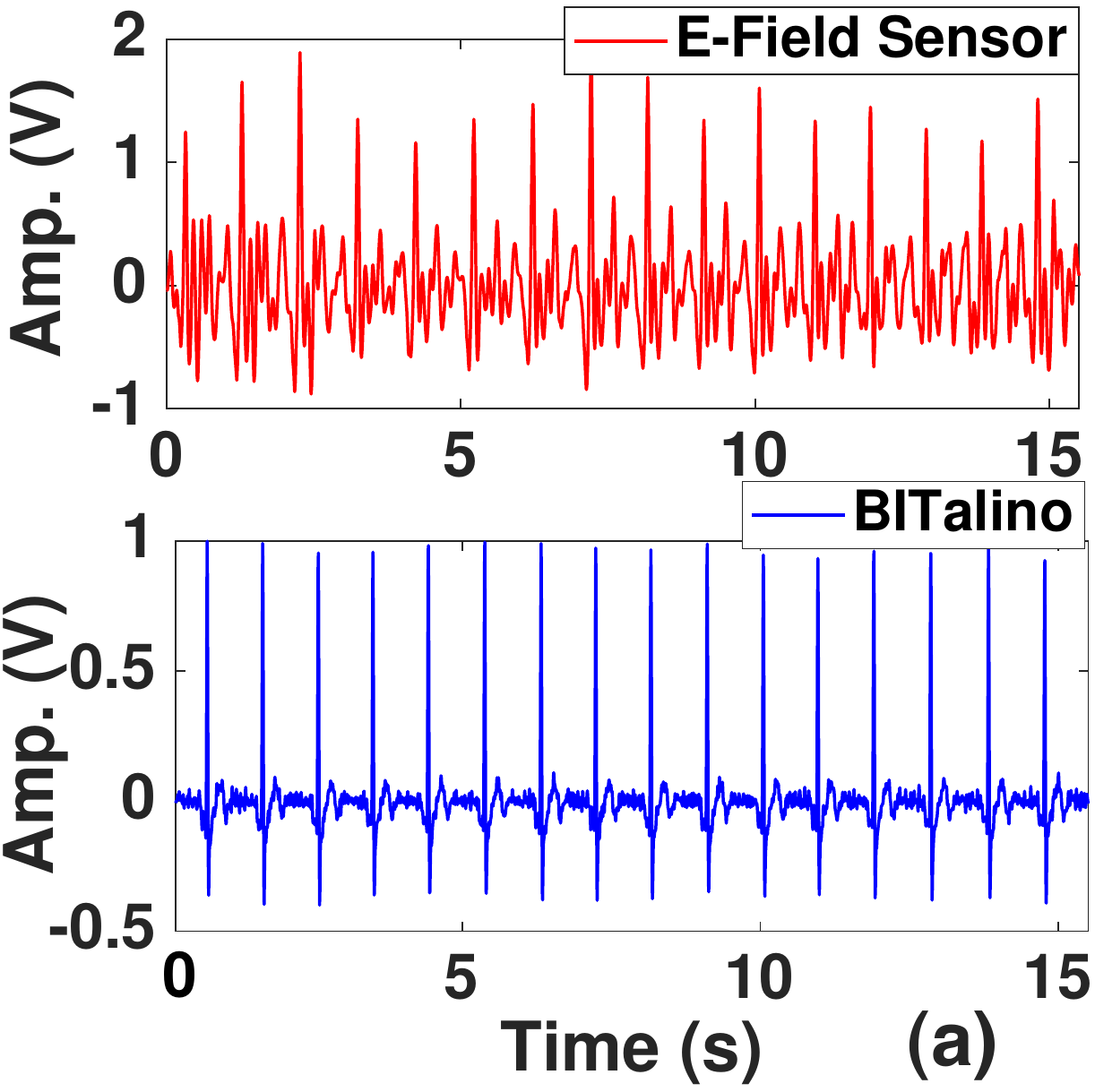}
	\includegraphics[width = 0.49\columnwidth]{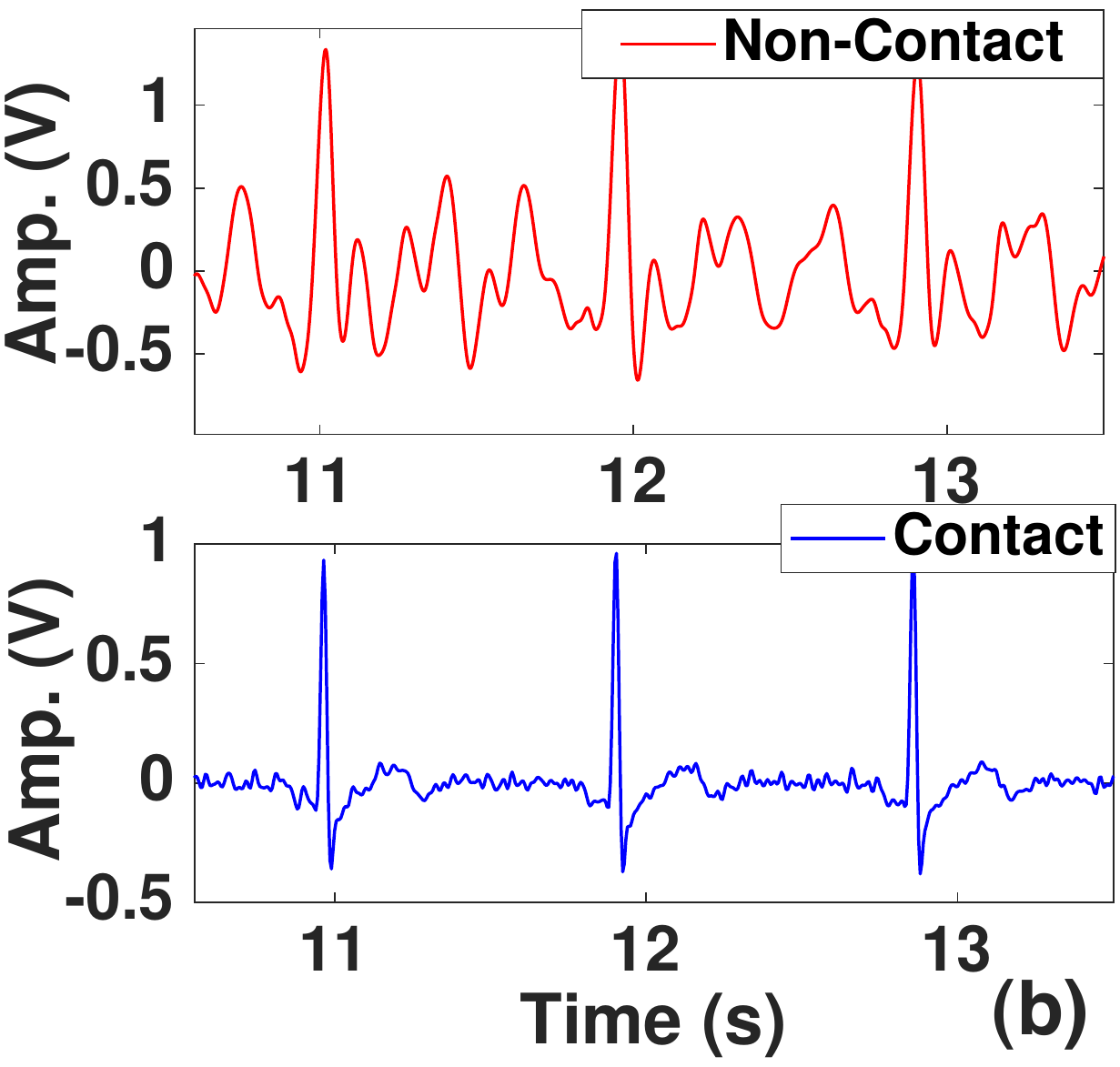}
	\caption{(a) Synchronization of two ECG recordings from the reference contact sensor (BITalino) and our custom E-Field sensor; (b) a zoomed-in view of the two recordings.}
	\label{fig:remote_ecg_com}
\end{figure}

Fig.~\ref{fig:remote_ecg_com}(a) shows two typical recordings from the reference sensor and the EPS (at $d=15$~cm) that were synchronized using a body motion event. The zoomed-in view in Fig.~\ref{fig:remote_ecg_com}(b) shows that QRS complexes from the non-contact sensor are i) clearly observed with good SNR, and ii) well-aligned in the time domain with those from the reference sensor.

\begin{figure}[htbp]
	\centering
	\includegraphics[width = 0.48\columnwidth]{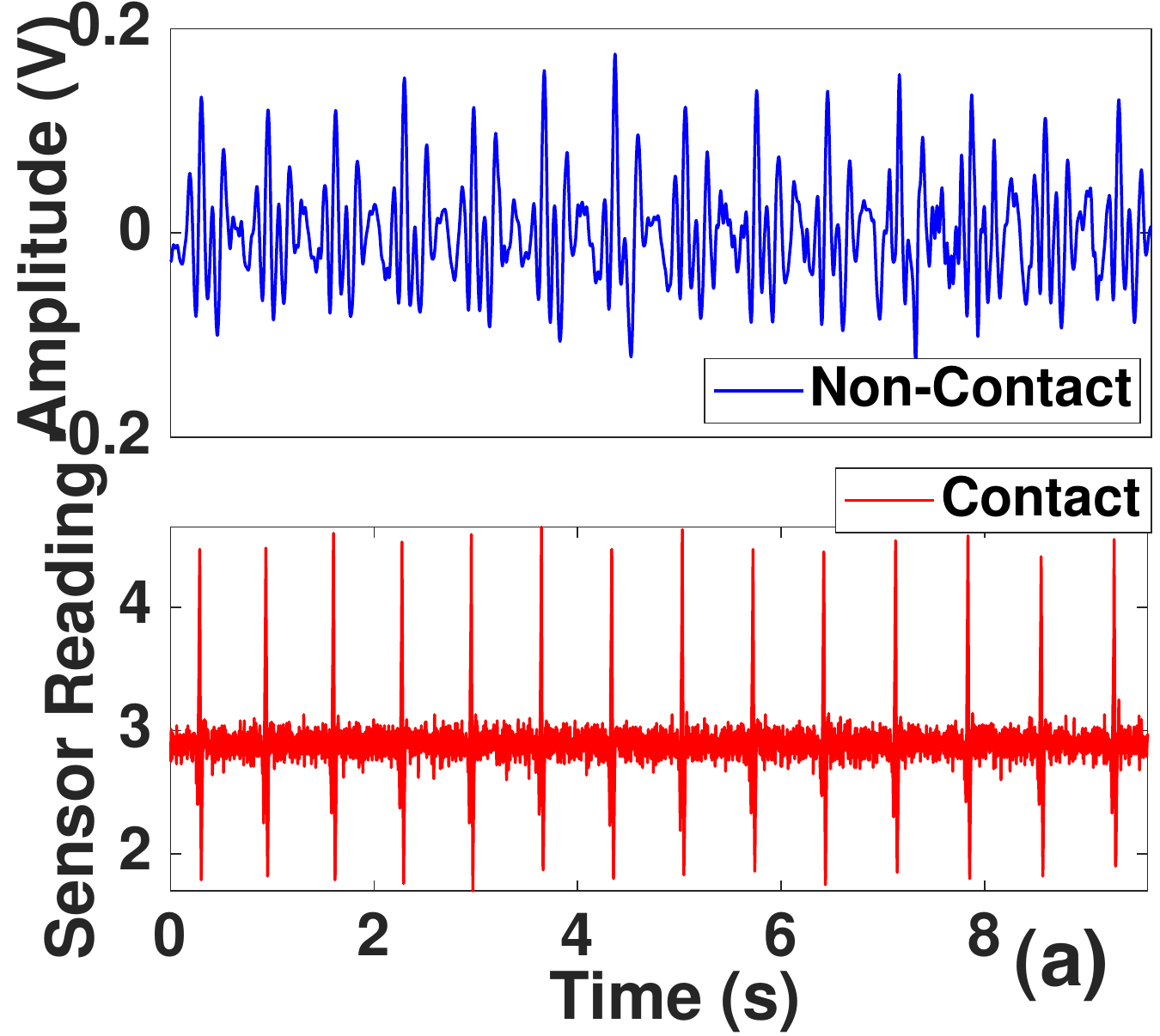}
	\includegraphics[width = 0.48\columnwidth]{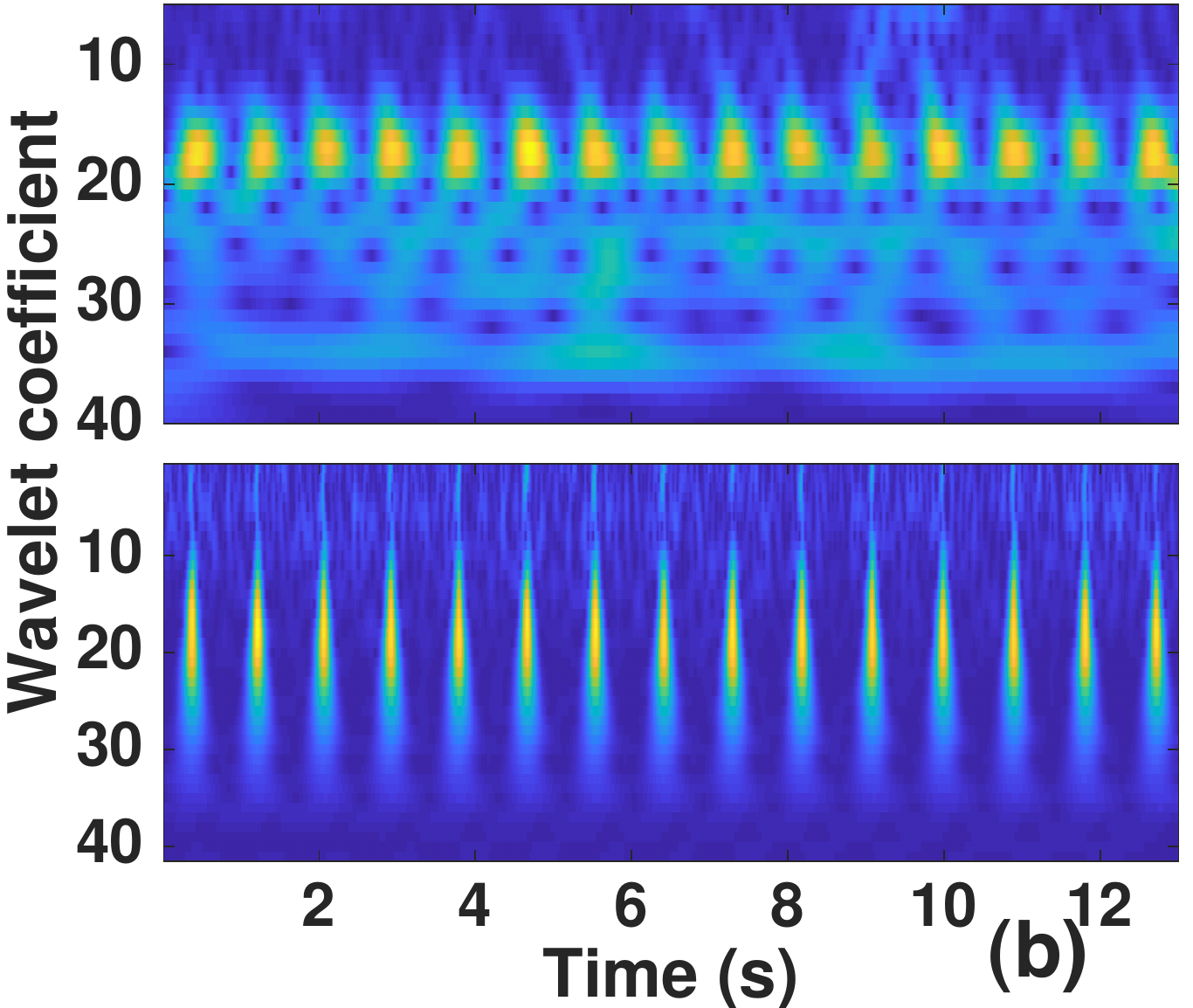}
	\caption{(a) Typical synchronized recordings from our non-contact sensor (at $d=40$~cm) and the reference contact sensor; (b) power spectrum of two synchronized ECG recordings using CWT.}
	\label{fig:remote_ecg_com2}
\end{figure}

Fig.~\ref{fig:remote_ecg_com2}(b) shows the power spectrum of two ECG recordings (top: from the custom EPS at $d=40$~cm; bottom: from the reference contact sensor) estimated using a CWT. The power spectra are in good agreement with each other, as for the time-domain data shown in Fig.~\ref{fig:remote_ecg_com2}(a). No statistically significant differences were observed between the HR values estimated from these two synchronized recordings.

\begin{figure}[htbp]
	\centering
	\includegraphics[width = 0.47\columnwidth]{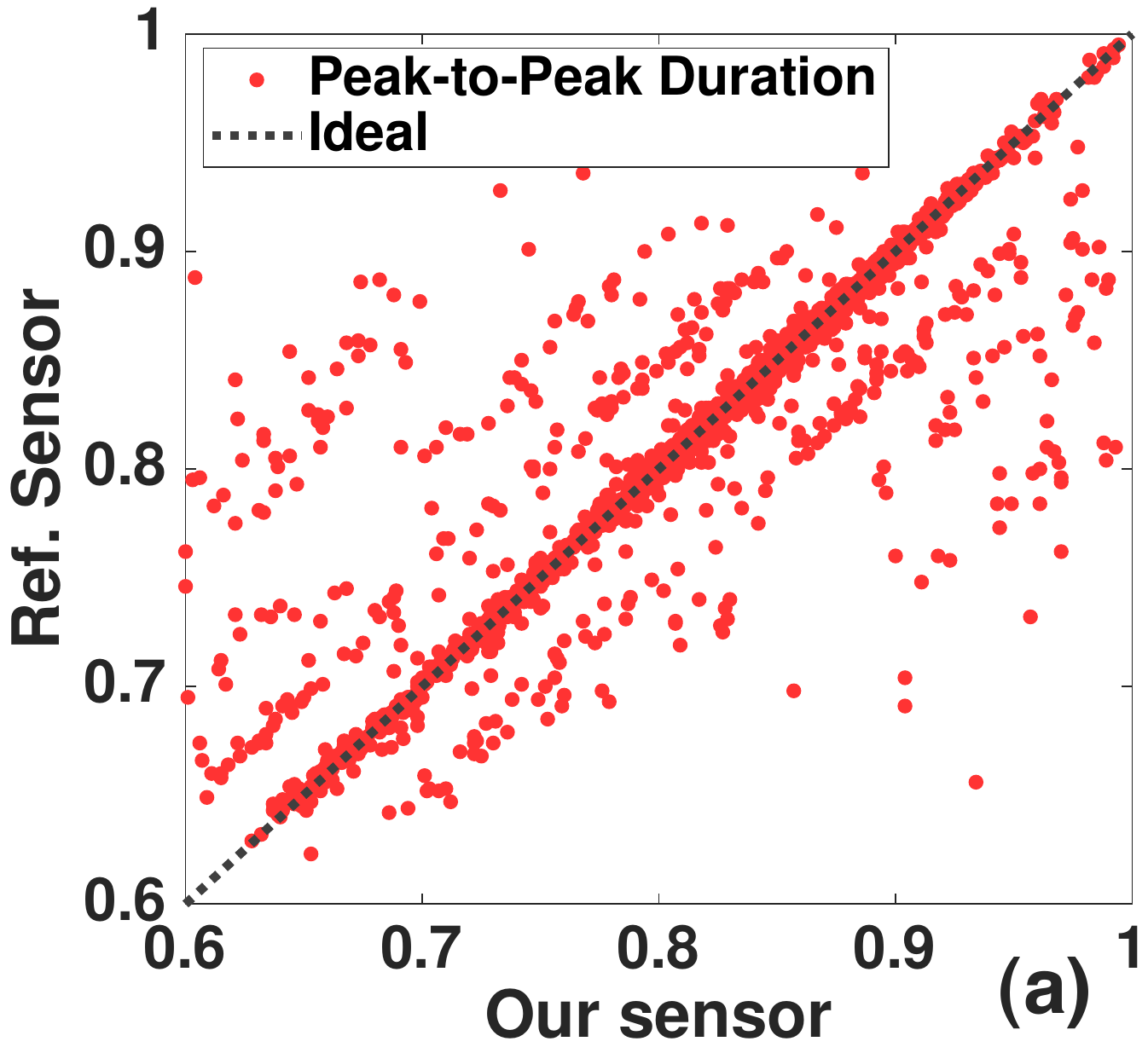}
	\includegraphics[width = 0.485\columnwidth]{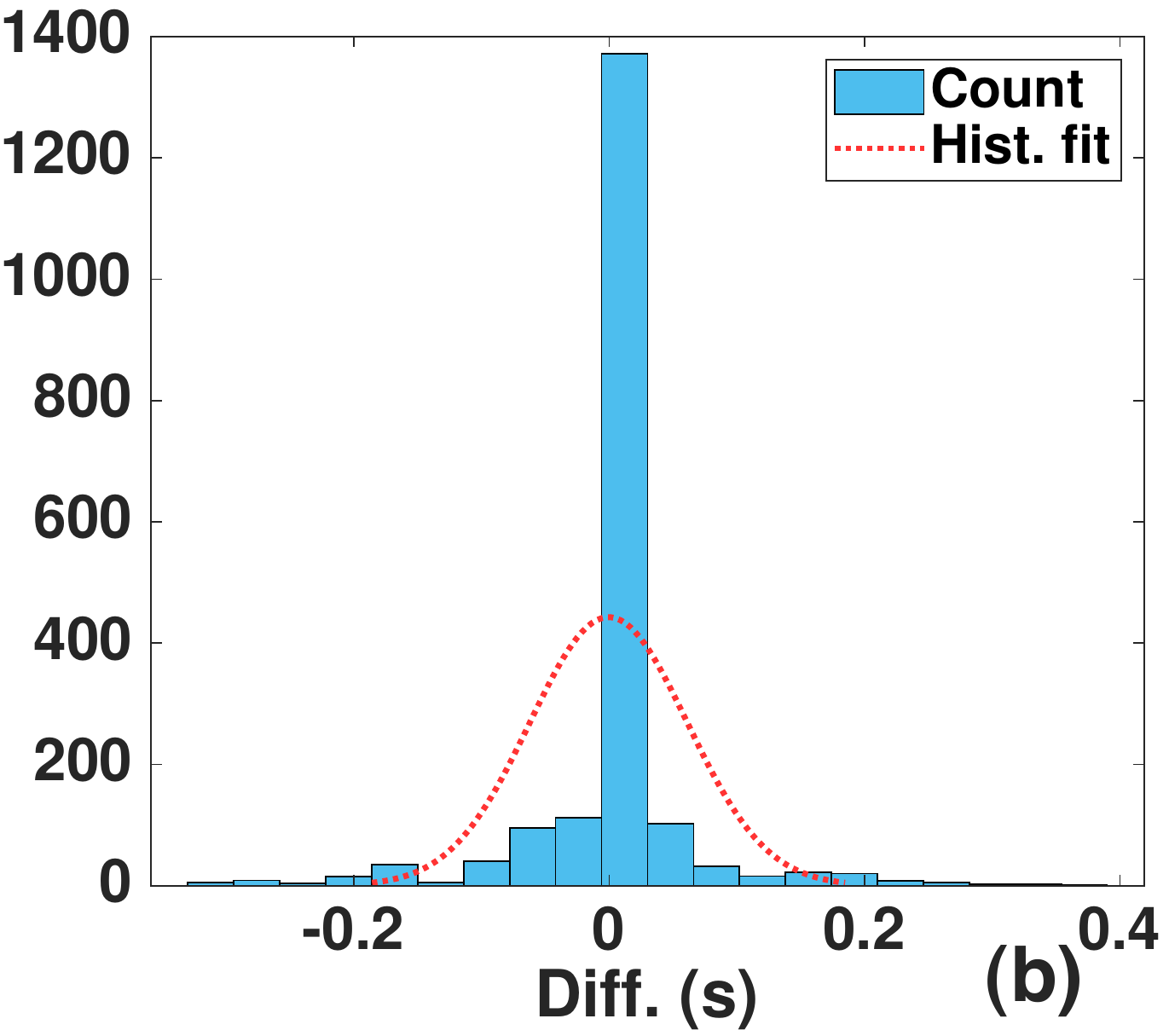}
	\caption{(a) Comparison of the extracted heart beat durations using the reference sensor and our custom sensor for sensing distances from 15-50 cm; (b) histogram of the differences between beat durations measured by the two sensors, along with a Gaussian fit to the data.}
	\label{fig:remote_ecg_stat}
\end{figure}

Beat durations (durations between peaks of the QRS complexes) were estimated to evaluate agreement between the contact and non-contact recordings as a function of time. Fig.~\ref{fig:remote_ecg_stat}(a) compares extracted beat durations using the reference sensor and the EPS ($d=15$~cm to 50~cm) over 48 recordings ($\sim$1900 beat durations). The two are strongly correlated, with a mean timing difference of $-0.528$~ms and a standard deviation of 3.8~ms. Fig.~\ref{fig:remote_ecg_stat}(b) shows the histogram of the differences between measured beat durations (i.e., measurement errors for the EPS, assuming the contact sensor as a reference) and a Gaussian fit to this data.

\begin{figure}[htbp]
	\centering
	\includegraphics[width = 0.49\columnwidth]{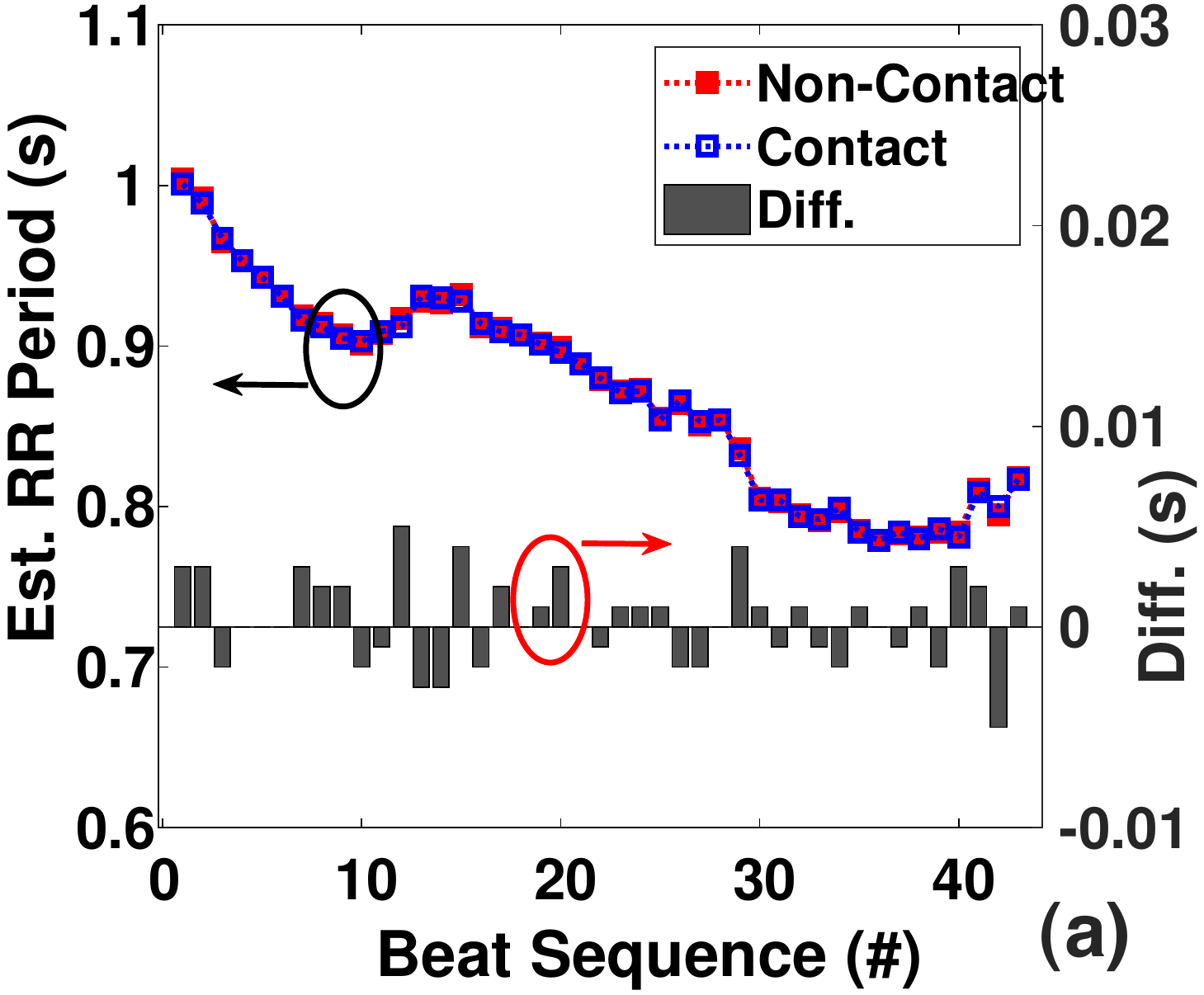}
	\includegraphics[width = 0.49\columnwidth]{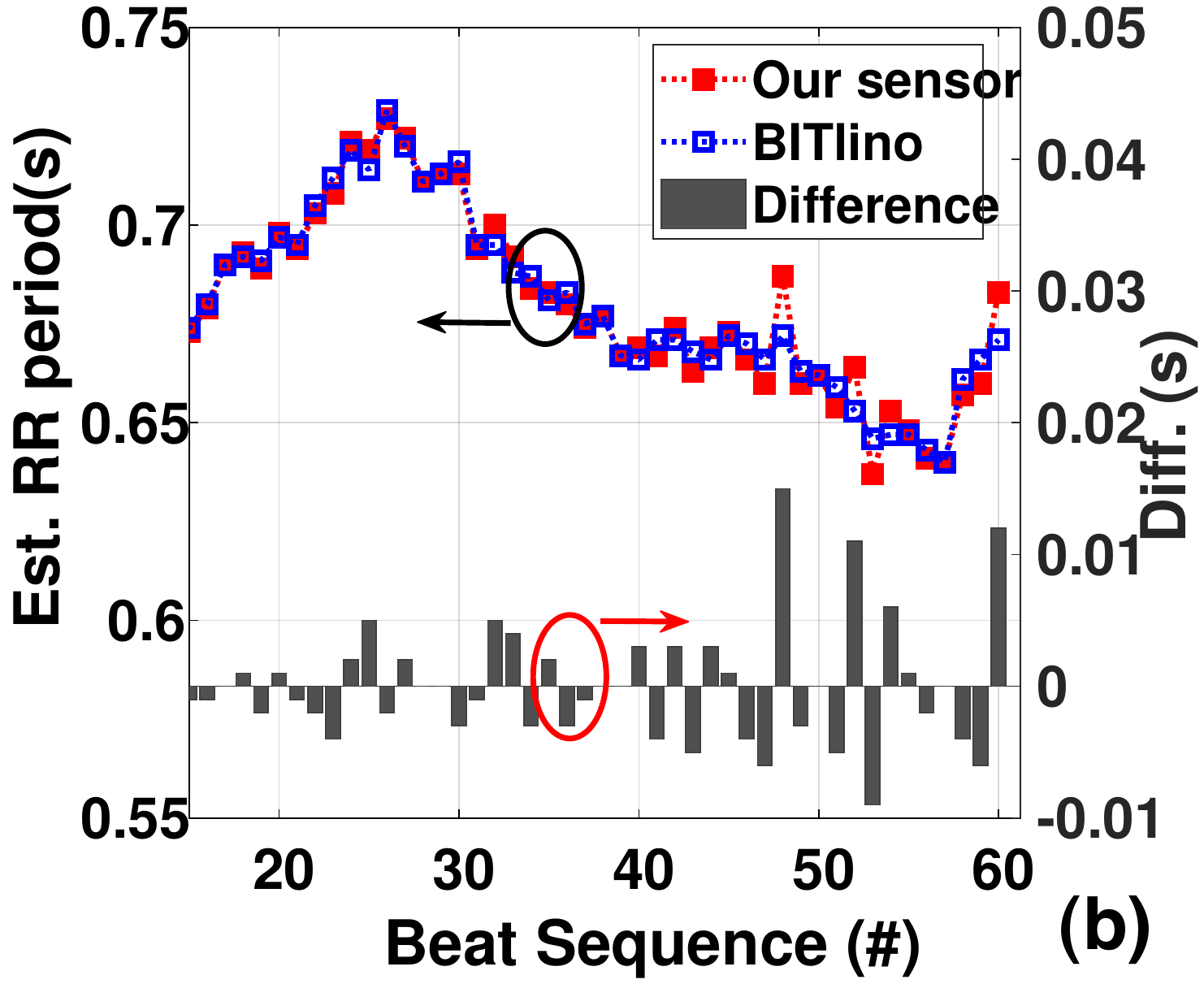}
	\caption{Beat duration (R-R interval) comparison of the reference contact sensor and our proposed non-contact sensor along two typical heart beat sequences at (a) $d=30$~cm, and (b) $d=40$~cm. The corresponding maximum timing difference are $<5$~ms and $<15$~ms, respectively.}
	\label{fig:remote_ecg_seq}
\end{figure}

Fig.~\ref{fig:remote_ecg_seq} shows two examples (at $d=30$~cm and 40~cm) of measured beat durations (R-R intervals) along a typical beat sequence (shown as time steps). The sequences measured by the two sensors are in good agreement. The bin graphs show the timing differences between the two sequences, which are larger for Fig.~\ref{fig:remote_ecg_seq}(b) due to the decrease in SNR with distance. 

\begin{figure}[htbp]
\centering
\includegraphics[width = 0.48\columnwidth]{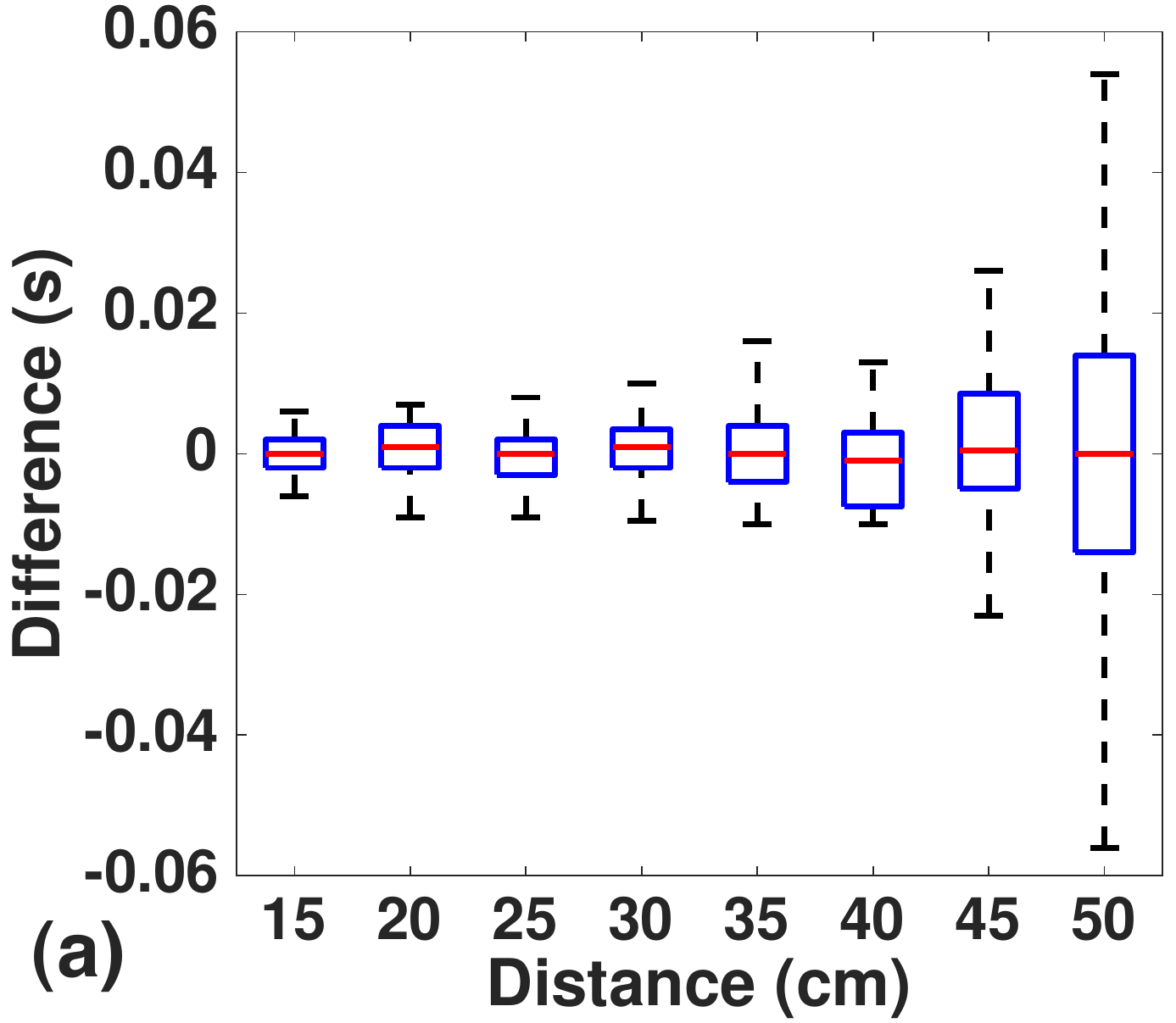}
\includegraphics[width = 0.48\columnwidth]{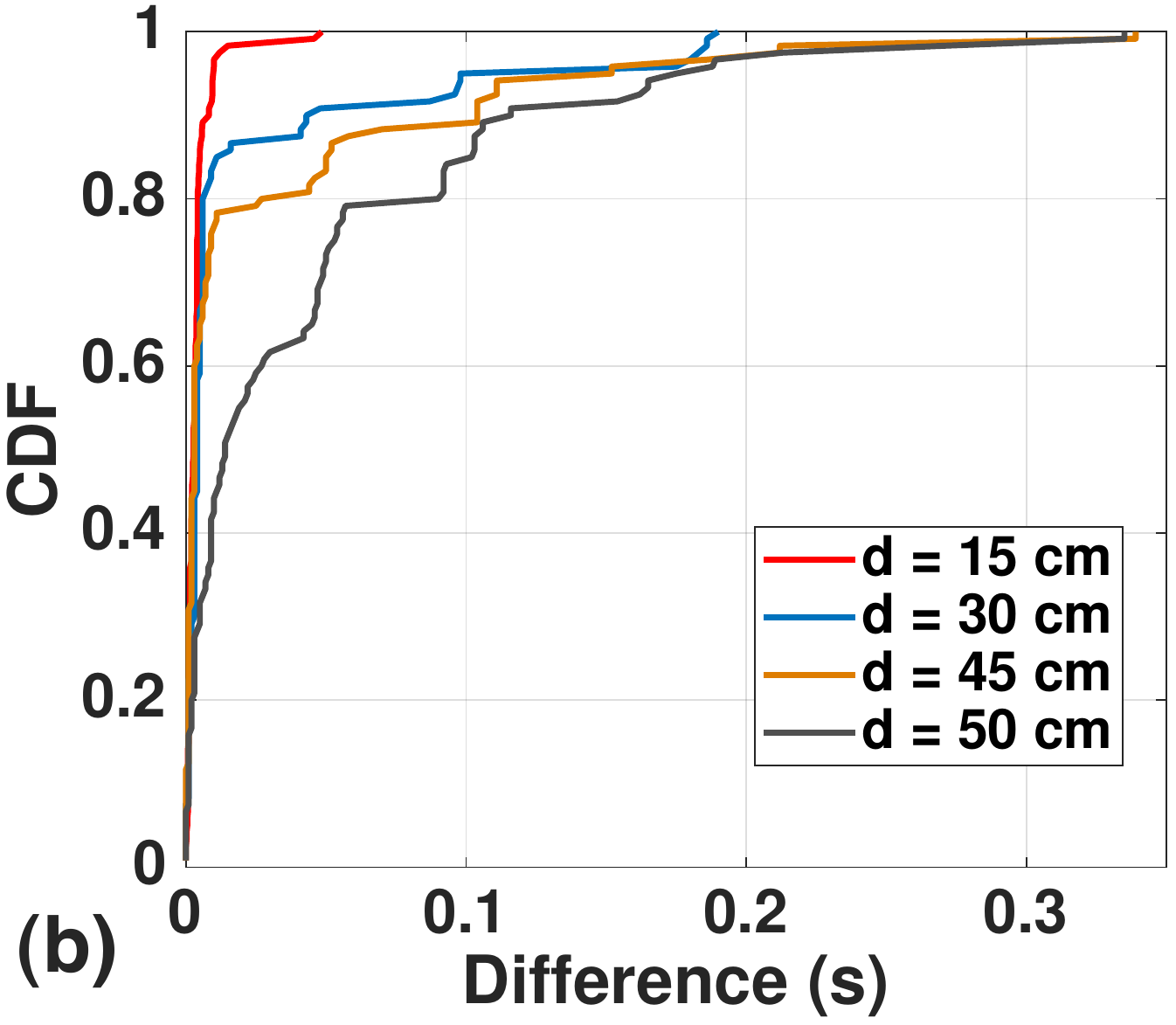}
\caption{(a) Heart beat duration errors obtained from measured ECG data when the human subject sits at different distances, outliers are not included; (b) selected cumulative density functions (CDFs) of the duration errors.}
\label{fig:remote_ecg_dist}
\end{figure}

Fig.~\ref{fig:remote_ecg_dist}(a) confirms that the timing differences increase as $d$ increases; this is because the signal strength decays as $1/d^{3}$, leading to larger errors in duration estimates. The corresponding cumulative density function (CDF) of the differences in beat duration is shown in Fig.~\ref{fig:remote_ecg_dist}(b) for different values of $d$. Considering  the  typical cardiac cycle  duration of $\sim$1.1~sec, the figure suggests that the proposed EPS shows promise for measuring HRV at distances up to at least 50~cm. HRV is widely used to characterize the many physiological factors modulating the normal rhythm of the heart, and is a useful indicator of both current and impending cardiac diseases~\cite{acharya2006heart}.

\subsection{Non-contact Respiratory Cycle (RC) Monitoring}
Non-contact RC signals were measured at different sensor-subject distances. Specifically, the subject was located at distances of $d = 10$~cm to 100~cm (with a step of 10~cm) from the surface of the electrode. Each experimental setting was repeated 5 times to reduce the effects of sensor-subject distance variations during the experiments. 

\begin{figure}[htbp]
	\centering
	\includegraphics[width = 0.49\columnwidth]{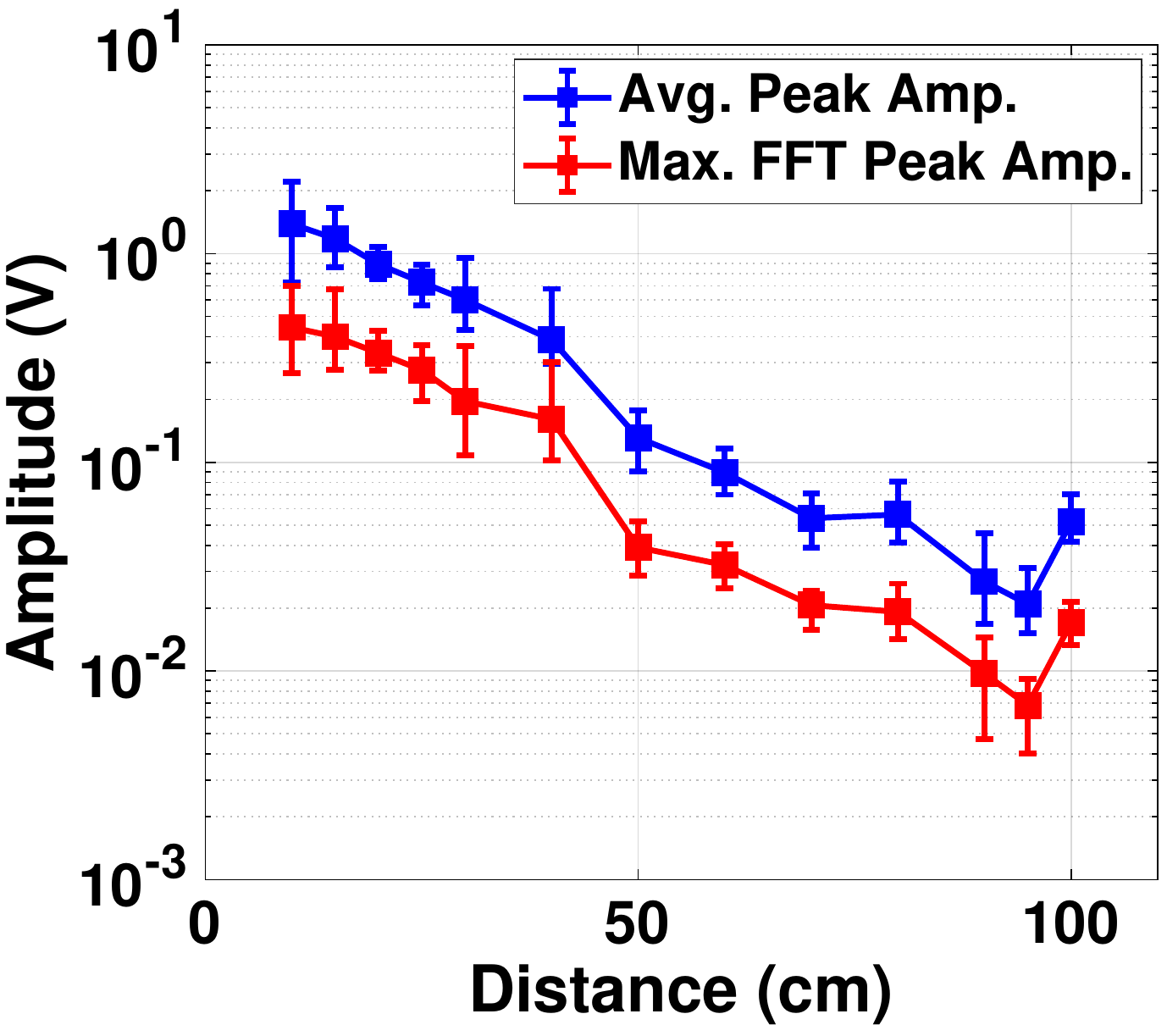}
	\includegraphics[width = 0.45\columnwidth]{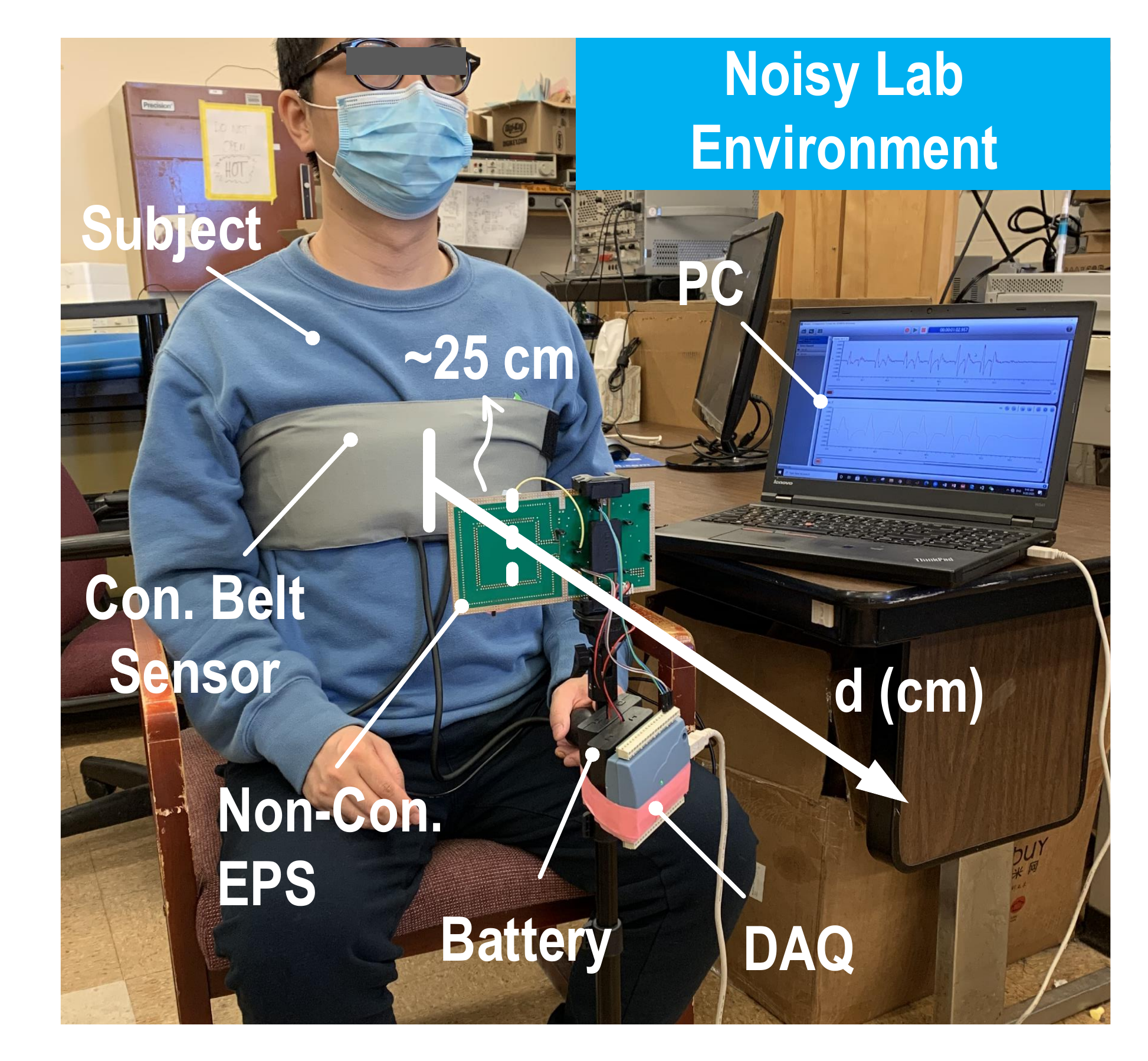}
	\caption{(a) Measured respiration cycle (RC) amplitudes at different sensor-subject distances (10 cm to 100 cm); and (b) the experimental setup.}
	\label{fig:remote_rr_snr}
\end{figure}

Fig.~\ref{fig:remote_rr_snr}(a) shows that the measured RC signal amplitude decays with distance $d$ approximately as $s(d)=k/d^{2}$. The measured decay rate can be explained as follows. RC sensing is based on detecting changes in the coupling capacitance $C_{c}$ due to chest wall motion (denoted by $\Delta C_{c}$). In particular, the fact that $C_{c}$ varies with time induces current variations ($I_{in}=dQ_c/dt=V_{body}dC_{c}/dt$) at the electrode assuming that human body potential $V_{body}$ remains approximately constant. If we model the field sharing effects by assuming that $C_{c}\propto 1/d$ (as in a parallel-plate capacitance model), the amplitude of the resulting AC current scales as $\Delta C_{c} \propto 1/d^{2}$:
\begin{align}
\nonumber    I_{in}&=V_{body}\frac{d}{dt}\left(\frac{\epsilon_0 A}{d_0+\Delta d\sin(\omega_{RR} t)}\right)\\
    &\approx -V_{body}\left(\frac{\epsilon_0 A}{d_0^2}\right)\omega_{RR}\Delta d \cos(\omega_{RR}t),\quad \Delta d\ll d_0
\end{align}
where $\epsilon_{0}$ is permittivity of air, $d_0$ is the average distance from chest to sensing electrode, $A$ is the equivalent plate area of the coupling capacitor (somewhat larger than the physical electrode area $w^2$ due to fringe fields), and for simplicity we have assumed a sinusoidal motion of the chest wall with amplitude $\Delta d$ and frequency $\omega_{RR}$. The resulting TIA output voltage ($V_{out}=I_{in}\times R_f$) is proportional to the derivative of the chest wall motion. It can thus be integrated with time to obtain a waveform that is directly proportional to the motion:
\begin{equation}
V_{int} = -R_{f} \int_{0}^{t}{I_{in}(t)dt}\approx V_{body}\left(\frac{\epsilon_0 A}{d_0^2}\right)\Delta d \sin(\omega_{RR}t).
\label{eq:remote_eq3}
\end{equation}

\begin{figure}[htbp]
	\centering
	\includegraphics[width = 0.48\columnwidth]{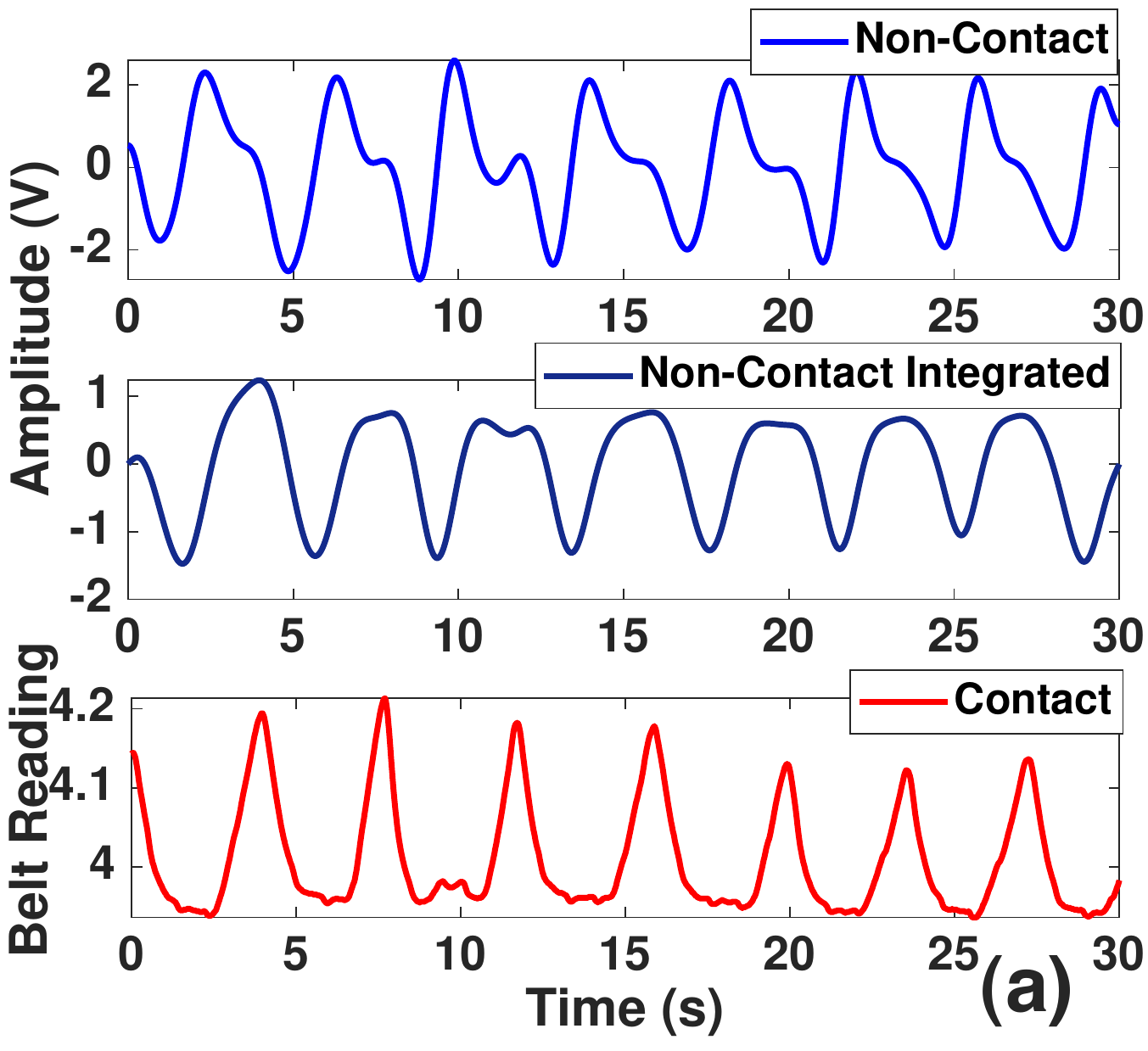}
	\includegraphics[width = 0.495\columnwidth]{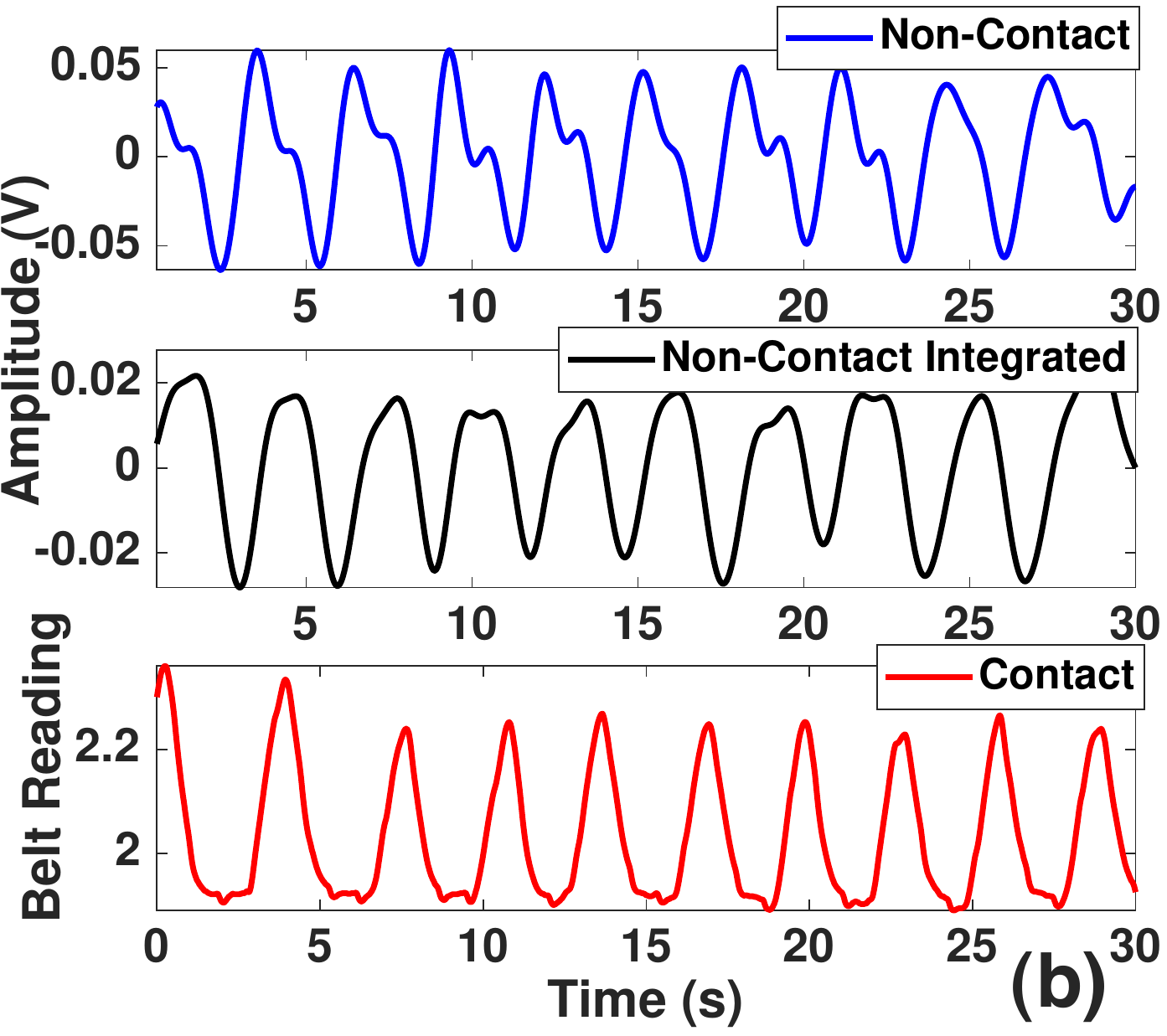}
	\caption{Two sets of synchronized respiration recordings from the reference contact sensor and our non-contact EPS at sensor-subject distances of 10~cm (a) and 100~cm (b). The middle row shows the non-contact measurements after integrating the received signal along time.}
	\label{fig:remote_rr_trans}
\end{figure}

The non-contact EPS was validated using the reference contact sensor (NeuLog USB respiration monitoring belt), as shown in Fig.~\ref{fig:remote_rr_snr}(b). Fig.~\ref{fig:remote_rr_trans} shows typical recordings from the two sensors that were synchronized using a cross-correlation method; the sensor-subject distances were (a) $d=10$~cm, and (b) $d=100$~cm, respectively. Note that the integrated non-contact signal $V_{int}$ is in phase with the recording from the belt sensor. As shown in eqn.~(\ref{eq:remote_eq3}), the AC component of $V_{int}(t)$ is in phase with the chest motion, and thus the belt sensor reading. In particular, the peaks of both signals correspond to peak lung pressure. Fig.~\ref{fig:remote_rr_trans} confirms excellent correlation of peak-to-peak durations and peak locations between the integrated signals and the contact belt readings.

\begin{figure}[htbp]
	\centering
	\includegraphics[width = 0.88\columnwidth]{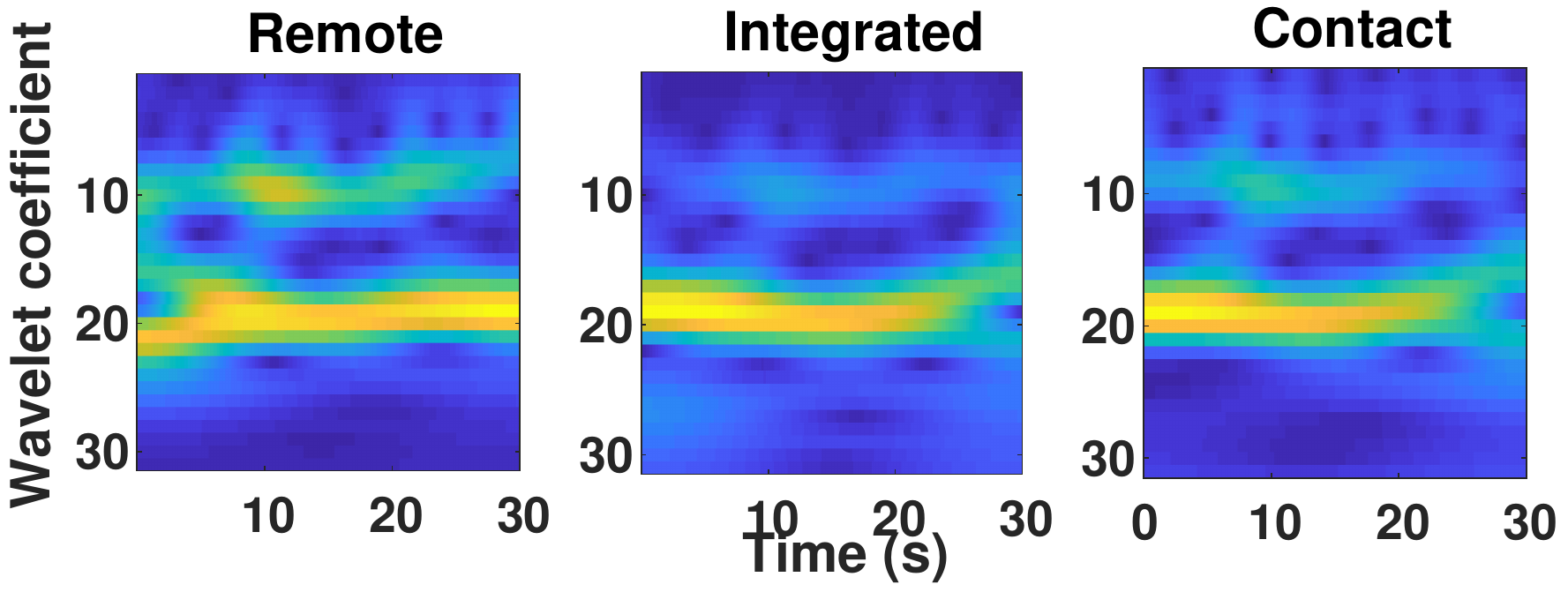}
	\includegraphics[width = 0.88\columnwidth]{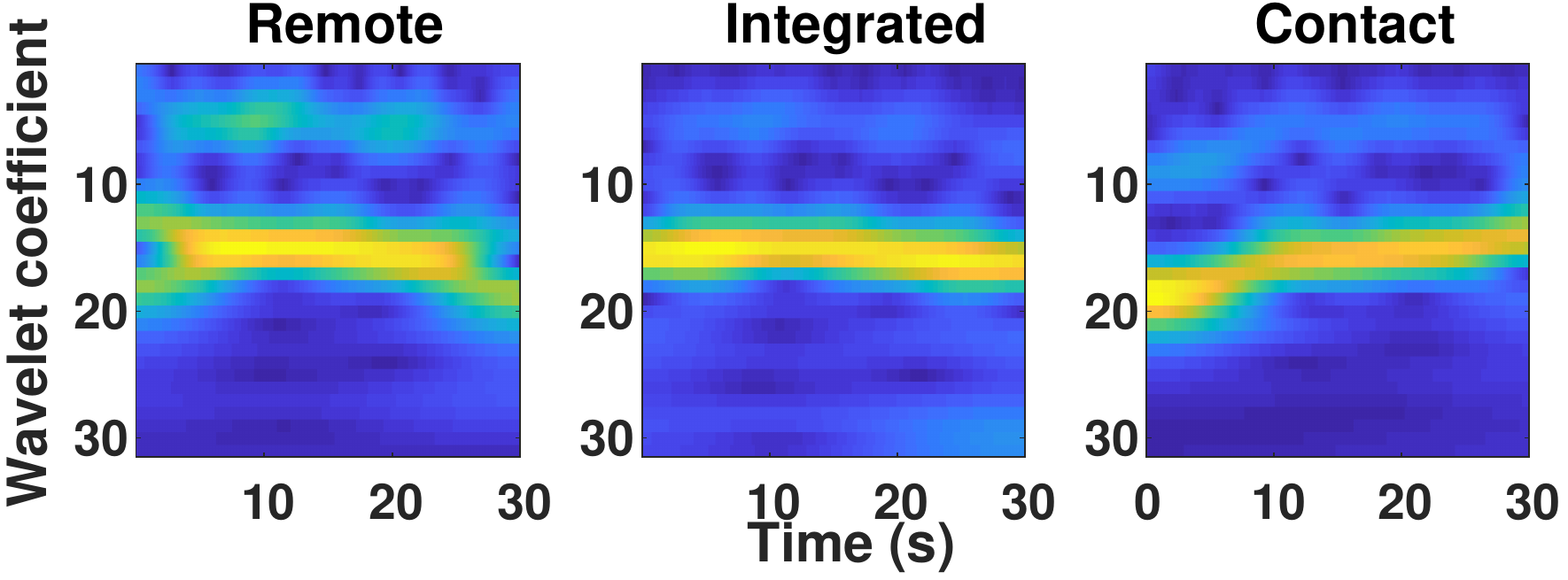}
	\caption{Power spectra of two synchronized respiration recordings (from Fig.~\ref{fig:remote_rr_trans} ) using CWT when the human subject is sitting at distances of 10 cm (top) and 100 cm (bottom) from the sensors.}
	\label{fig:remote_rr_sp}
\end{figure}

Fig.~\ref{fig:remote_rr_sp} plots the power spectra of the data shown in Fig.~\ref{fig:remote_rr_trans}, as estimated using a CWT; the spectra of the integrated EPS signals are in good agreement with those from the contact sensor. Also, there are no statistically significant differences between the RR values estimated from the two recordings.

\begin{figure}[htbp]
	\centering
	\includegraphics[width = 0.45\columnwidth]{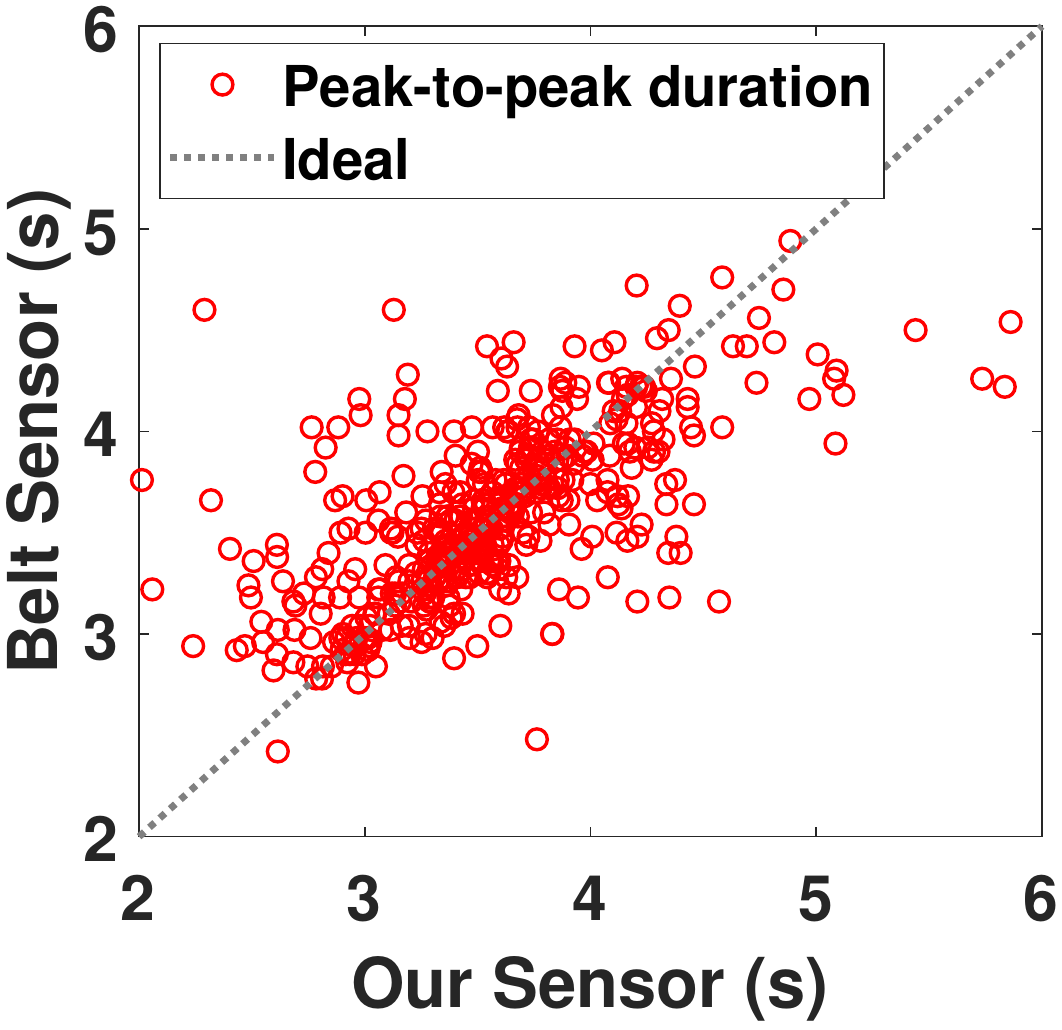}
	\includegraphics[width = 0.49\columnwidth]{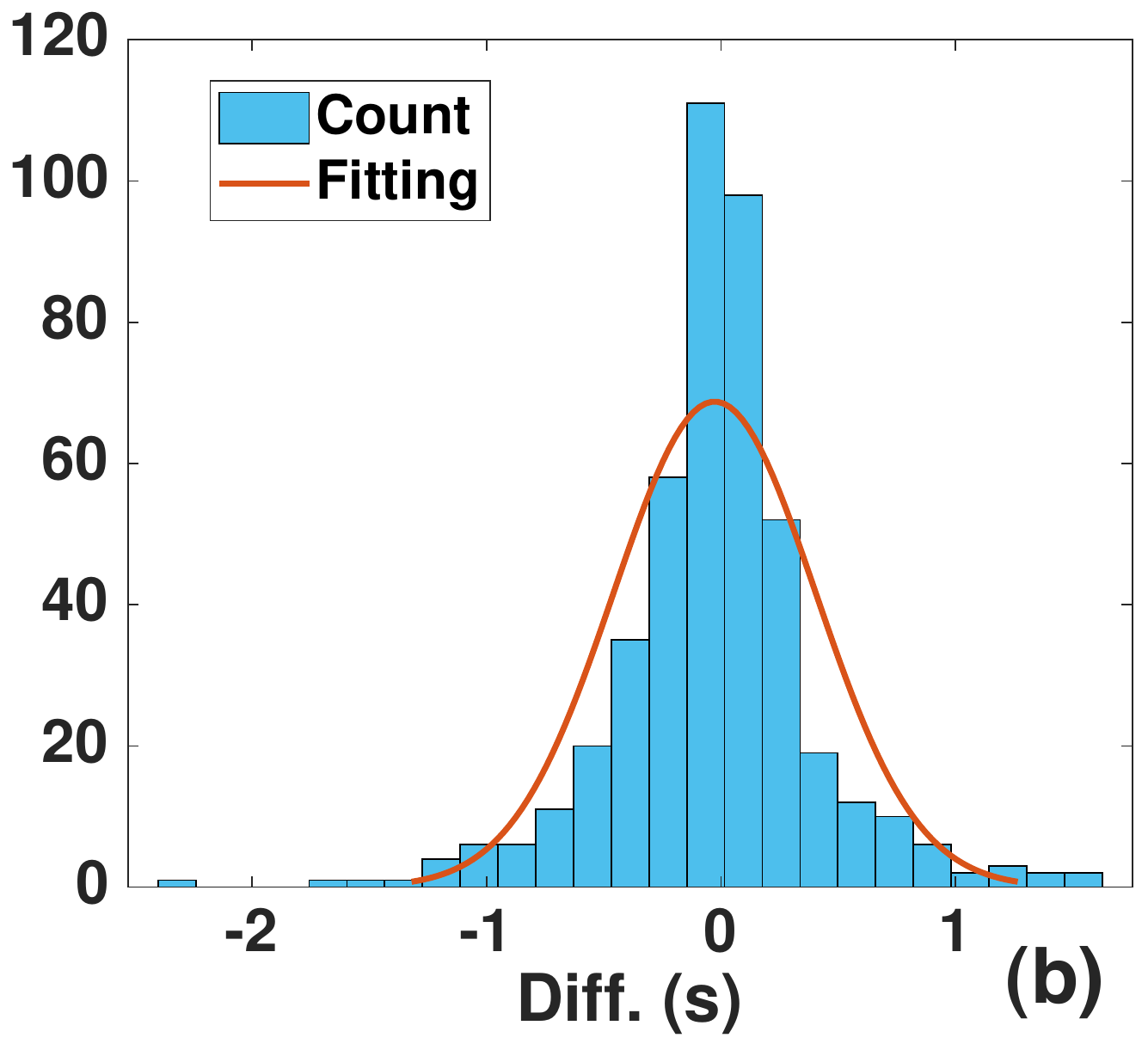}
	\caption{(a) Comparison of extracted peak-to-peak RC durations using the reference sensor and the EPS positioned at distances from $d=10$ to 100~cm; (b) histogram of the differences between RC durations measured by the two sensors, along with a Gaussian fit to the data.}
	\label{fig:remote_rr_stat}
\end{figure}

Fig.~\ref{fig:remote_rr_stat}(a) compares peak$-$to$-$peak durations extracted from multiple recordings (65 recordings, $\sim$461 durations) made using the reference sensor and the EPS (distance $d=10$~cm to 100~cm). The two are strongly correlated, with a mean timing difference of $-0.0273$~sec and a standard deviation of 0.185~sec. Fig.~\ref{fig:remote_rr_stat}(b) shows the histogram of the differences between measured peak$-$to$-$peak durations (i.e., measurement errors for the EPS, assuming the contact sensor as a reference) and a Gaussian fit to this data.

\begin{figure}[htbp]
	\centering
	\includegraphics[width = 0.49\columnwidth]{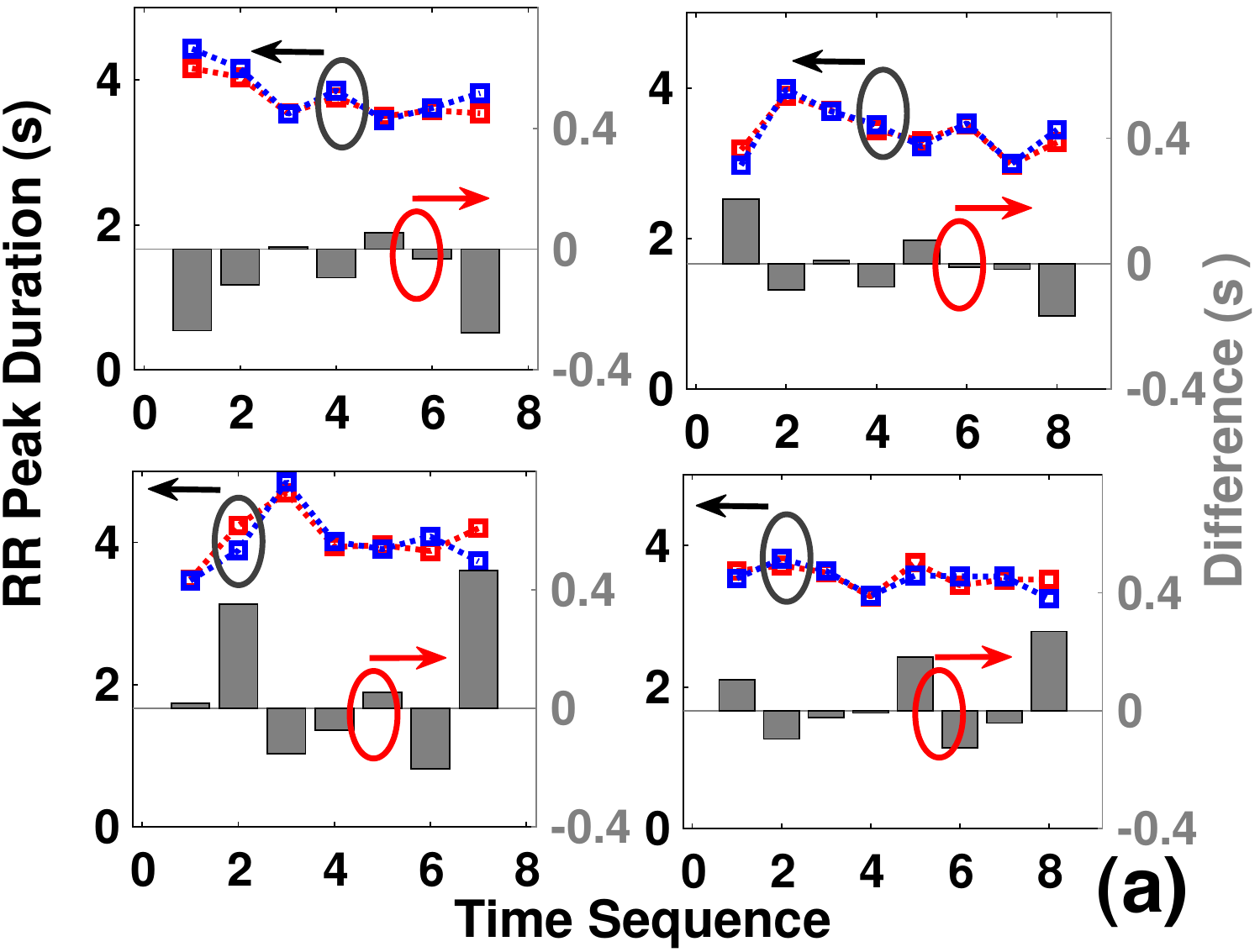}
	\includegraphics[width = 0.475\columnwidth]{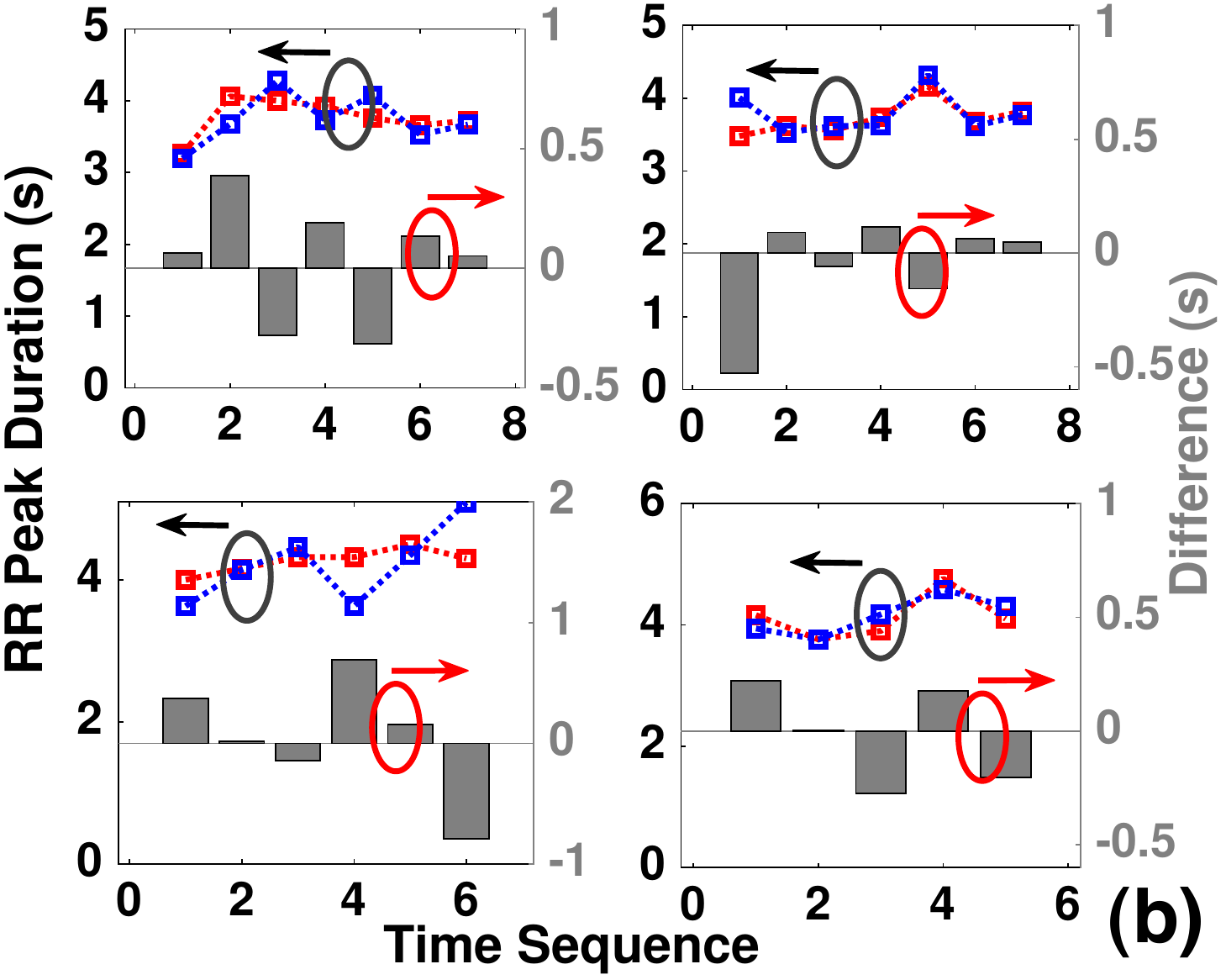}
	\caption{Peak-to-peak duration comparisons of the reference sensor and our custom sensor along several typical respiration cycle sequences, with human-sensor distances of 20~cm (a) and 70~cm (b).}
	\label{fig:remote_rr_seq}
\end{figure}

Fig.~\ref{fig:remote_rr_seq} shows two examples ($d=20$~cm and 70~cm) of measured peak-to-peak durations along a typical RC sequence (shown as time steps). The two sequences are in excellent agreement with each other. The bin graphs show the timing differences between the two sequences, which are larger for Fig.~\ref{fig:remote_rr_seq}(b) due to the decrease in SNR with distance.

\begin{figure}[htbp]
	\centering
	\includegraphics[width = 0.48\columnwidth]{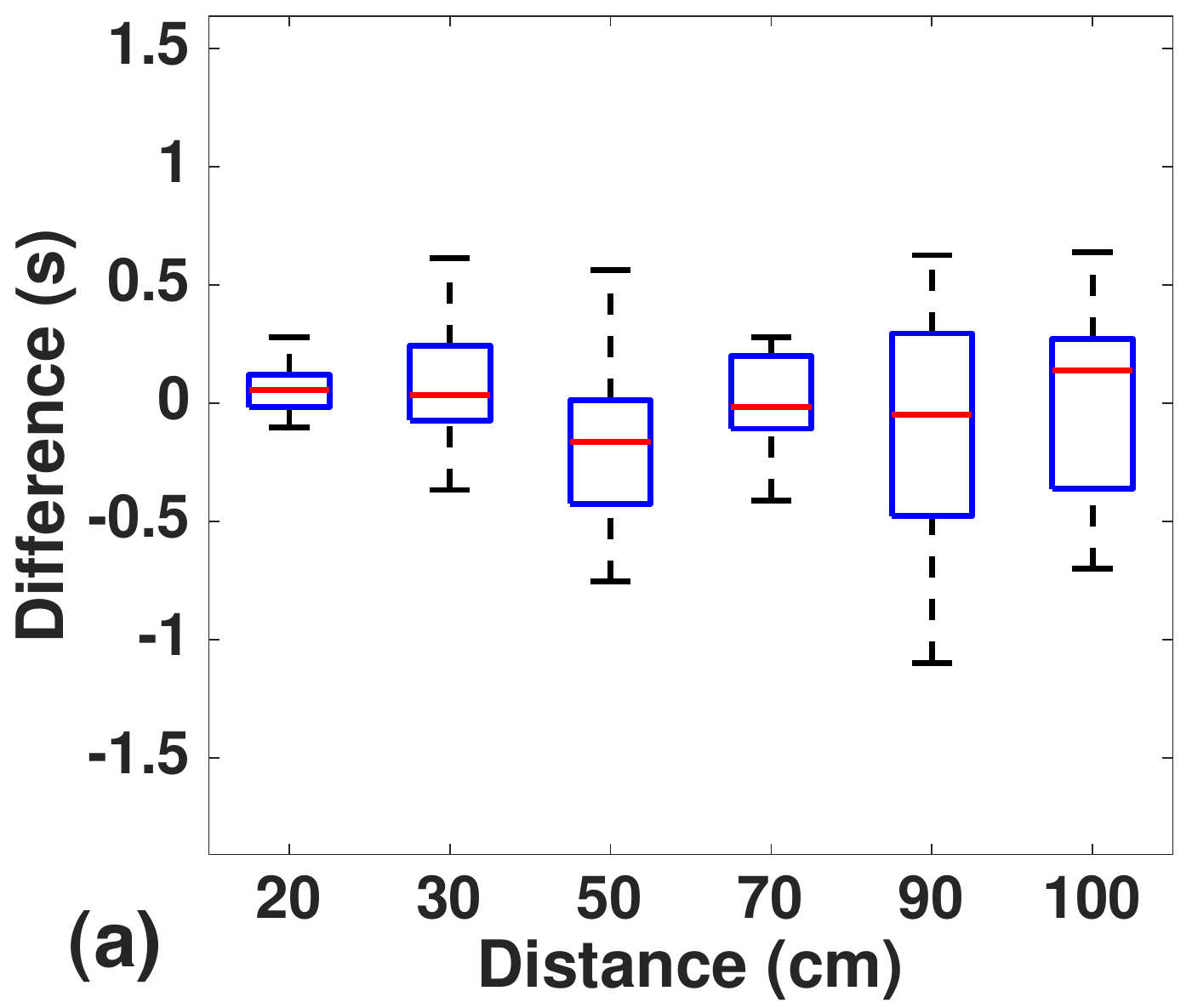}
	\includegraphics[width = 0.46\columnwidth]{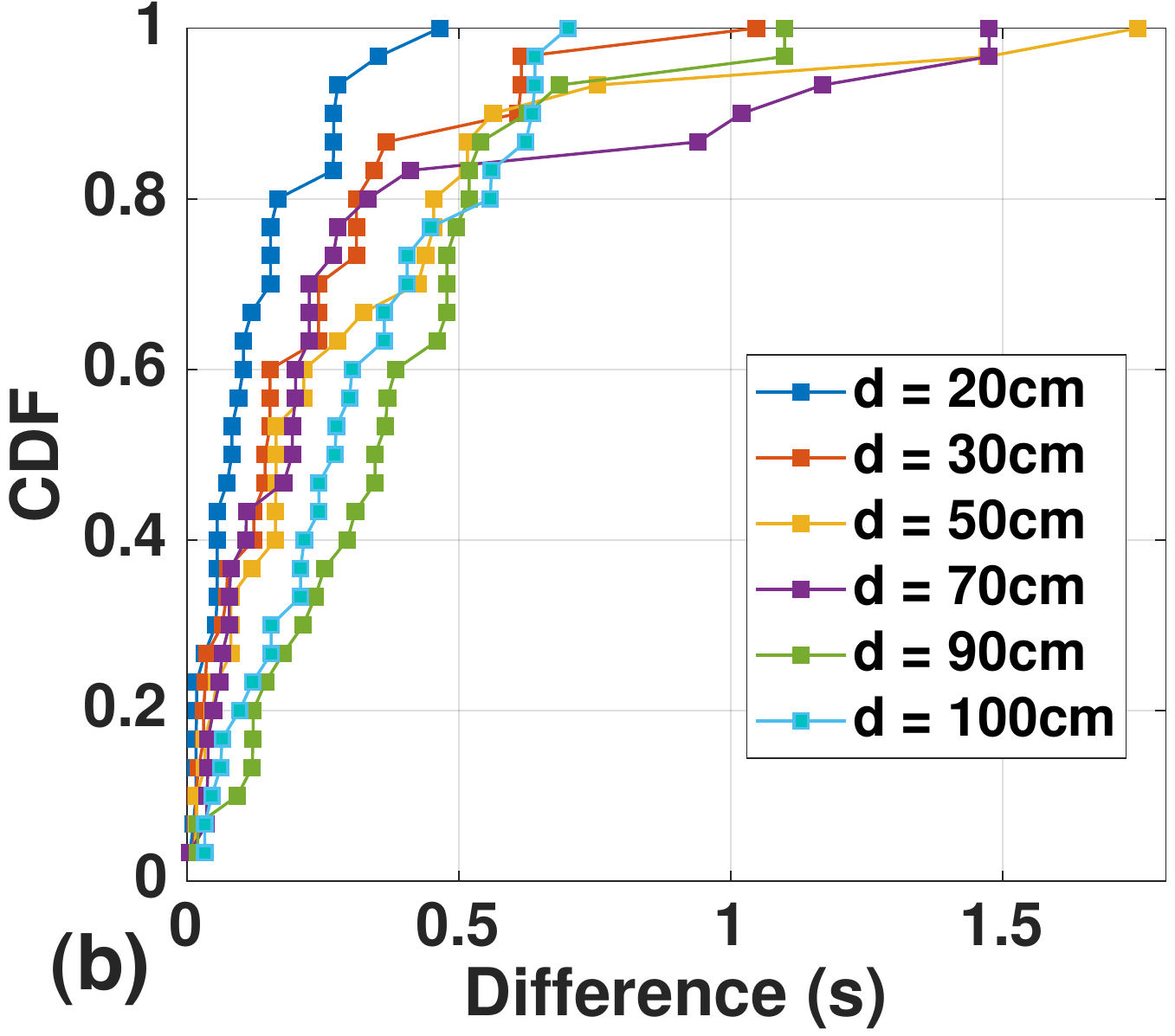}
	\caption{(a) Respiration cycle duration comparison when the human subject sits at different distances from the sensor, outliers are not shown; (b) cumulative density functions (CDF) of the duration errors.}
	\label{fig:remote_rr_dist}
\end{figure}

Fig.~\ref{fig:remote_rr_dist}(a) shows that timing differences increase with $d$ since the signal strength decays as $1/d^2$, leading to larger duration estimation errors. Cumulative density functions (CDFs) of the timing differences are shown in Fig.~\ref{fig:remote_rr_dist}(b) for different values of $d$. Considering the typical RC cycle duration of $\sim$5~sec, these results suggest that the EPS can be used to reliably monitor RR for distances up to $d=100$~cm.

\begin{figure}[htbp]
	\centering
	\includegraphics[width = 0.40\columnwidth]{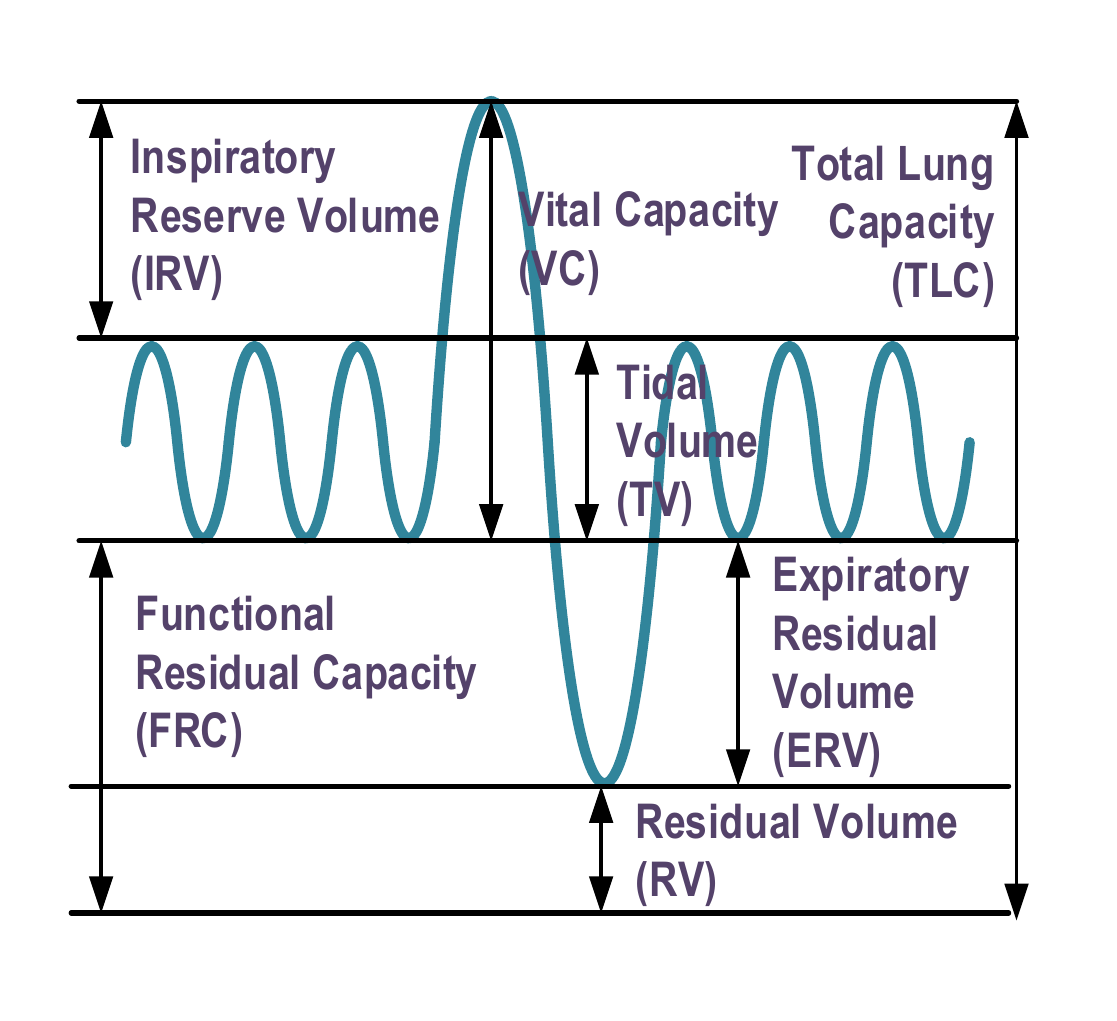}
	\includegraphics[width = 0.58\columnwidth]{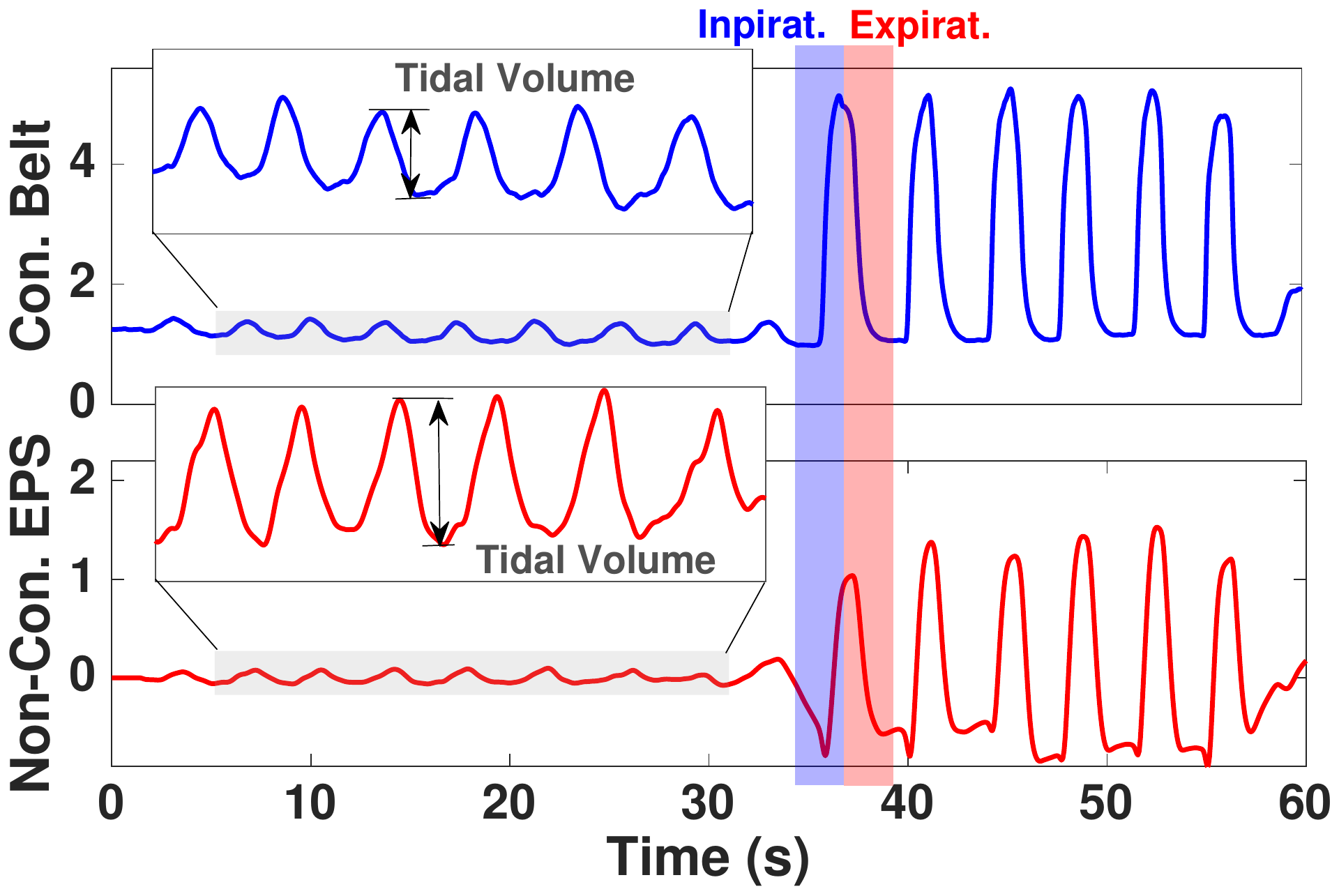}
	\caption{(a) Parameter definitions for a typical spirometry measurement; and (b) their corresponding transit waveforms during shallow breathing (left) and deep forced breathing (right).}
	\label{fig:remote_rr_fev_m}
\end{figure}

Spirometry is a common pulmonary function test (PFT) for assessing breathing patterns that identify conditions such as asthma or pulmonary fibrosis~\cite{miller2005standardisation}. Fig.~\ref{fig:remote_rr_fev_m}(a) shows a typical spirometry test and its parameter definitions, while Fig.~\ref{fig:remote_rr_fev_m}(b) compares the measured respiratory waveforms for shallow breathing and forced inspiration/expiration using the contact belt sensor and the EPS; the two are in good agreement.

\begin{figure}[htbp]
	\centering
	\includegraphics[width = 0.47\columnwidth]{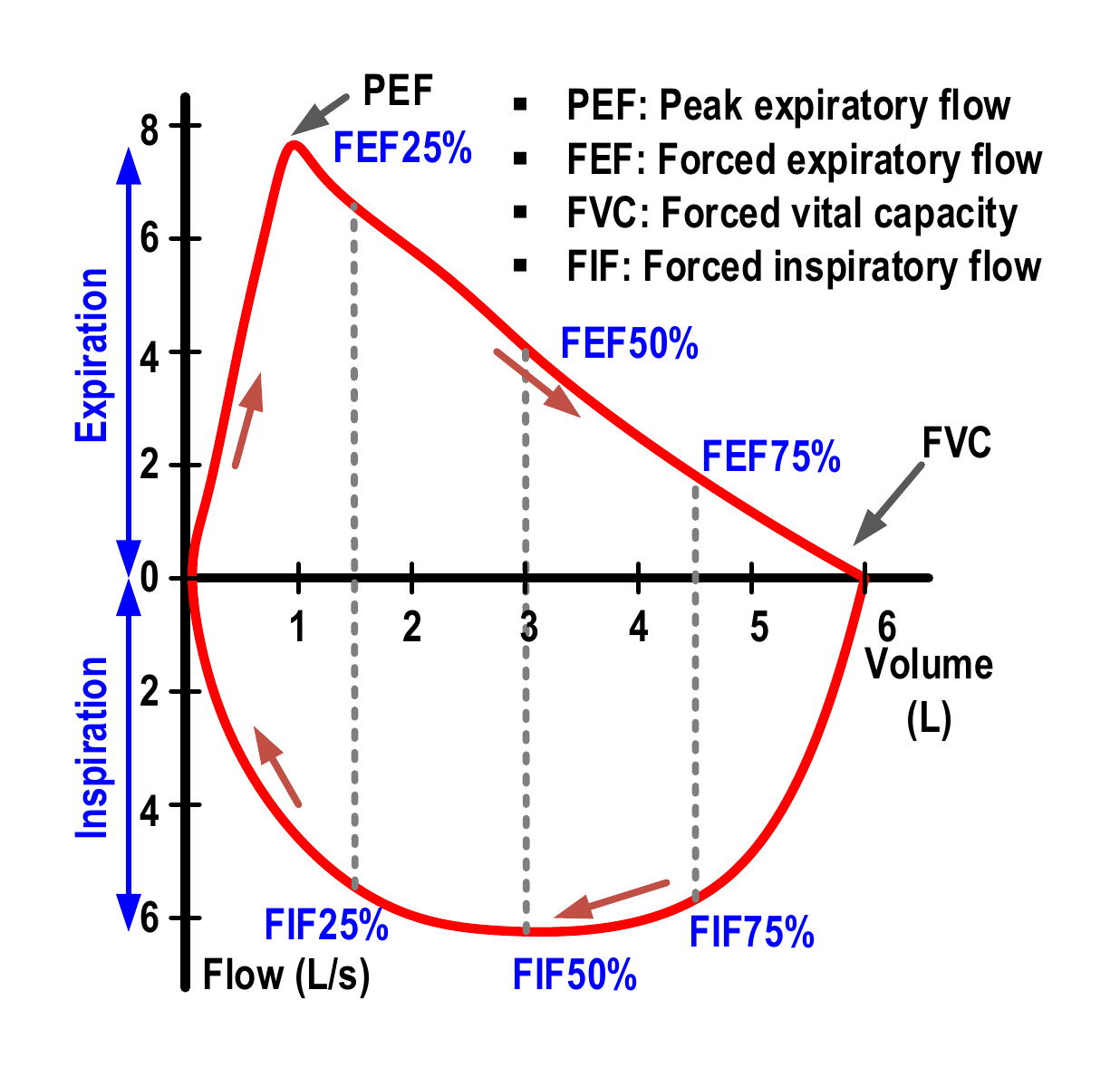}
	\includegraphics[width = 0.46\columnwidth]{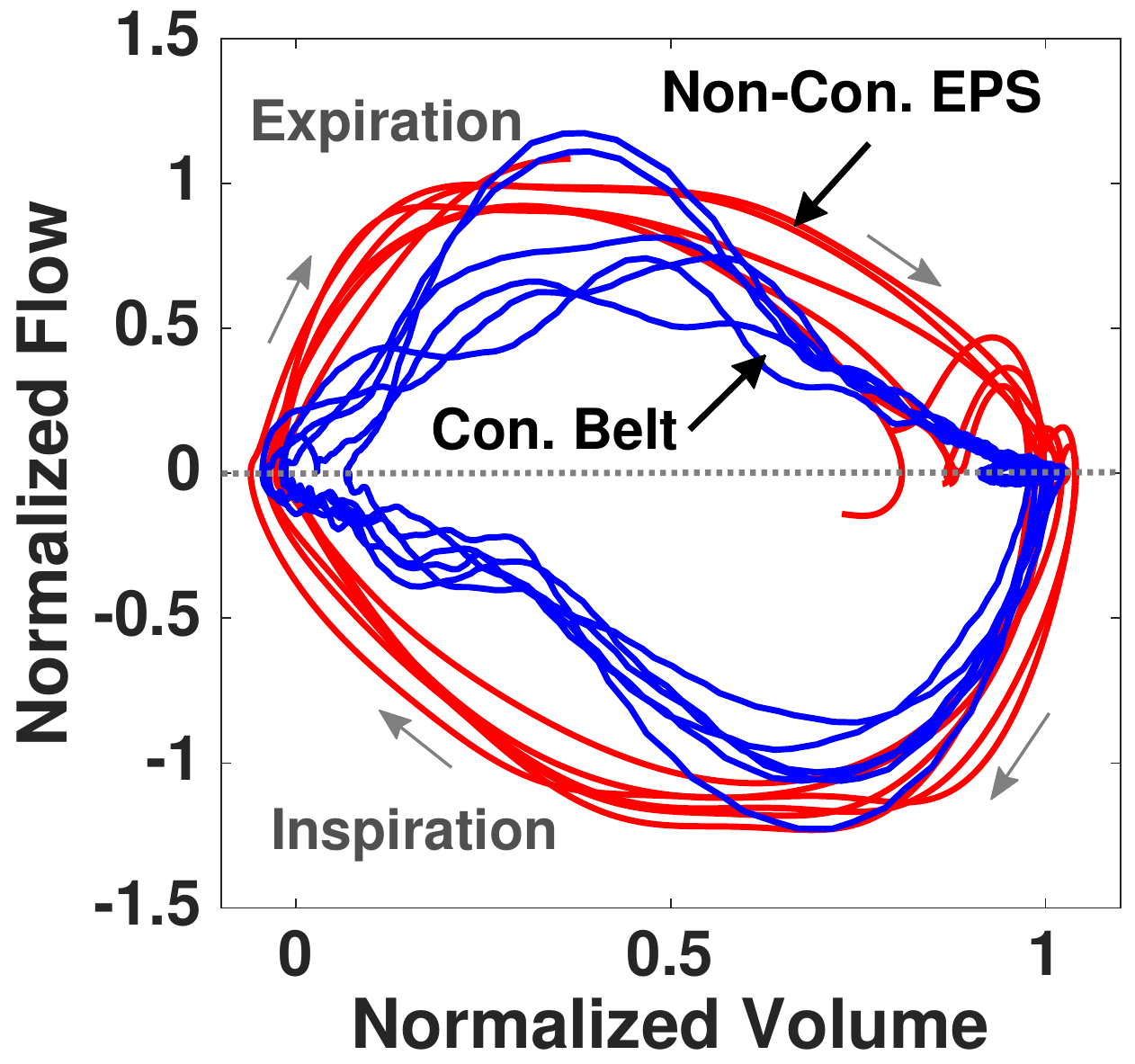}
	\caption{(a) Flow-volume (F-V) loop showing a normal FVC maneuver; and (b) its estimation using the contact belt sensor and the non-contact EPS.}
	\label{fig:remote_rr_fev}
\end{figure}

Fig.~\ref{fig:remote_rr_fev}(a) shows the flow-volume (F-V) loop during a successful forced vital capacity (FVC) maneuver, as extracted from Fig.~\ref{fig:remote_rr_fev_m}(a). Positive and negative ``flow'' values represent expiration and inspiration, respectively, while the ``volume'' axis represents volume in the spirometer. The flow trace moves clockwise, starting at the FVC point during inspiration, and rapidly mounts to a peak during expiration. Forced expiratory flow (FEF) is the flow during the middle portion of a forced expiration, and $25-75\%$ FEF appears to be a sensitive parameter for detecting obstructive small airway disease~\cite{marseglia2007role}. Instead of using a traditional spirometer for validation, here we demonstrate that the contact sensor and our EPS provide similar F-V loops and thus FEF values, as shown in Fig.~\ref{fig:remote_rr_fev}(b).

\subsection{Non-contact Electroencephalogram (EEG) Sensing}
\begin{figure}[htbp]
	\centering
	\includegraphics[width = 1\columnwidth]{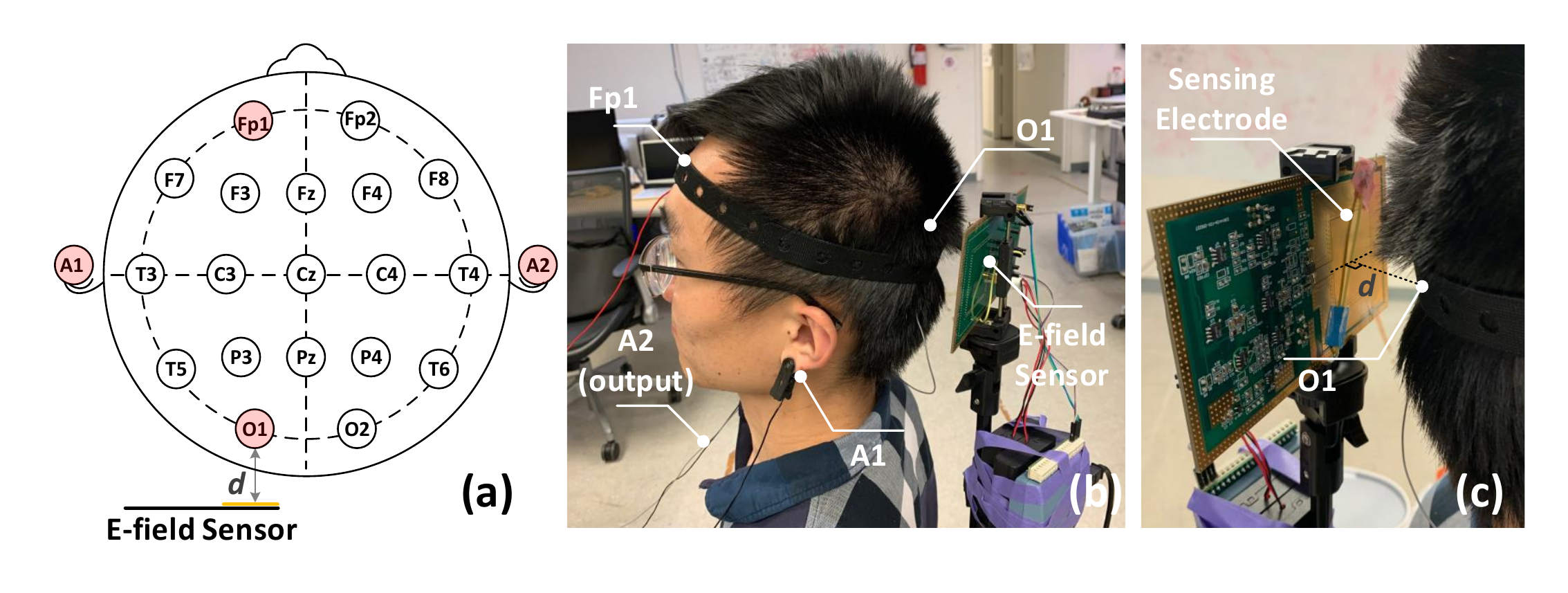}
	\caption{(a) Sketch view of 10-20 EEG electrode placement and E-field sensor location, (b) the corresponding experimental setup, where only 4 electrodes are implemented, and (c) zoom-in view for contact electrode (O1) sensing and non-contact E-field sensing.}
	\label{fig:remote_eeg_setup}
\end{figure}

\begin{figure}[htbp]
	\centering
	\includegraphics[width = 0.49\columnwidth]{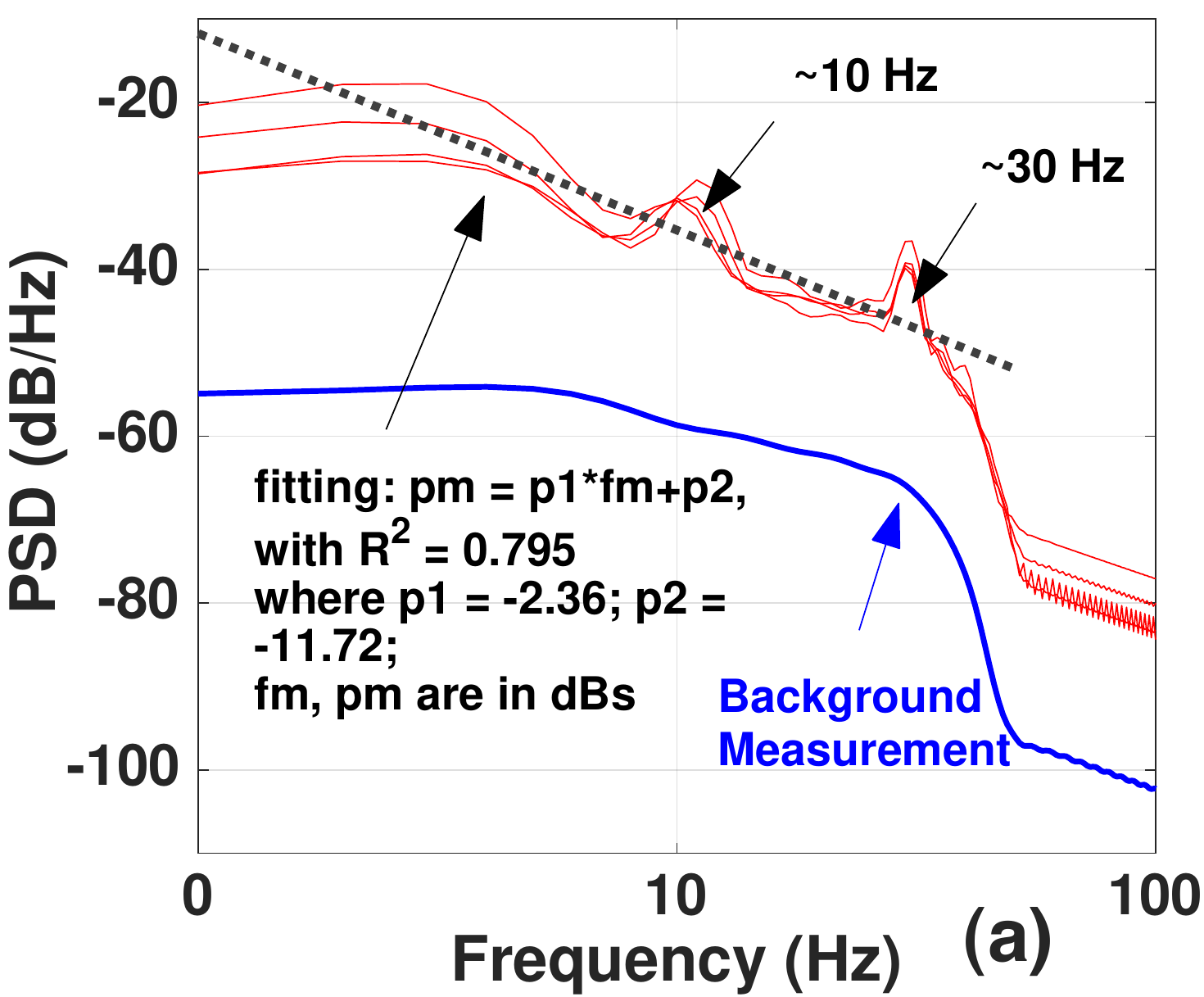}
	\includegraphics[width = 0.48\columnwidth]{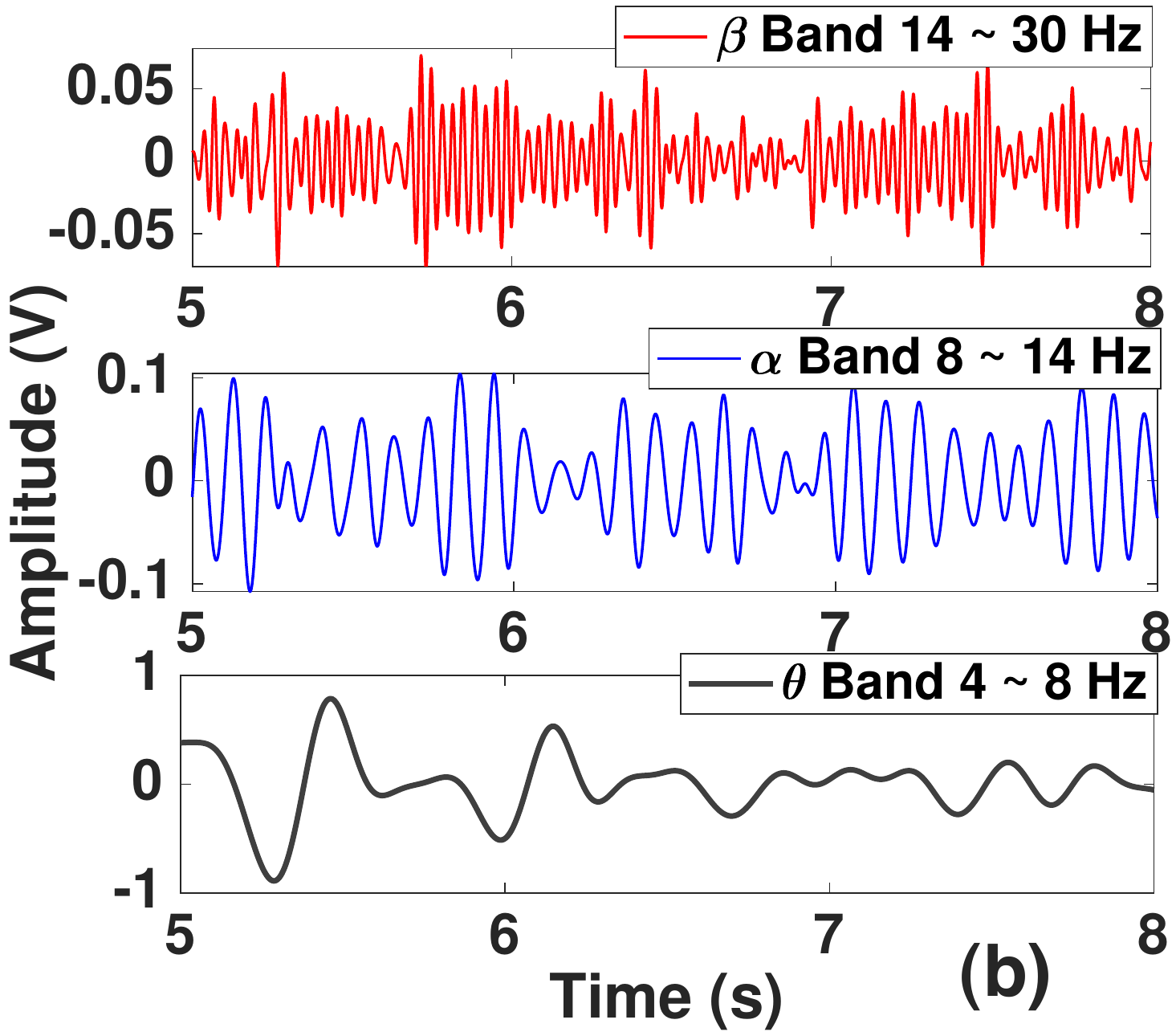}
	\caption{(a) PSD of measured EEG waveforms using the EPS, with the subject’s forehead $\sim$5~cm away from the sensing electrode, and (b) the filtered EEG waveforms within different filtering bands.}
	\label{fig:remote_eeg_sp}
\end{figure}

This subsection describes non-contact EEG measurement results using the ECG channel of the EPS. The experimental setup is shown in Fig.~\ref{fig:remote_eeg_setup}, with the back of the subject's head (occipital lobe) being $d\approx 5$~cm away from the sensing electrode. Fig.~\ref{fig:remote_eeg_sp}(a) shows power spectra for 4 typical non-contact EEG recordings from an awake subject using Welch’s power spectral density estimate, as calculated using MATLAB; this plot shows the well-known scale-invariance properties of EEG spectra, i.e., a power-law spectrum of the form $1/f^{\gamma}$, where $\gamma$ is the fitting slope ($\gamma = 2.36$ in this case). In addition, peaks within the $\beta$ band (at $\sim$30~Hz) and $\alpha$ band (at $\sim$10~Hz) can be clearly observed; these peaks are also visible in the filtered time domain waveforms, as shown in Fig.~\ref{fig:remote_eeg_sp}(b).

\begin{figure}[htbp]
	\centering
	\includegraphics[width=0.48\columnwidth]{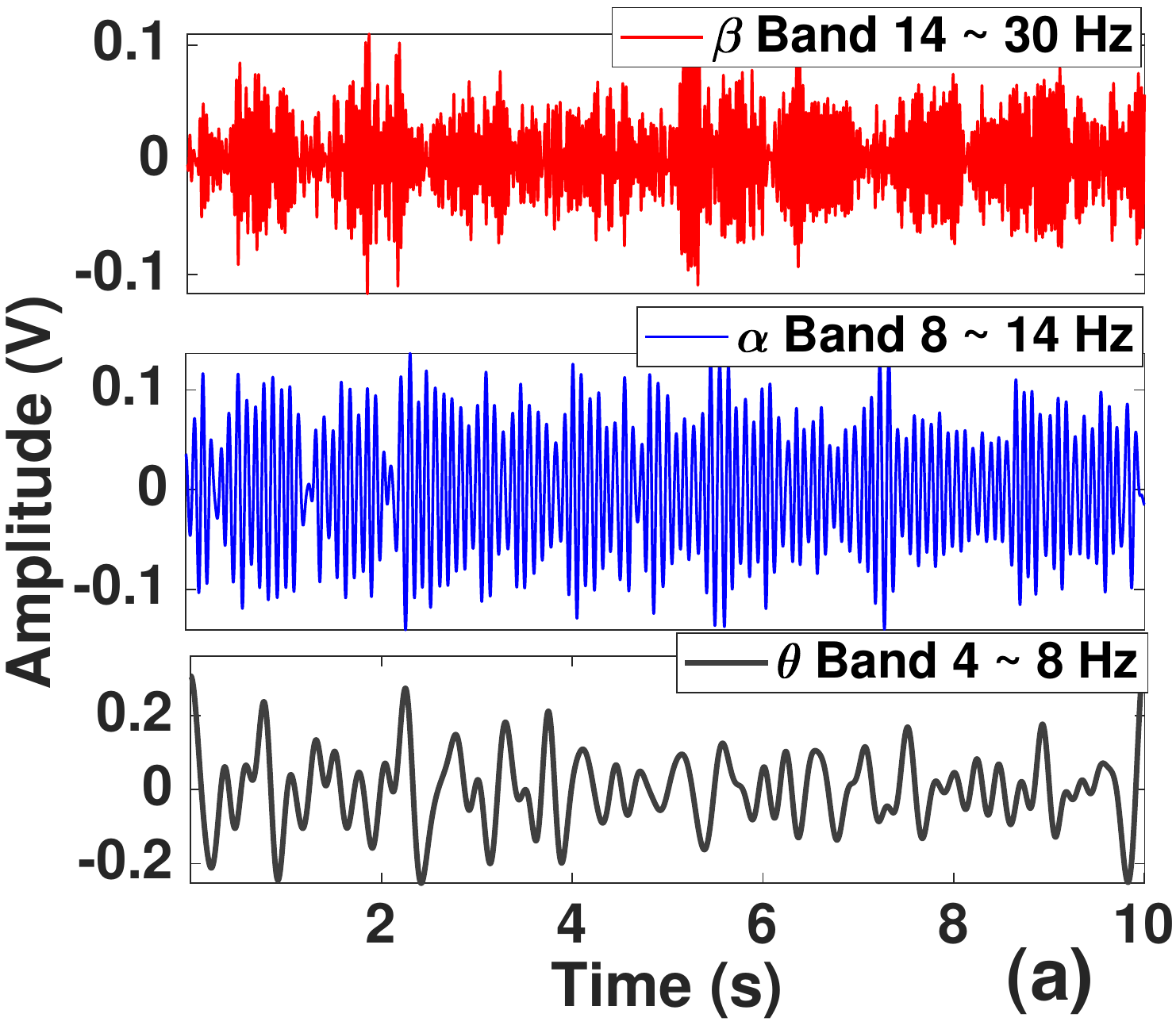}
	\includegraphics[width=0.48\columnwidth]{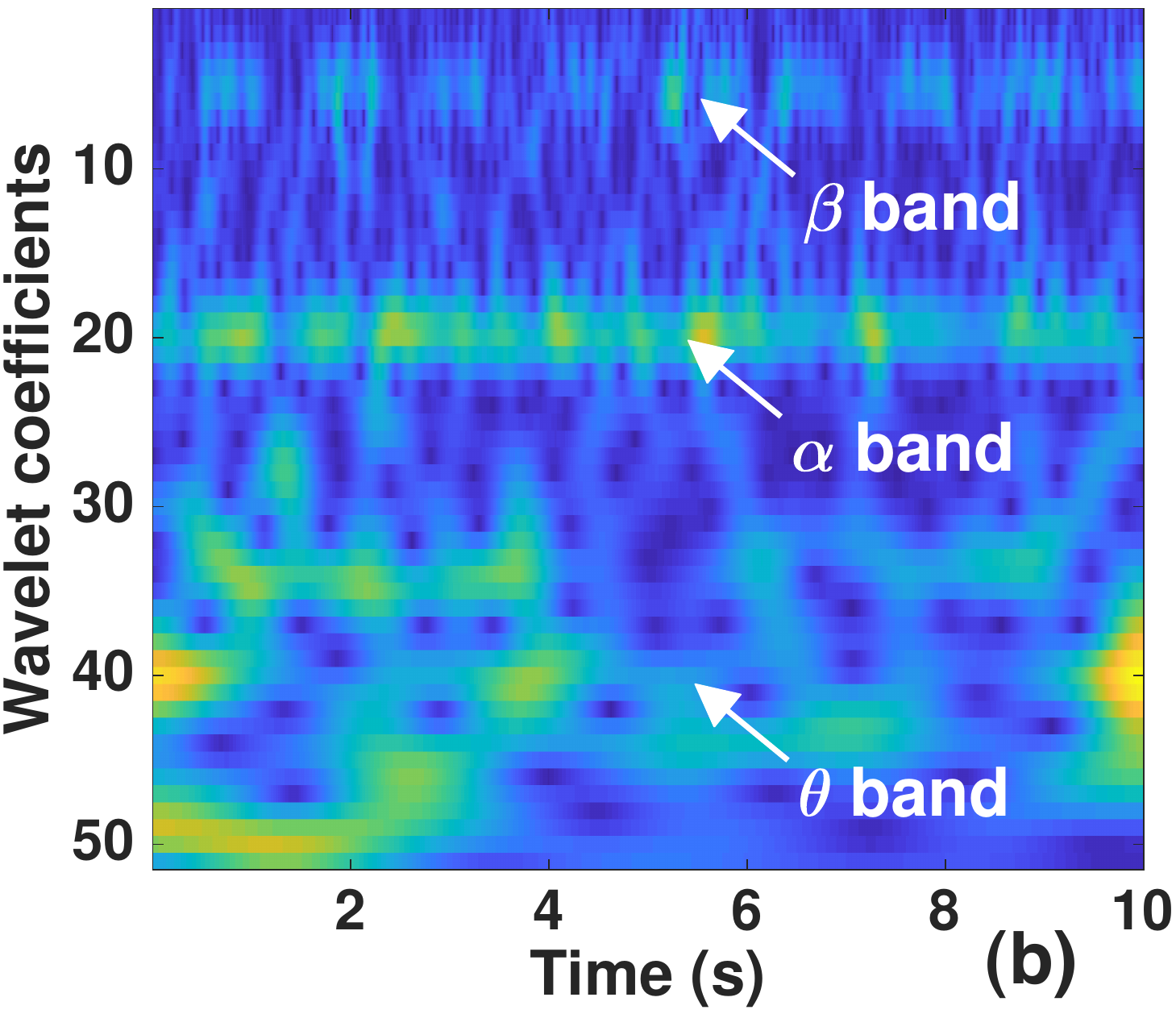}
	\caption{(a) Typical recording of EEG waveforms at different filtering bands, and (b) their corresponding power spectra, estimated using a CWT.}
	\label{fig:remote_eeg_trans}
\end{figure}

\begin{figure}[htbp]
	\centering
	\includegraphics[width=0.475\columnwidth]{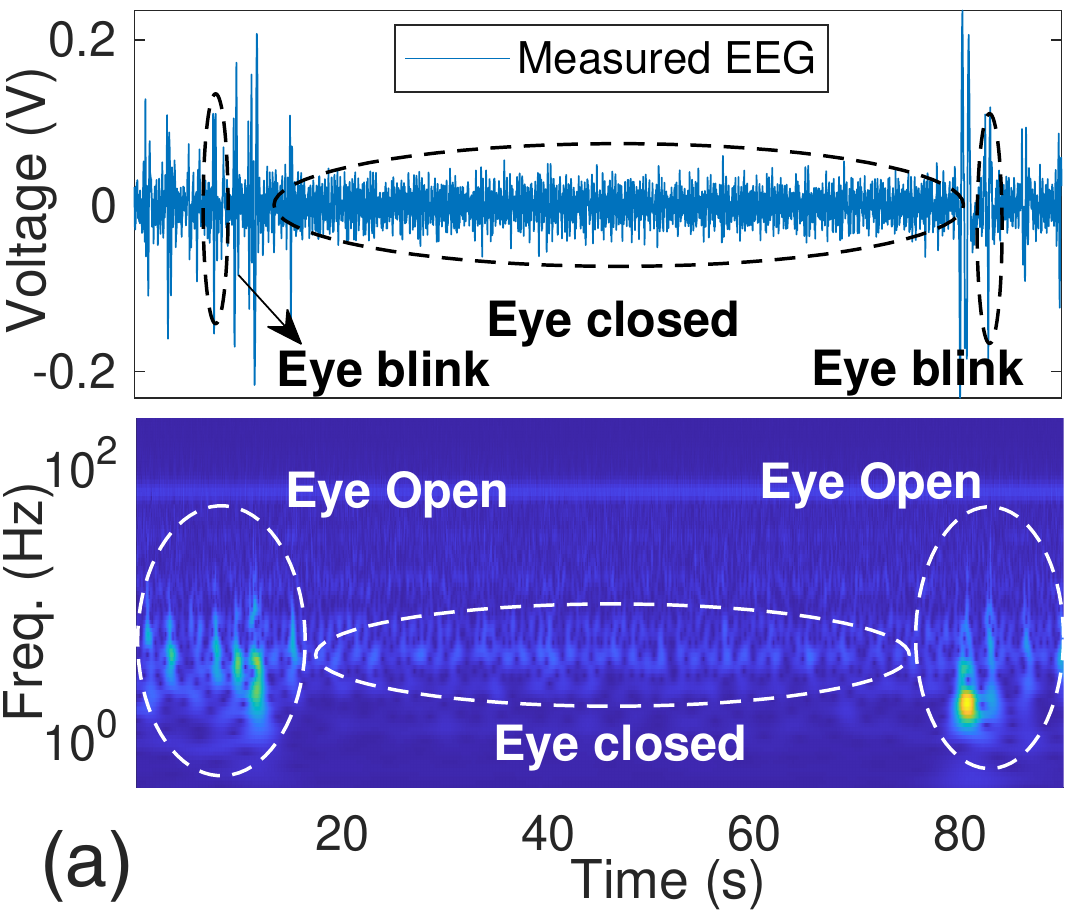}
	\includegraphics[width=0.505\columnwidth]{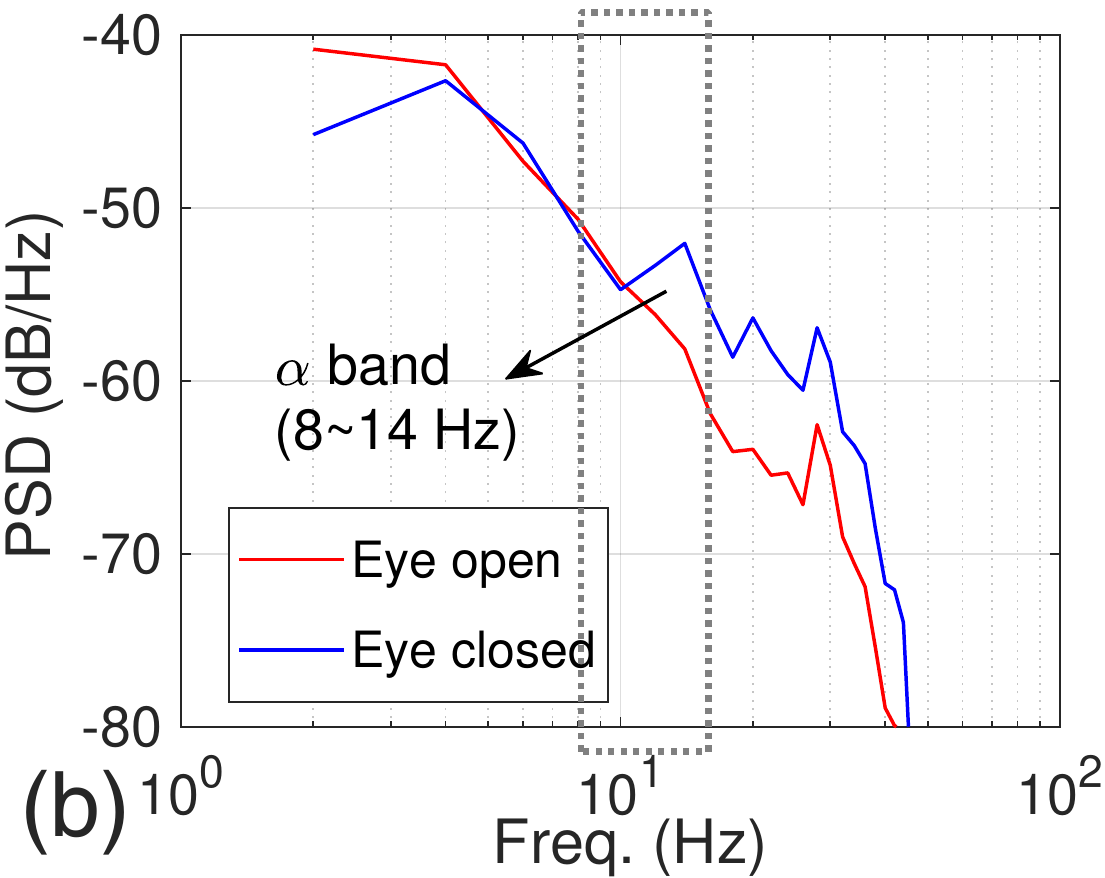}
	\caption{{(a) Typical time- and frequency-domain EEG waveforms measured when the eyes are closed (in between eye blinks), and (b) the corresponding power spectra using Welch's PSD estimate.}}
	\label{fig:eeg_eye}
\end{figure}

Fig.~\ref{fig:remote_eeg_trans}(a) illustrates another example of EEG waveforms measured from an awake subject using the EPS; the $\beta$ band (14-30~Hz), $\alpha$ band (8-14~Hz), and $\theta$ band (4-8~Hz) were extracted using a filter bank. Fig.~\ref{fig:remote_eeg_trans}(b) shows the corresponding power spectrum, as estimated using a CWT: peaks are visible in these three bands, as expected. In additional experiments, as shown in Fig.~\ref{fig:eeg_eye}, we compare the EEG spectra measured with the subject's eyes open and closed, respectively. A significant increase in $\alpha$ band activity is observed with the eyes closed, as expected. In addition, artifacts from eye blinks, which are common in EEG recordings~\cite{hoffmann2008correction}, are also clearly observed.

\subsection{Physiological Monitoring {in Sleep-Like Postures}}
\begin{figure}[htbp]
	\centering
	\includegraphics[width=0.88\columnwidth]{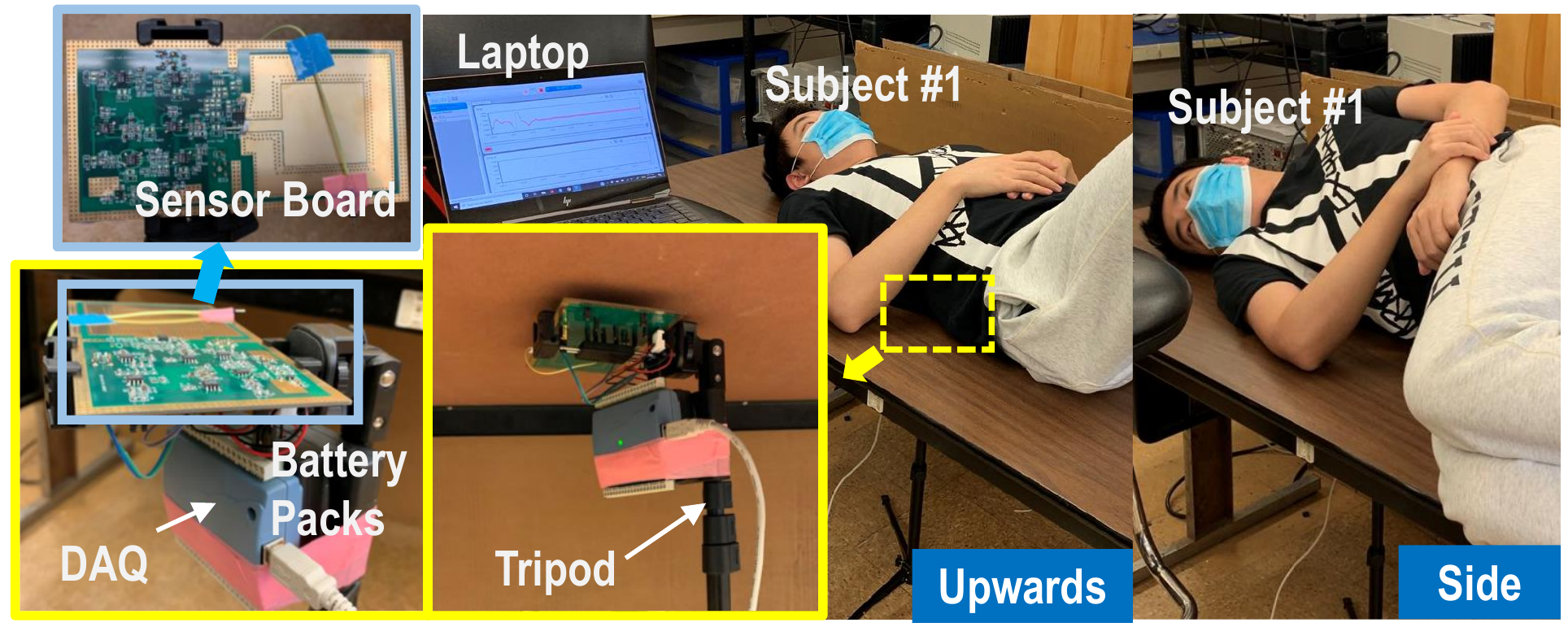}
	\caption{Experimental setup for non-contact sensing of cardiopulmonary signals (RC and ECG) in sleep-like postures.}
	\label{fig:remote_Setup2}
\end{figure}

\begin{figure}[htbp]
	\centering
	\includegraphics[width=0.48\columnwidth]{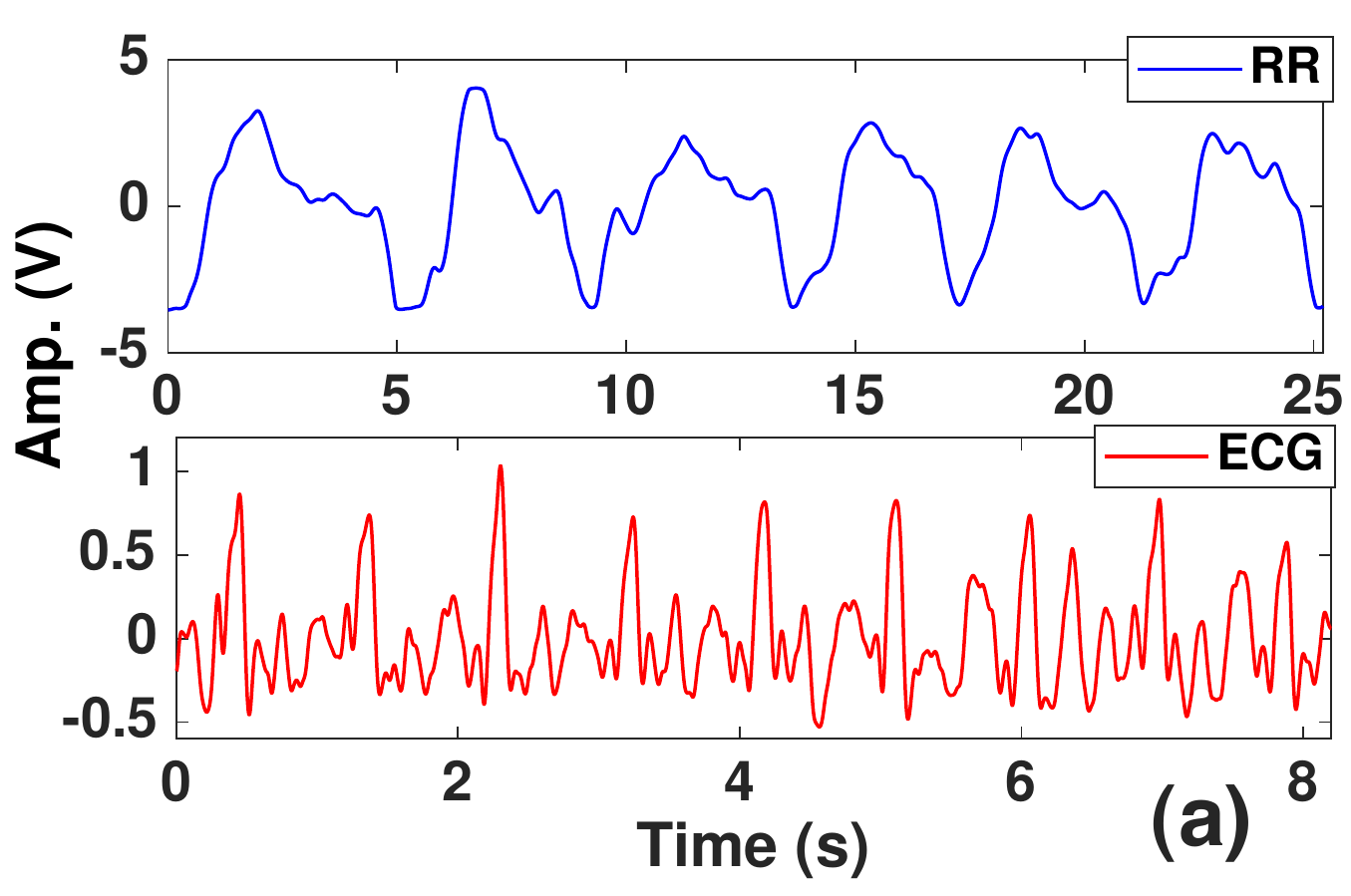}
	\includegraphics[width=0.49\columnwidth]{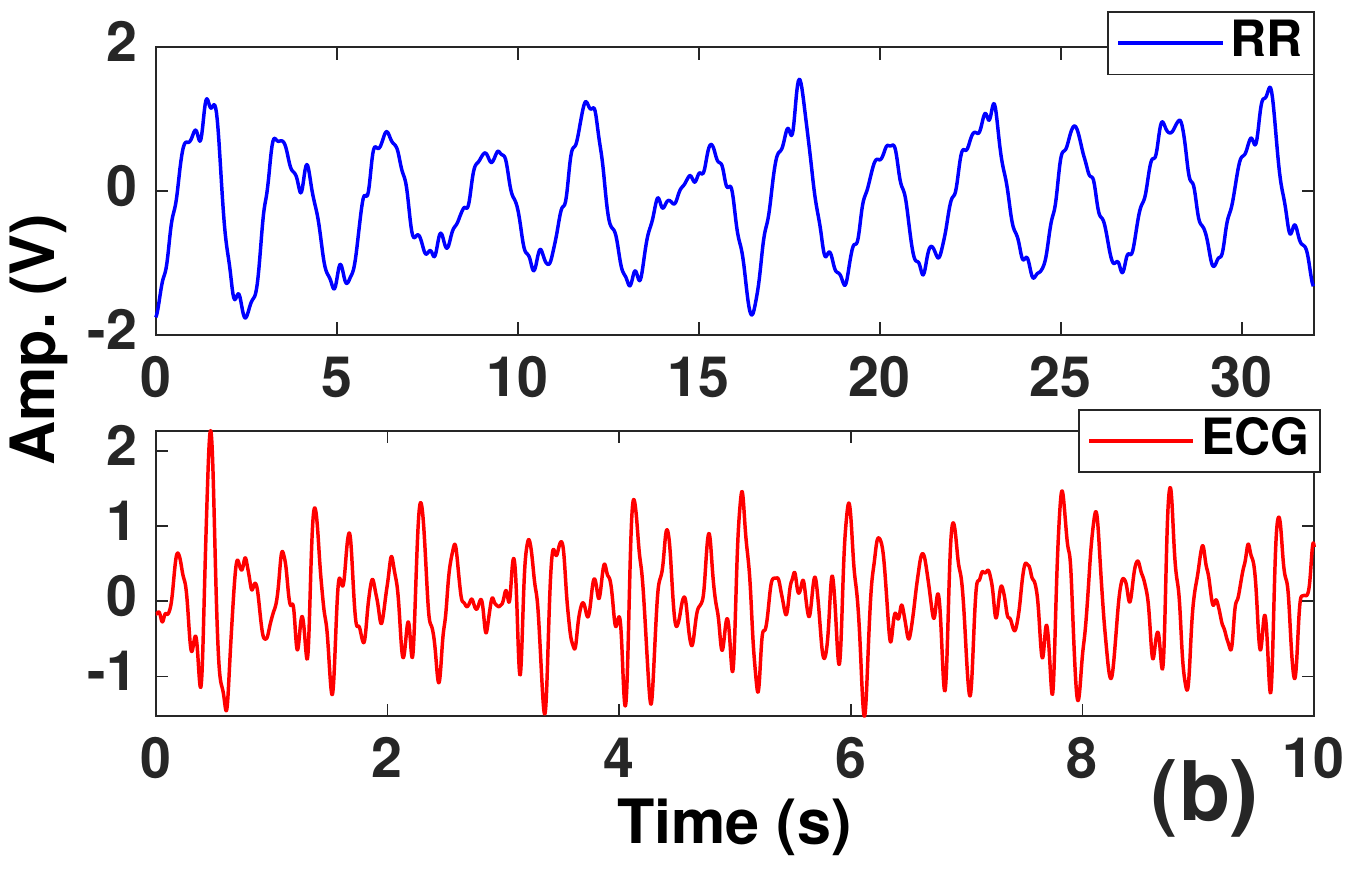}
	\caption{Non-contact sensing of RC and ECG signals during different sleep-like postures: (a) upward-facing, and (b) side-facing.}
	\label{fig:remote_up_side}
\end{figure}

Non-contact cardiovascular and EEG sensing in home settings has the potential to reduce the cost of sleep studies compared to conventional data collection in a sleep lab~\cite{flemons2003home}. Fig.~\ref{fig:remote_Setup2} illustrates a typical example of non-contact vital signal monitoring during sleep-like conditions using our custom EPS. The EPS was placed $\sim$2~cm under a wooden table ($3$~cm thick) and fixed on a tripod, while the subject lay down on the table with different postures: upward- and side-facing, respectively. The measured RC and ECG waveforms are shown in Fig.~\ref{fig:remote_up_side}. Both signals are clearly observed during different sleep-like postures. Note that the measured QRS-complex features in the side-facing case (a) die down faster than that in the upward-facing case (b); this is because the heart-vector projection is further attenuated in the first case. Adding another EPS near the head (e.g., mounted on a headboard) would allow EEG to be simultaneously monitored for sleep studies.

\subsection{Wireless Measurements}
The EPS nodes were integrated with off-the-shelf wireless modules to realize a self-contained sensing solution. Wireless links were set up using a low-power API and protocol (known as EasyLink) supported by the chosen MCU (CC2650, Texas Instruments). Each sensor node was configured to simultaneously sample two channels at 1~kS/s and transmit 100 samples per data packet to a remote base station. {Relevant signal features (e.g., HR and RR) can be extracted from the raw data prior to transmission to reduce the wireless data rate and power consumption, but was not implemented in this work.} The base station time-stamps the received data using an on-board GPS module and then communicates with a secondary MCU board (Teensy 3.6) to save the time-stamped data to a SD card. In addition, the base station can live-steam the measured waveforms (100 samples per frame) on a built-in screen. Wireless ECG and RC recording at distances of several meters were successfully demonstrated using this setup.

\subsection{Comparison with Prior Work}
Table~\ref{table:remote_table_com} compares our work with recent literature on non-contact detection of cardiopulmonary signals and EEG. Earlier non-contact sensing systems have relied on differential voltage sensing using instrumentation amplifiers (IAs) with double electrodes~\cite{harland2001electric, TJS2007,Yu2014}, whereas our proposed system relies on current sensing using a TIA with a single electrode to obtain very high sensitivity. Meanwhile, most of the non-contact monitoring systems~\cite{Yu2014,TJS2007} have limited sensing range (only up to mm). Radar-based sensors (active sensing) such as~\cite{Adib2015} have the longest detection range, but are limited to sensing cardiac and respiratory rates (HR and RR, respectively). Also the detection median accuracies for HR and RR in ~\cite{Adib2015} are 98$\%$ and 99$\%$, respectively, and are degraded along distance and sensor orientation. Normally, there is no accuracy issues in passive non-contact sensing methods since the timing error is negligible. In~\cite{Adib2015} the system is quite bulky and power hungry compared with other work in the Table~\ref{table:remote_table_com}. Our proposed work can simultaneously monitoring ECG and RC, while others can only measure ECG at one time. Thus, our work combines the advantages of relatively long sensing range with suitability for multi-modal sensing (ECG, RC, and EEG).

\begin{table*}[htbp]
\centering
	\caption{Comparison with prior work on non-contact sensing of cardiopulmonary signals}
	\begin{tabularx}{0.98\linewidth}{ c c c l l l c l c}
		\hline
		\hline
		\multirow{2}{*}{Ref.} & \multirow{2}{*}{Size/Power} & \multirow{2}{*}{Dis.} & \multicolumn{3}{l}{Circuit Design} & \multicolumn{2}{l}{Characteristics} & \multirow{2}{*}{Applications} \\ \cline{4-8}
		&  &                        & AFE           & Bias                & Strategies                         & $R_{IN}$                & Noise          &                                      \\ \hline
		2002~\cite{harland2001electric}   & cm/mW               & 0.4 m                 & EPS-IA     & Guard    & AS; DE         & $\sim10^{15}\Omega$           & 4~$\mu$V/$\sqrt{\mathrm{Hz}}^{*}$   & ECG       \\
		2007~\cite{TJS2007}              & mm/mW    & \textless{}0.3 cm     & IA-based            & Reset               & AS; FL    & -                  & 2~$\mu$V$_{rms}^{\dagger}$            & ECG/EEG                      \\
		2014~\cite{Yu2014}              & mm/mW    & \textless{}0.3 cm      & IA-based            & Insulation, leakage & AS; DE &  $>10^{15}\Omega$    & 3.8~$\mu$V            & ECG/EEG                      \\
		2015~\cite{Adib2015}            & m/W      & 6 m                   & -             & -                   & Radar-based                        & -                  & -              & HR/RR                   \\
		This work             & cm/mW     & \textbf{0.5/1/0.05 m}                 & TIA-based     & Guard    & AS; SE & -                  & 0.07~fA/$\sqrt{\mathrm{Hz}}$ @ 25$^{\circ}$C          & \textbf{ECG/RC/EEG}               \\ \hline\hline
	\end{tabularx}
	\vskip 0.5ex
	{\textbf{AS}:~Active shielding;~\textbf{FL}:~Feedback loop;~\textbf{SE}:~Single electrode;~\textbf{DE}:~Double electrode;~\textbf{IA}:~Instrumentation amplifier;~$*$: @1~Hz;~$\dagger$: 1 - 100 Hz. \par}
	\label{table:remote_table_com}
\end{table*}

\section{Conclusion}
\label{sec:conclusion}
This paper has proposed passive non-contact $E$-field monitoring for human-aware smart environments. Specifically, we have successfully demonstrated non-contact sensing of ECG and RC up to distances of 0.5~m and 1.0~m, respectively in noisy unshielded environments by using a custom EPS. We have also demonstrated EEG sensing at distances of $\sim$5~cm from the forehead. Applications of non-contact EPS to coarse estimation of spirometry parameters and sleep monitoring were also proposed. Our work outperforms the current state of art on non-contact ECG/RC detection, for which the sensing range is only up to $\sim0.3$~m in well-shielded rooms. Future work will focus on signal processing and machine learning methods to extract relevant biomarkers from the acquired data.





\section*{Acknowledgment}
The authors would like to thank Mohammad S. Islam and Jifu Liang for assistance with the experiments. 

\ifCLASSOPTIONcaptionsoff
  \newpage
\fi


\bibliographystyle{IEEEtran}
\bibliography{IEEEabrv,references}

%








\end{document}